\documentclass[10pt]{amsart}
\pdfoutput=1
\usepackage{amsmath,amsthm,amssymb,latexsym} 
\usepackage{fullpage} 
\usepackage{upgreek} 
\usepackage{enumerate} 
\usepackage{graphicx} 
\usepackage{color} 
\usepackage[all]{xy} 
\usepackage{accents} 
\usepackage[pdftex,bookmarks,colorlinks,breaklinks]{hyperref}  
\definecolor{dullmagenta}{rgb}{0.4,0,0.4}   
\definecolor{darkblue}{rgb}{0,0,0.4}
\hypersetup{linkcolor=red,citecolor=blue,filecolor=dullmagenta,urlcolor=darkblue} 

\newtheorem{theorem}{Theorem}[section]
\newtheorem{lemma}[theorem]{Lemma}

\theoremstyle{definition}

\newtheorem{assumption}[theorem]{Assumption}

\theoremstyle{remark}
\newtheorem{remark}[theorem]{Remark}

\numberwithin{equation}{section}

\begin{document}

\title[Discrete Schlesinger Transformations]{Discrete Schlesinger Transformations, their Hamiltonian Formulation, 
and Difference Painlev\'e Equations}

\author{Anton Dzhamay}
\address{School of Mathematical Sciences\\ 
The University of Northern Colorado\\ 
Campus Box 122\\ 
501 20th Street\\ 
Greeley, CO 80639, USA}
\email{\href{mailto:adzham@unco.edu}{\texttt{adzham@unco.edu}}}

\author{Hidetaka Sakai}
\address{Graduate School of Mathematical Sciences\\ 
The University of Tokyo\\
3--8--1 Komaba Meguro-ku\\ 
Tokyo 153-8914, Japan}
\email{\href{mailto:sakai@ms.u-tokyo.ac.jp}{\texttt{sakai@ms.u-tokyo.ac.jp}}}

\author{Tomoyuki Takenawa}
\address{Faculty of Marine Technology\\
Tokyo University of Marine Science and Technology\\
2-1-6 Etchu-jima, Koto-ku\\ 
Tokyo, 135-8533, Japan}
\email{\href{mailto:takenawa@kaiyodai.ac.jp}{\texttt{takenawa@kaiyodai.ac.jp}}}


\date{}

\begin{abstract}
	Schlesinger transformations are algebraic transformations of 
	a Fuchsian system that preserve its monodromy representation
	and act on the characteristic indices of the system by integral 
	shifts. One of the important reasons to study such transformations  
 	is the relationship between Schlesinger transformations and  
	discrete Painlev\'e equations; this is also the main theme 
	behind our work. We derive \emph{discrete Schlesinger evolution equations}
	describing discrete dynamical systems generated by elementary
	Schlesinger transformations and give their discrete Hamiltonian 
	description w.r.t.~the standard symplectic 
	structure on the space of Fuchsian systems. 
	As an application, we compute explicitly 
	two examples of reduction from Schlesinger transformations to difference 
	Painlev\'e equations. The first example, d-$P\big(D_{4}^{(1)}\big)$ 
	(or difference Painlev\'e V), corresponds to B\"acklund transformations
	for continuous $P_{\text{VI}}$. The second example, d-$P\big(A_{2}^{(1)*}\big)$
	(with the symmetry group $E_{6}^{(1)}$), is purely discrete.  
	We also describe the role played by the geometry
	of the Okamoto space of initial conditions 
	in comparing 
	different equations of the same type.
\end{abstract}

\maketitle

\section{Introduction} 
\label{sec:introduction}

In the theory of ordinary linear differential equations on a complex domain, and
in particular in the theory of Fuchsian systems, an important characteristic 
of the equation is its configuration of singularities and the characteristic
indices at these singular points. Associated to this data is the notion of
the monodromy representation of the equation. Roughly speaking, this representation 
describes how the
fundamental solution matrix of the equation changes under analytic continuation along 
the closed paths around the singular points and it gives a significant 
insight into the global behavior of solutions of the equation. 

The fruitful idea
of deforming the equation by moving the location of critical points into
different configurations, or by changing the characteristic indices, 
without changing its monodromy representation
goes back
to B.~Riemann, but the actual foundations of this theory 
of \emph{isomonodromic deformations} in the Fuchsian case were laid down in the works of 
R. Fuchs \cite{Fuc:1907:ULHDZOMDIEGWSS},
L. Schlesinger \cite{Sch:1912:UEKDBOFKP}, and 
R. Garnier \cite{Gar:1926:SDPDRPLSDLDSO}.
An extension of the theory to 
the non-Fuchsian case 
was done much later in the series of papers by M.~Jimbo, T.~Miwa, and
K.~Ueno,
\cite{JimMiwUen:1981:MPDLODEWRCGTF,JimMiw:1982:MPDLODEWRC,JimMiw:1981:MPDLODEWRC}.
At present, the theory of isomonodromic deformations is a very active 
research field with deep connections to other areas such as
the theory of integrable systems, classical theory of differential equations, and
differential and algebraic geometry.

It is important to distinguish between continuous
and discrete isomonodromic deformations. In the continuous case 
deformation parameters are
locations of singular points of the system. 
The resulting isomonodromic flows on the space of coefficients
of the Fuchsian system are given by Schlesinger \emph{equations}.
In \cite{JimMiwMorSat:1980:DMIBFPT}  M.~Jimbo, T.~Miwa, Y.~M\^ori, and M.~Sato 
showed that these equations, when considered as a dynamical 
system on the decomposition space of the Fuchsian system, 
can be written in the Hamiltonian form. For a 
$2\times 2$ Fuchsian system with $n=4$ poles, Schlesinger equations
reduce to the most general of the Painlev\'e equations, Painlev\'e VI (this result
goes back to the work of R.~Fuchs, \cite{Fuc:1907:ULHDZOMDIEGWSS}).
Painlev\'e equations are also Hamiltonian, due to the work of Okamoto,
\cite{Oka:1980:PHAWPEIDESBPH}, and in fact
the reduction from Schlesinger equations to $P_{\text{VI}}$ can be carried over on the level of 
Hamiltonians, see \cite{Sak:2010:IDA4PTE}.
Fuchsian systems with more than two accessory parameters give rise to 
higher order analogues of Painlev\'e equations, see the survey
\cite{KawNakSak:2013:TACOFPE} for the four-dimensional case.

It is also possible to deform the characteristic indices of the system, 
but in that case the isomonodromy
condition requires that the indices change by integral shifts, and so the
resulting dynamic is \emph{discrete} and is expressed in the form
of \emph{difference} equations called Schlesinger \emph{transformations}. 
Similarly to the continuous case, these transformations reduce to 
discrete analogues of Painlev\'e-type equations and the study 
of this correspondence has been a major research topic in the field 
of discrete integrable systems over the last twenty years,
\cite{RamGraHie:1991:DVOTPE,PapNijGraRam:1992:IDPFDAOPE}.
We need to remark here that Schlesinger transformations correspond to 
\emph{difference} Painlev\'e equations, but the other two types
of \emph{discrete} Painlev\'e equations,
\emph{$q$-difference} and \emph{elliptic-difference} Painlev\'e equations, 
can also be considered in a modification of this approach.

Discrete Painlev\'e equations share many properties with the differential 
Painlev\'e equations, e.g., the existence of special solutions 
such as algebraic solutions or solutions that can be expressed in terms of special functions, 
affine Weyl group symmetries, and the geometric classification of
equations in terms of rational surfaces. 
Our goal for this paper is to describe, very explicitly, some aspects of 
such correspondence for the isomonodromic approach, following the outline 
above for the continuous case and adapting it to the so-called \emph{elementary} 
Schlesinger transformations. We do this in three steps. 

First, we give \emph{explicit}
evolution equations for the discrete dynamical system on the space of 
coefficient matrices that is generated by elementary Schlesinger transformations. We 
call such equations \emph{difference Schlesinger evolution equations}. 
These equations make it easy to compute
Schlesinger transformations and consequently to study reductions to difference
Painlev\'e equations, since the usual computation of  Schlesinger
transformations from the compatibility condition between the Fuchsian
equation and the deformation equation is often quite complicated.
We should also mention here 
that in \cite{Bor:2004:ITLSDE} A.~Borodin considered difference Schlesinger 
equations for isomonodromic deformations of \emph{difference} linear systems. He showed that,
under a certain continuous limit, such equations converge to implicit equations 
describing Schlesinger transformations of the corresponding continuous system. 

Next, we use these equations to define the evolution dynamic on the 
decomposition space of the Fuchsian system and give the discrete Hamiltonian 
description of such flows. Since the Hamiltonian framework is not
yet well-developed in the discrete case, having examples of such description is
essential to better understanding of this formalism. Further, this is an essential 
first step towards developing the Hamiltonian theory of discrete Painlev\'e
equations, as well as the Hamiltonian reduction procedure in the discrete
case. 

Finally, we consider explicit examples of reductions from 
Schlesinger transformations to difference Painlev\'e equations.
In Sakai's geometric approach to Painlev\'e equations,
difference equations are classified by the surface type 
(or symmetry type) as follows:
\begin{center}
	\begin{tabular}{rccccccccccc}
		Surface: & $A_{0}^{(1)**}$ & $A_{1}^{(1)*}$ & $A_{2}^{(1)*}$ & $D_{4}^{(1)}$ & $D_{5}^{(1)}$ & $D_{6}^{(1)}$ & $D_{7}^{(1)}$ 
		& $D_{8}^{(1)}$ & $E_{6}^{(1)}$ & $E_{7}^{(1)}$ & $E_{8}^{(1)}$ \\
		Symmetry: & $E_{8}^{(1)}$ & $E_{7}^{(1)}$ & $E_{6}^{(1)}$ & $D_{4}^{(1)}$ & $A_{3}^{(1)}$ & $(A_{1} + A_{1})^{(1)}$ & $A_{1}^{(1)}$
		& --- & $A_{2}^{(1)}$ & $A_{1}^{(1)}$ & ---\\
	\end{tabular}	
\end{center}
Difference Painlev\'e equations corresponding to surfaces of types $D_{l}$ and $E_{l}$ appear as B\"acklund transformations of differential
Painlev\'e equations, but equations corresponding to surfaces of type $A_{l}$ are purely discrete. Thus, we are particularly interested
in the left side of the above list. In \cite{Sak:2007:PDPEATLF}
H.~Sakai posed a problem of representing difference Painlev\'e equations with surface types
$A_{0}^{(1)**}$, $A_{1}^{(1)*}$, and $A_{2}^{(1)*}$ as reductions of Schlesinger
transformations of Fuchsian systems. 
In \cite{Boa:2009:QADPE}
P.~Boalch described the corresponding Fuchsian systems without computing the Schlesinger transformations and their 
reductions explicitly. Also, in \cite{AriBor:2006:MSDDPE} D.~Arinkin and A.~Borodin obtained difference Painlev\'e 
equation d-$P(A_2^{(1)*})$ from isomonodromy transformations of \emph{difference} Fuchsian systems. 
In this paper we explicitly compute reductions from Schlesinger transformations 
to difference Painlev\'e equations d-$P(D_4^{(1)})$ (also known as difference Painlev\'e-V
equation d-$P_{\text{V}}$) and d-$P(A_2^{(1)\ast})$. We start with a suitable parametrization for 
the space of Fuchsian system of the appropriate type (which for these examples is two-dimensional,
and so we denote the parameters by $x$ and $y$)
and use the discrete Schlesinger evolution equations to explicitly compute the evolution 
$\bar{x} = \bar{x}(x,y)$ and $\bar{y} = \bar{y}(x,y)$. For this map we construct the Okamoto space of initial
conditions that resolves the indeterminate points of the dynamic, verify that this surface is indeed
of the required type, and then explicitly find the translation direction in the symmetry sub-lattice. 
We also compare the dynamic given by discrete Schlesinger transformations to other examples
of difference Painlev\'e equations of the same type by first finding the isomorphism of blow-down 
structures of their Okamoto surfaces, and then comparing the corresponding translation direction.

The interesting feature of the d-$P(D_4^{(1)})$ example is that we can compute, for the same Fuchsian 
system and using the same coordinates, continuous deformations
that are described by the Painlev\'e-VI equation $P_{\text{VI}}$, and discrete 
deformations that are described by the difference Painlev\'e-V
equation d-$P_{\text{V}}$, which corresponds to the B\"acklund transformations of 
$P_{\text{VI}}$. We use this as a test case for our approach. The d-$P(A_2^{(1)*})$ example is new and is a partial 
answer to Sakai's question (we plan to consider reductions to 
difference Painlev\'e equations d-$P(A_1^{(1)*})$ and d-$P(A_0^{(1)**})$ in the subsequent publication).
A particularly interesting feature of the second example, that we explain in detail in the main text, 
is that its translation direction is different from that of the standard form for this equation.

The text is organized as follows. In the next section we briefly set up 
the context of the problem and describe in more details some of the main objects.
In Section~\ref{sec:schlesinger_transformations} we derive the equations of the elementary Schlesinger 
transformations and in Section~\ref{sec:hamiltonian_formulation} we show how to write them in the Hamiltonian form.
Section~\ref{sec:examples} is dedicated to the d-$P(D_4^{(1)})$ and d-$P(A_2^{(1)*})$ examples,
and the final section is conclusions and discussion of the results.

\textbf{Acknowledgements}: This work started at the \emph{Discrete Integrable Systems}
program at the \emph{Isaac Newton Institute for Mathematical Sciences} in Cambridge, UK. 
The authors are very grateful to the Institute for its hospitality and the stimulating working 
environment. 
H.S.~was supported by Grant-in-Aid no. 24540205 of the Japan Society for the Promotion of Science.
Part of this work was done when A.D.~visited H.S.~and T.T~in
Tokyo, and he thanks the \emph{Tokyo University of Marine
Science and Technology} and the \emph{University of Tokyo} for their hospitality.
A.D.~would also like to acknowledge the generous travel support by the 
\emph{University of Northern Colorado} Provost Fund which was very important for the 
success of this collaboration.


\section{Preliminaries} 
\label{sec:preliminaries}
	
\subsection{Fuchsian Systems} 
Schlesinger transformations that we study originate from the deformation theory of Fuchsian systems. Recall that a \emph{Fuchsian system}
(or a \emph{Fuchsian equation}) is a matrix linear differential equation on the Riemann sphere $\mathbb{CP}^{1}$ such that all of its singular points 
are \emph{regular singular points}. We consider a generic case when a Fuchsian system can be written in the 
\emph{Schlesinger normal form}, i.e., when its coefficient matrix is a rational function with at most simple 
poles at some (distinct) points $z_{1},\dots, z_{n}$ (and possibly at the point $z_{0} = \infty$),
\begin{equation}
	\frac{d \mathbf{Y}}{dz} = \mathbf{A}(z) \mathbf{Y} = \left( \sum_{i=1}^{n} \frac{ \mathbf{A}_{i} }{ z-z_{i} }  \right) \mathbf{Y},
	\label{eq:fuchs-1}
\end{equation}
Here $\mathbf{A}_{i} = \operatorname{ res }_{z_{i}} \mathbf{A}(z)\, dz$ are constant $m\times m$ matrices. We also put 
$\mathbf{A}_{\infty} = \operatorname{res}_{\infty} \mathbf{A}(z)\, dz = -\sum_{i=1}^{n} \mathbf{A}_{i}$ and so this equation is regular 
at infinity iff $\mathbf{A}_{\infty} = \mathbf{0}$.
\label{sub:equations}

\begin{assumption}\label{assume:diagonal}
	From now on we make an additional semi-simplicity assumption 
	that the coefficient matrices $\mathbf{A}_{i}$ are \emph{diagonalizable}.
\end{assumption}	

\subsection{Spectral Type and Accessory Parameters} 
\label{sub:accessory_parameters}

Geometrically, Schlesinger (or isomonodromic) dynamic takes place on the space of coefficients $\mathbf{A}(z)$ of the Fuchsian system
(\ref{eq:fuchs-1}), considered modulo gauge transformations. Here we briefly outline the description 
of this space following \cite{Sak:2010:IDA4PTE}.

First, given $\mathbf{A}(z)$, we separate the locations of singular points $z_{i}$ 
(that we think of as parameters of the dynamic)
from the residue matrices $\mathbf{A}_{i}$ at those points. Thus, we define
\begin{equation*}
	\tilde{\mathcal{F}}(\mathbf{z}) = \tilde{\mathcal{F}}(z_{1},\dots,z_{n}) =  
	\{(\mathbf{A}_{1},\dots,\mathbf{A}_{n})\mid \mathbf{A}_{i}\in \operatorname{End}(\mathbb{C}^{m})\}
\end{equation*}
to be the set of all Fuchsian systems with possible singularities at the points $z_{i}$.
Further, since Schlesinger dynamic either preserves (in the continuous case) 
or shifts (in the discrete case) the eigenvalues of $\mathbf{A}_{i}$, we treat the eigenvalues as the parameters 
of the dynamic as well. Thus, the appropriate phase space is a quotient of the fiber of the eigenvalue map by gauge
transformations. 
Local coordinates on the phase space are called \emph{accessory parameters}. Note that the space $\mathcal{A}$ of accessory
parameters is quite complicated and its dimension depends on the spectral type of $\mathbf{A}(z)$ that we define next. 

Spectral type of $\mathbf{A}(z)$ encodes the degeneracy of eigenvalues $\theta_{i}^{j}$ of the coefficient matrices $\mathbf{A}_{i}$
($\theta_{i}^{j}$ are also called the \emph{characteristic indices} of the system) via
partitions $m = m_{i}^{1} + \cdots + m_{i}^{l_{i}}$, $m_{i}^{1}\geq\cdots\geq m_{i}^{l_{i}}\geq 1$,
where $m$ is the matrix size and $m_{i}^{j}$ denotes the multiplicities of the eigenvalues of $\mathbf{A}_{i}$. 
The \emph{spectral type} $\mathfrak{m}$ of $\mathbf{A}(z)$ is then defined to be the collection of these partitions for all indices 
(including $\infty$):
\begin{equation*}
	\mathfrak{m} = m_{1}^{1}\cdots m_{1}^{l_{1}},	m_{2}^{1}\cdots m_{2}^{l_{2}}, \cdots, 
	m_{n}^{1}\cdots m_{n}^{l_{n}}, 	m_{\infty}^{1}\cdots m_{\infty}^{l_{\infty}}.
\end{equation*}
It can be used to classify Fuchsian systems up to isomorphisms and the operations of addition
and middle convolution introduced by N.~Katz \cite{Kat:1996:RLS}, as in the recent work by T.~Oshima \cite{Osh:2013:COFSATCP}.

\begin{assumption}\label{assume:rank-reduce}
	Since for a Fuchsian system we can use scalar local gauge transformation of the form $\tilde{\mathbf{Y}}(z) = w(z)^{-1} \mathbf{Y}(z)$,
	where $w(z)$ is a solution of the \emph{scalar} equation
	\begin{equation*}
		\frac{ dw }{ dz } = \sum_{i=1}^{n} \frac{ \theta_{i}^{j}}{ z - z_{i} }w,
	\end{equation*} 
	to change the residue matrices by $\tilde{\mathbf{A}}_{i} = \mathbf{A}_{i} - \theta_{i}^{j} \mathbf{1}$, we can assume, 
	without loss of generality,  that one of the eigenvalues $\theta_{i}^{j} = 0$. Thus, we always assume that the 
	eigenvalue  $\theta_{i}^{j}$ of	the highest multiplicity $m_{i}^{1}$ is zero and we denote the corresponding 
	subset of $\tilde{\mathcal{F}}(\mathbf{z})$ of \emph{reduced} Fuchsian equations 
	by $\mathcal{F}(\mathbf{z})$. This amounts to choosing a representative in the quotient 
	space of all Fuchsian equations by the group of local \emph{scalar} gauge transformations.
\end{assumption}	

In view of \textbf{Assumptions~\ref{assume:diagonal}} and~\textbf{\ref{assume:rank-reduce}}, 
$\mathbf{A}_{i}$ is similar to a diagonal matrix $\operatorname{diag}\{\theta_{i}^{1},\dots, \theta_{i}^{r_{i}},0,\dots,0\}$,
where  $r_{i} = \operatorname{rank}(\mathbf{A}_{i})$. 
Omitting the zero eigenvalues, we put 
\begin{equation*}
\mathbf{\Theta}_{i} = \operatorname{diag}\{\theta_{i}^{1},\dots, \theta_{i}^{r_{i}}\}.\label{eq:theta-i}	
\end{equation*}
Denote by $\tilde{\mathcal{T}}$ the set of all possible diagonal matrices of the spectral type $\mathfrak{m}$ with the highest degeneracy 
eigenvalues at \emph{finite} points set to $0$ and omitted, as above. Thus,
$\tilde{\mathcal{T}} = \{\mathbf{\Theta} = (\mathbf{\Theta}_{1},\dots,\mathbf{\Theta}_{n};\mathbf{\Theta}_{\infty})\}$ and
$r_{i} = m - m_{i}^{1}$. Then we have the eigenvalue map $\operatorname{Sp}: \mathcal{F}(\mathbf{z})\to \mathcal{T}$ from the
set of reduced Fuchsian equations to the set of characteristic indices, where $\mathcal{T}\subset\tilde{\mathcal{T}}$ is the subset of 
$\mathbf{\Theta}$ satisfying the \emph{Fuchs relation} (or the trace condition)
\begin{equation*}\label{eq:fuchs-relation}
	\sum_{i=1}^{n,\infty}\sum_{j=1}^{r_{i}} \theta_{i}^{j} = 0.
\end{equation*} 
With these
definitions, given $\mathbf{\Theta}\in \mathcal{T}$, the fiber of the eigenvalue map is 
$\mathcal{F}(\mathbf{z})_{\mathbf{\Theta}} = \operatorname{Sp}^{-1}(\mathbf{\Theta})$.

We still need to take into account global similarity transformations. It is convenient to do so in two steps. 
First we \emph{normalize} our equation at infinity by reducing 
$\mathbf{A}_{\infty}$ to a particular form, e.g., making it diagonal, and second we take the 
quotient by the stabilizer subgroup $G_{\mathbf{A}_{\infty}}$ of the group $\mathbb{GL}_{m}$ of global similarity transformations.
Thus, for a given $\mathbf{\Theta}\in \mathcal{T}$, we \emph{fix} $\mathbf{A}_{\infty}$ such that 
$\operatorname{Sp}(\mathbf{A}\infty) = \mathbf{\Theta}_{\infty}$ and denote by 
$\mathcal{F}(\mathbf{z})_{(\mathbf{\Theta},\mathbf{A}_\infty)}$ the subset of all Fuchsian equations in 
$\mathcal{F}(\mathbf{z})_{\mathbf{\Theta}}$ satisfying  the condition $\mathbf{A}_{1} + \cdots + \mathbf{A}_{n} = - \mathbf{A}_{\infty}$.
Then, finally, our phase space is 
\begin{equation*}
	\mathcal{A} = \mathcal{A}_{(\mathbf{\Theta},\mathbf{A}_\infty)} = 
	\mathcal{F}(\mathbf{z})_{(\mathbf{\Theta},\mathbf{A}_\infty)}/G_{\mathbf{A}_{\infty}}.
\end{equation*}

When $\mathbf{A}_{\infty}$ is diagonal, which is the case that we are mainly interested in, the dimension of this space 
$\mathcal{A}$ of accessory parameters of the spectral type $\mathfrak{m}$ is given 
by the  formula
\begin{equation*}
	\dim \mathcal{A} 
	= (n-1)m^{2} - \sum_{i=1}^{n,\infty} \left( \sum_{j=1}^{l_{i}} (m_{i}^{l_{i}})^{2} \right) + 2.
\end{equation*}


\subsection{The Decomposition Space} 
\label{sub:the_decomposition_space}
To study the Hamiltonian structure of Schlesinger equations, and also of Schlesinger transformations, it is convenient to 
consider a larger space that we call the \emph{decomposition space}. It is defined as follows. 

In view of  \textbf{Assumption~\ref{assume:diagonal}}, there exist full sets of \emph{right} eigenvectors 
$\mathbf{b}_{i,j}$, $\mathbf{A}_{i} \mathbf{b}_{i,j} = \theta_{i}^{j} \mathbf{b}_{i,j}$, 
and \emph{left} eigenvectors $\mathbf{c}_{i}^{j\dag}$, $\mathbf{c}_{i}^{j\dag} \mathbf{A}_{i} = \theta_{i}^{j} \mathbf{c}_{i}^{j\dag}$, (here 
we use the $\dag$ symbol to indicate a \emph{row}-vector).  In the matrix form, omitting vectors with indices 
$j>r_{i}$ that are in the kernel of $\mathbf{A}_{i}$, we can write 
\begin{equation*}
	\mathbf{B}_{i} = \begin{bmatrix}	 \mathbf{b}_{i,1} \cdots \mathbf{b}_{i,r_{i}}	\end{bmatrix}, \quad
	\mathbf{A}_{i} \mathbf{B}_{i} = \mathbf{B}_{i} \mathbf{\Theta}_{i},\quad 
	\mathbf{C}_{i}^{\dag}  = \begin{bmatrix} \mathbf{c}_{i}^{1\dag}\\ \vdots \\ \mathbf{c}_{i}^{r_{i}\dag}	\end{bmatrix}, \quad
	\mathbf{C}_{i}^{\dag} \mathbf{A}_{i} =  \mathbf{\Theta}_{i} \mathbf{C}_{i}^{\dag},
\end{equation*}
with $\mathbf{\Theta}_{i}$ defined by (\ref{eq:theta-i}). Then we have a decomposition
$\mathbf{A}_{i} = \mathbf{B}_{i} \mathbf{C}_{i}^{\dag}$, provided that 
$\mathbf{C}_{i}^{\dag} \mathbf{B}_{i} = \mathbf{\Theta}_{i}$. This
condition is related to the normalization ambiguity of the eigenvectors.

Thus, given $\mathbf{A}_{i}$, we can construct (in a non-unique way) a corresponding decomposition pair 
$(\mathbf{B}_{i}, \mathbf{C}_{i}^{\dag})$.
The space of all such pairs for all finite indices $1\leq i\leq n$, without any additional conditions, 
is our \emph{decomposition space}. We denote it as 
\begin{equation*}
	\mathcal{B}\times \mathcal{C} = 
	(\mathbb{C}^{r_{1}}\times\cdots\times \mathbb{C}^{r_{n}})\times ((\mathbb{C}^{r_{1}})^{\dag}\times \cdots\times (\mathbb{C}^{r_{n}})^{\dag})\simeq 
	(\mathbb{C}^{r_{1}}\times(\mathbb{C}^{r_{1}})^{\dag})\times\cdots\times (\mathbb{C}^{r_{n}}\times (\mathbb{C}^{r_{n}})^{\dag}),
\end{equation*}
since it is convenient to write an element $(\mathbf{B},\mathbf{C}^{\dag})$ of this space as a list of $n$ pairs 
$(\mathbf{B}_{1}, \mathbf{C}_{1}^\dag; \cdots; \mathbf{B}_{n}, \mathbf{C}_{n}^{\dag})$. 
The decomposition space can be equipped with the following natural symplectic structure. We define the symplectic form as
\begin{equation*}
	\omega = \sum_{i=1}^{n} \operatorname{tr}(d \mathbf{C}_{i}^\dag \wedge d \mathbf{B}_{i}),
\end{equation*}
and so the canonical coordinates are the matrix elements of $\mathbf{B}_{i}$ and $\mathbf{C}_{i}^{\dag}$.

\begin{remark}\label{rem:coords}
	To simplify the notation we use the same symbols, e.g.,  
	$\mathbf{B}_{i}$ or $\mathbf{b}_{i,j}$, to denote both the canonical 
	coordinate system on the decomposition space and an actual point in the space; the exact meaning of the symbol
	is always clear from the context.
\end{remark}

The idea of using the decomposition space $\mathcal{B}\times \mathcal{C}$ to study the Hamiltonian structure of isomonodromic 
deformations goes back to the paper by M.~Jimbo, T.~Miwa, Y.~M\^ori, and M.~Sato \cite{JimMiwMorSat:1980:DMIBFPT}.
Since then it has been used by many other researchers studying isomonodromic deformations, 
most notably by J.~Harnad, see, e.g., \cite{Har:1994:DIDAMMTLA}.

\begin{remark}\label{rem:transf}
	There are two natural actions on the decomposition space $\mathcal{B}\times \mathcal{C}$. First, the group  
	$\mathbb{GL}_{m}$ of global gauge transformations of the Fuchsian system induces the following
	action. Given $\mathbf{P}\in \mathbb{GL}_{m}$, 
	we have the action $\mathbf{A}_{i}\mapsto \mathbf{P} \mathbf{A}_{i} \mathbf{P}^{-1}$ which translates into the action
	$(\mathbf{B}_{i},\mathbf{C}_{i}^{\dag}) \mapsto 
	(\mathbf{P}\mathbf{B}_{i},\mathbf{C}_{i}^{\dag} \mathbf{P}^{-1})$. We refer to such transformations as
	\emph{similarity transformations}. It is often necessary to restrict this action to the subgroup 
	$G_{\mathbf{A}_{\infty}}$ preserving the form of $\mathbf{A}_{\infty}$. Second, for any pair
	$(\mathbf{B}_{i}, \mathbf{C}_{i}^{\dag})$ the pair
	$(\mathbf{B}_{i}\mathbf{Q}_{i}, \mathbf{Q}_{i}^{-1}\mathbf{C}_{i}^{\dag})$  
	determines the same matrix $\mathbf{A}_{i}$ for $\mathbf{Q}_{i}\in \mathbb{GL}_{r_{i}}$. 
	The condition 
	$\mathbf{Q}_{i}^{-1}\mathbf{C}_{i}^{\dag} \mathbf{B}_{i}\mathbf{Q}_{i} = 
	\mathbf{Q}_{i}^{-1}\mathbf{\Theta}_{i}\mathbf{Q}_{i} = \mathbf{\Theta}_{i}$
	restricts $\mathbf{Q}_{i}$ to the stabilizer subgroup $G_{\mathbf{\Theta}_{i}}$ of $\mathbb{GL}_{r_{i}}$.
	In particular, when all $\theta_{i}^{j}$ are distinct, 
	$\mathbf{Q}_{i}$ has to be a diagonal matrix.
	We refer to such transformations as \emph{trivial transformations}.  These two actions obviously commute
	with each other.
\end{remark}

We are now ready to give an alternative description of the space of accessory parameters 
$\mathcal{A}_{(\mathbf{\Theta},\mathbf{A}_\infty)}$. Given the pair $(\mathbf{\Theta}, \mathbf{A}_{\infty})$ 
as in the previous section, let
\begin{align*}
	(\mathcal{B}\times \mathcal{C})_{(\mathbf{\Theta},\mathbf{A}_{\infty})} &= 
	\{(\mathbf{B}_{1}, \mathbf{C}_{1}^\dag; \cdots; \mathbf{B}_{n}, 
	\mathbf{C}_{n}^{\dag})\in \mathcal{B}\times \mathcal{C}
	\mid \mathbf{C}_{i}^{\dag} \mathbf{B}_{i} = \mathbf{\Theta}_{i},
	\sum_{i=1}^{n} \mathbf{B}_{i} \mathbf{C}_{i}^{\dag} = - \mathbf{A}_{\infty}\} 
	\subset \mathcal{B}\times \mathcal{C}.
	\intertext{Then}
	\mathcal{A} &= \mathcal{A}_{(\mathbf{\Theta},\mathbf{A}_\infty)} = (\mathcal{B}\times \mathcal{C})_{(\mathbf{\Theta},\mathbf{A}_{\infty})}/
	\big((G_{\mathbf{\Theta}_{1}}\times\cdots \times G_{\mathbf{\Theta_{n}}})\times G_{\mathbf{A}_{\infty}}\big).
\end{align*}

The following diagram illustrates the relationship between different spaces defined in these two sections:
\begin{figure}[ht]
	\centering
		\includegraphics[width=3in]{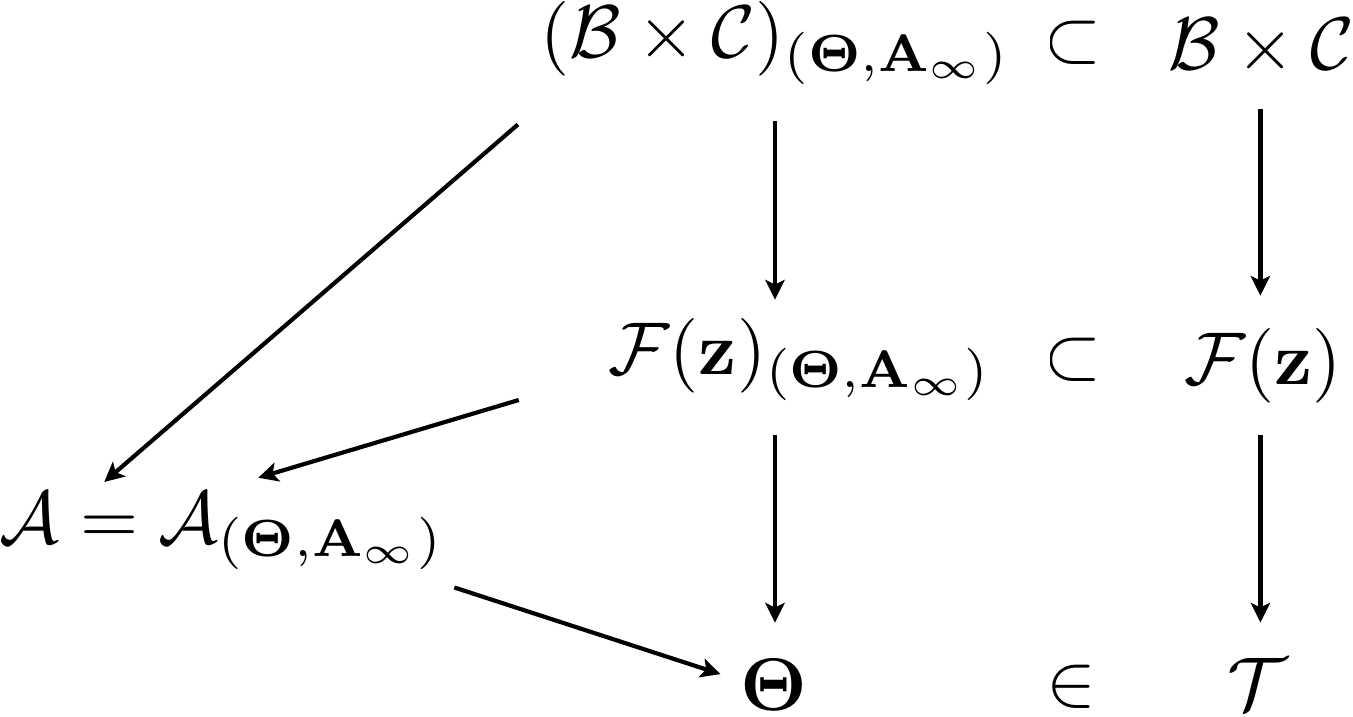}
	\caption{The decomposition space and the space of accessory parameters}
	\label{fig:spaces}
\end{figure}


\subsection{Schlesinger Equations and Schlesinger Transformations} 
\label{sub:schlesinger_equations_and_schlesinger_transformations}
Consider now isomonodromic deformations of a Fuchsian system. For the 
continuous case
we take the deformation parameters to be locations of critical points. Thus, we let 
$\mathbf{A} = \mathbf{A}(z;\mathbf{z}) = \mathbf{A}(z;z_{1},\dots,z_{n})$ 
(i.e., $\mathbf{A}_{i} = \mathbf{A}_{i}(\mathbf{z})$). In \cite{Sch:1912:UEKDBOFKP} L.~Schlesinger 
showed that isomonodromic deformations are
described, if we use the modern terminology, by the following \emph{differential-differential Lax pair}:
	\begin{equation*}
		\left\{ 
		\begin{aligned}
			\frac{d \mathbf{Y}}{dz} &= \mathbf{A}(z;\mathbf{z}) \mathbf{Y} = 
			\left( \sum_{i=1}^{n} \frac{ \mathbf{A}_{i}(\mathbf{z}) }{ z-z_{i} }  \right) \mathbf{Y}. \\
			\frac{\partial \mathbf{Y}}{\partial z_{i}} &= - \frac{ \mathbf{A}_{i}(\mathbf{z}) }{ z-z_{i} }   \mathbf{Y}, \qquad
			i=1,\dots,n.
		\end{aligned}
		\right.
	\end{equation*}

Compatibility conditions for these deformation equations are called the \emph{Schlesinger equations}. These are 
partial differential equations on the coefficient matrices $\mathbf{A}_{i}(\mathbf{z})$ and they have the form
\begin{equation}
	\frac{ \partial \mathbf{A}_{j}}{ \partial z_{i} } = \frac{ [\mathbf{A}_{j},\mathbf{A}_{i}] }{ z_{j} - z_{i} }\quad (j\neq i),\qquad
	\frac{ \partial \mathbf{A}_{i}}{ \partial z_{i} } = - \sum_{j\neq i}\frac{ [\mathbf{A}_{j},\mathbf{A}_{i}] }{ z_{j} - z_{i} }.
	\label{eq:Sch-eqs}
\end{equation}

In \cite{JimMiwMorSat:1980:DMIBFPT} Jimbo~\emph{et.al} showed that  Schlesinger equations can be written as a Hamiltonian 
system on the decomposition space $\mathcal{B}\times \mathcal{C}$,
\begin{equation*}
	\frac{ \partial \mathbf{B}_{i} }{ \partial z_{j} } = \frac{ \partial \mathcal{H}_{j} }{ \partial \mathbf{C}_{i}^{\dag} },\qquad
	\frac{ \partial \mathbf{C}_{i}^{\dag} }{ \partial z_{j} } = - \frac{ \partial \mathcal{H}_{j} }{ \partial \mathbf{B}_{i} },
\end{equation*}
with the Hamiltonian 
\begin{equation*}
	\mathcal{H}_{j} = \mathcal{H}_{j}(\mathbf{B},\mathbf{C}^{\dag}) =
	 \sum_{i\neq j} \frac{ \operatorname{tr} (\mathbf{A}_{j} \mathbf{A}_{i}) }{ z_{j} - z_{i} },\qquad
	\mathbf{A}_{i} = \mathbf{B}_{i} \mathbf{C}_{i}^{\dag}.\label{eq:cont-Ham}
\end{equation*}
Schlesinger equations admit reductions to Painlev\'e-type nonlinear differential equations, which remains valid
on the level of the Hamiltonians as well; we review an example of $P_{VI}$ in Section~\ref{sec:examples}. The main question 
that we consider is what happens to this picture in the \emph{discrete} case.

The discrete counterparts of Schlesinger equations are Schlesinger transformations. Those are rational 
transformations preserving the singularity structure and the monodromy data of the Fuchsian system (\ref{eq:fuchs-1}), 
except for the integral shifts in the characteristic indices $\theta_{i}^{j}$. The study of such transformations again 
goes back to 
Schlesinger \cite{Sch:1912:UEKDBOFKP}. This, and the more general case of irregular singular points, 
was considered in great detail in \cite{JimMiw:1981:MPDLODEWRC}. Schlesinger transformations 
are given by the following \emph{differential--difference Lax Pair}:
\begin{equation*}
	\left\{ 
	\begin{aligned}
		\frac{d \mathbf{Y}}{dz} 
		& = \mathbf{A}(z;\mathbf{\Theta}) \mathbf{Y} = 
		\left( \sum_{i=1}^{n} \frac{ \mathbf{A}_{i}(\mathbf{\Theta}) }{ z-z_{i} }  \right) \mathbf{Y}, \\
		\bar{\mathbf{Y}}(z) &= \mathbf{R}(z) \mathbf{Y}(z). 
	\end{aligned}
	\right.,
\end{equation*}
where 
$\mathbf{R}(z)$ is a specially chosen rational matrix function called the \emph{multiplier} of the transformation.
The coefficient matrix $\mathbf{A}(z;\mathbf{\Theta})$ of our Fuchsian system then transforms to $\bar{\mathbf{A}}(z;\mathbf{\Theta})$ 
that is  related to $\mathbf{A}(z;\mathbf{\Theta})$ by the equation
\begin{equation}
	\bar{\mathbf{A}}(z;\mathbf{\Theta})\mathbf{R}(z) = \mathbf{R}(z) \mathbf{A}(z;\mathbf{\Theta}) + \frac{d \mathbf{R}(z)}{dz}. \label{eq:sch-transf}
\end{equation}
In this paper we focus on 
the \emph{elementary} Schlesinger transformations that only change two of the characteristic indices by 
unit shifts, i.e., $\bar{\theta}_{\alpha}^{\mu} = \theta_{\alpha}^{\mu} - 1$ and 
$\bar{\theta}_{\beta}^{\nu} = \theta_{\beta}^{\nu} + 1$, $\alpha\neq \beta$. 
For the examples section, we also assume $\theta_{\alpha}^{\mu}$ 
and $\theta_{\beta}^{\nu}$ have no multiplicities. In \cite{JimMiw:1981:MPDLODEWRC} 
such elementary transformation are denoted by $\left\{\begin{smallmatrix}
	\alpha&\beta\\\mu&\nu
\end{smallmatrix}\right\}$. For such transformations we solve the equations~(\ref{eq:sch-transf}) to obtain 
explicit discrete evolutions equations for the coefficient matrices $\mathbf{A}_{i}(\mathbf{\Theta})$, we call
such equations \emph{discrete Schlesinger evolution equations} since they are the exact discrete analogue of 
equations~(\ref{eq:Sch-eqs}), see Theorem~\ref{thm:schlesinger}. In Theorem~\ref{thm:evolution} we 
lift these equations to get the dynamic on the decomposition space and in Theorem~\ref{thm:Hamiltonian}
we represent these equation as a discrete Hamiltonian system, in the sense explained in the next section, using the same canonical 
coordinates as \cite{JimMiwMorSat:1980:DMIBFPT}. We explicitly construct a discrete Hamiltonian function 
$\mathcal{H}^{+}(\mathbf{B}, \bar{\mathbf{C}}^{\dag})$ 
of the decomposition space that generates our dynamic. Finally, we study reductions of this dynamic
to the space of accessory parameters. The reduced dynamic is very complicated and is given by Painlev\'e-type equations.
Schematically, the situation is as in Figure~\ref{fig:sch-trans}:

\begin{figure}[h]
	\centering
		\includegraphics[width=3in]{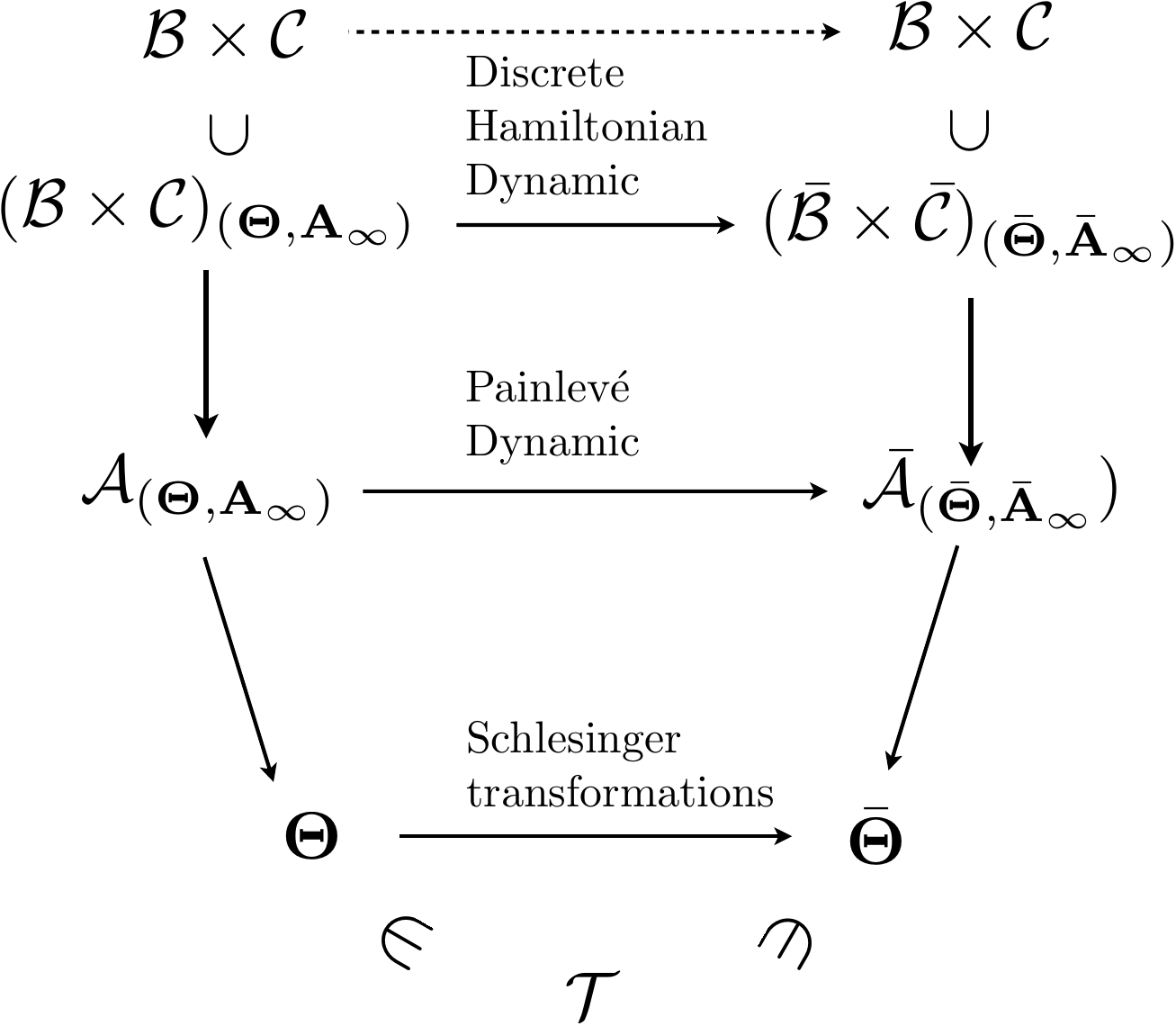}
	\caption{Schlesinger Transformations}
	\label{fig:sch-trans}
\end{figure}

Unfortunately, at present we do not have a discrete version of the Hamiltonian 
reduction procedure from the decomposition space to the space of accessory parameters. 
This is a very important problem for future research.


\subsection{Discrete Lagrangian and Hamiltonian Formalism} 
\label{sub:discrete_lagrangian_and_hamiltonian_formalism}
In developing a discrete version of the Hamiltonian formalism, one approach is to 
discretize the continuous dynamic preserving its integrability properties, see the recent 
encyclopedic book by Y.~Suris, \cite{Sur:2003:PIDHA}, but it requires underlying
continuous dynamical system. An alternative and more direct procedure is to develop a 
discrete version based on the variational principles. In this case, the Lagrangian formalism
is more natural, but it is possible to extend it to include the discrete version 
of the Hamiltonian formalism as well. This approach has its origin in the optimal
control theory (see, e.g., B.~Jordan and E.~Polak \cite{JorPol:1964:TOACODOCS}, and J.~Cadzow \cite{Cad:1970:DCOV})
and mechanics (see e.g., J.~Logan \cite{Log:1973:FIITDVC} and S.~Maeda \cite{Mae:1982:LFODSACODS}). In the theory
of integrable systems it was first used in the foundational works by A.~Veselov \cite{Ves:1991:ILRFMP,Ves:1988:ISWDTDO} and 
A.~Veselov and J.~Moser \cite{MosVes:1991:DVSCISFMP}. A lot of recent work in this field is motivated by the 
development of very effective \emph{symplectic integrators}, see an excellent review paper by  
J.~Marsden and M.~West \cite{MarWes:2001:DMVI} (and the references therein), as well as more recent works by 
S.~Lall and M.~West \cite{LalWes:2006:DVHM} and A.~Bloch, M.~Leok, and T.~Oshawa \cite{OhsBloLeo:2011:DHT}
that emphasize the Hamiltonian aspects of the theory. Below we give a \emph{very} brief outline of this approach
closely following \cite{OhsBloLeo:2011:DHT}, we refer to this and other publications above for details.

Let $\mathcal{Q}$ be our configuration space. Then, in the discrete case, the \emph{Lagrangian} 
$\mathcal{L} \in \mathcal{F}(\mathcal{Q} \times \mathcal{Q})$
is a function
on the \emph{state space} $\mathcal{Q}\times \mathcal{Q}$. For the discrete time parameter 
$k\in \mathbb{Z}$, a trajectory of motion is a map $\mathbf{q}: \mathbb{Z}\to \mathcal{Q}$, $\mathbf{q}(k) =: \mathbf{q}_{k}$, or,
equivalently, a sequence $\{\mathbf{q}_{k}\in \mathcal{Q}\}$. The action functional on the space of such sequences
is given by $\mathfrak{S}(\{\mathbf{q}_{k}\}) = \sum_{k} \mathcal{L}(\mathbf{q}_{k}, \mathbf{q}_{k+1})$.
Using the variational principle $\delta\frak{S} = 0$ we obtain 
\emph{discrete Euler-Lagrange} equations that have the form
	\begin{equation*}
		D_{2}\mathcal{L}(\mathbf{q}_{k-1},\mathbf{q}_{k}) + D_{1} \mathcal{L}(\mathbf{q}_{k},\mathbf{q}_{k+1}) = \mathbf{0},	
	\end{equation*}
	where $D_{1}$ (resp.~$D_{2}$) denote vectors of partial derivatives w.r.t.~the first (resp.~second) sets of local
	coordinates on the state space $\mathcal{Q}\times \mathcal{Q}$. These equations then implicitly determine the map
	(or more precisely, the correspondence)	$\mathbf{q}_{k+1} = \phi(\mathbf{q}_{k-1},\mathbf{q}_{k})$, which then defines
	the discrete Lagrangian flow $F_{\mathcal{L}}: \mathcal{Q}\times \mathcal{Q} \to \mathcal{Q}\times \mathcal{Q}$ on the state space
	via $F_{\mathcal{L}}(\mathbf{q}_{k-1},\mathbf{q}_{k}) = (\mathbf{q}_{k},\mathbf{q}_{k+1})$. 
	This flow is symplectic w.r.t.~the discrete Lagrangian symplectic form
	\begin{equation*}
		\Omega_{\mathcal{L}}(\mathbf{q}_{k},\mathbf{q}_{k+1}) := d \vartheta^{+}_{\mathcal{L}} = d \vartheta^{-}_{\mathcal{L}}
		= D_{1} D_{2} \mathcal{L}(\mathbf{q}_{k},\mathbf{q}_{k+1}) d \mathbf{q}_{k} \wedge d\mathbf{q}_{k+1}, 
 	\end{equation*}
	$F_{\mathcal{L}}^{*} \Omega_{\mathcal{L}} = \Omega_{\mathcal{L}}$, where one-forms 
	$\vartheta^{\pm}_{\mathcal{L}}: \mathcal{Q}\times \mathcal{Q}  \to 	T^{*}(\mathcal{Q}\times \mathcal{Q})$
	are defined by 
	\begin{equation*}
		\vartheta^{+}_{\mathcal{L}}(\mathbf{q}_{k},\mathbf{q}_{k+1}) = D_{2} \mathcal{L}(\mathbf{q}_{k},\mathbf{q}_{k+1}) d \mathbf{q}_{k+1},\qquad
		\vartheta^{+}_{\mathcal{L}}(\mathbf{q}_{k},\mathbf{q}_{k+1}) = -D_{1} \mathcal{L}(\mathbf{q}_{k},\mathbf{q}_{k+1}) d \mathbf{q}_{k}.
	\end{equation*}
	
	If we introduce \emph{right} and \emph{left discrete Legendre transforms}  
	$\mathbb{F}\mathcal{L}^{\pm}: \mathcal{Q}\times \mathcal{Q}  \to 	T^{*}\mathcal{Q}$ 
	and the \emph{momenta} variables $\mathbf{p}_{k}$, $\mathbf{p}_{k+1}$ by
	\begin{align*}
		\mathbb{F}\mathcal{L}^{+} (\mathbf{q}_{k},\mathbf{q}_{k+1})  &= (\mathbf{q}_{k+1},\mathbf{p}_{k+1}) = 
		(\mathbf{q}_{k+1}, D_{2} \mathcal{L}(\mathbf{q}_{k},\mathbf{q}_{k+1})),
		\\
		\mathbb{F}\mathcal{L}^{-} (\mathbf{q}_{k},\mathbf{q}_{k+1})  &= (\mathbf{q}_{k},\mathbf{p}_{k}) = 
		(\mathbf{q}_{k}, -D_{1} \mathcal{L}(\mathbf{q}_{k},\mathbf{q}_{k+1})),
	\end{align*}
	we see that $\vartheta^{\pm}_{\mathcal{L}} = (\mathbb{F}\mathcal{L}^{\pm})^{*} \vartheta$ and 
	$\Omega_{\mathcal{L}} = (\mathbb{F}\mathcal{L}^{\pm})^{*}\Omega$, where $\vartheta$ and $\Omega$ are the standard 
	Liouville and symplectic forms on $T^{*}\mathcal{Q}$ respectively.
	
	We can then define the discrete Hamiltonian flow $\tilde{F}_{\mathcal{L}}:T^{*}\mathcal{Q}\to T^{*}\mathcal{Q}$ by
	$\tilde{F}_{\mathcal{L}}(\mathbf{q}_{k}, \mathbf{p}_{k}) = (\mathbf{q}_{k+1}, \mathbf{p}_{k+1})$, i.e., 
	$\tilde{F}_{\mathcal{L}} = \mathbb{F}\mathcal{L}^{+}\circ (\mathbb{F}\mathcal{L}^{-})^{-1}$.  But then the equations 
	\begin{equation*}
		\mathbf{p}_{k} = - D_{1}\mathcal{L}(\mathbf{q}_{k},\mathbf{q}_{k+1}),\qquad 
		\mathbf{p}_{k+1} = D_{2}\mathcal{L}(\mathbf{q}_{k},\mathbf{q}_{k+1}),
		\label{eq:lagr-gen-eqns}
	\end{equation*} 
	mean that $\mathcal{L}(\mathbf{q}_{k},\mathbf{q}_{k+1})$ is just the generating function 
	(of type one, see \cite{GolPooSaf:2000:CM} for the terminology) of the canonical transformation 
	$\tilde{F}_{\mathcal{L}}$. We can then define the \emph{right} and \emph{left discrete
	Hamiltonian functions}  as generating functions of this canonical transformation of type two and three respectively. Namely,  
	\emph{right discrete Hamiltonian} is
	\begin{equation*}
		\mathcal{H}^{+}(\mathbf{q}_{k},\mathbf{p}_{k+1}):= \mathbf{p}_{k+1} \mathbf{q}_{k+1} - \mathcal{L}(\mathbf{q}_{k},\mathbf{q}_{k+1}),
		\label{eq:hplus-gen-func}
	\end{equation*}
	and then the map $\tilde{F}_{\mathcal{L}}:T^{*}\mathcal{Q}\to T^{*}\mathcal{Q}$ is given (implicitly) by the 
	\emph{right discrete Hamiltonian equations}	
	\begin{equation*}
		\mathbf{q}_{k+1} = D_{2} \mathcal{H}^{+} (\mathbf{q}_{k}, \mathbf{p}_{k+1}),\qquad 
		\mathbf{p}_{k} = D_{1} \mathcal{H}^{+} (\mathbf{q}_{k}, \mathbf{p}_{k+1}).
		\label{eq:hplus-gen-eqns}
	\end{equation*}
	Similarly, \emph{left discrete Hamiltonian}
	\begin{equation*}
		\mathcal{H}^{-}(\mathbf{q}_{k+1},\mathbf{p}_{k}):= -\mathbf{p}_{k} \mathbf{q}_{k} - \mathcal{L}(\mathbf{q}_{k},\mathbf{q}_{k+1})
		\label{eq:hminus-gen-func}
	\end{equation*}
	gives \emph{left discrete Hamiltonian equations}	
	\begin{equation*}
		\mathbf{q}_{k} = -D_{1} \mathcal{H}^{-} (\mathbf{q}_{k+1}, \mathbf{p}_{k}),\qquad 
		\mathbf{p}_{k+1} = -D_{2} \mathcal{H}^{-} (\mathbf{q}_{k+1}, \mathbf{p}_{k}).
		\label{eq:hminus-gen-eqns}
	\end{equation*}
	For completeness, we mention that the generating function of type four,
	\begin{align*}
		\mathcal{R}(\mathbf{p}_{k}, \mathbf{p}_{k+1}):&= \mathbf{q}_{k} \mathbf{p}_{k} - \mathbf{q}_{k+1} \mathbf{p}_{k+1} + \mathcal{L}(q_{k},\mathbf{q}_{k+1})
		= \mathbf{q}_{k} \mathbf{p}_{k} - \mathcal{H}^{+}(\mathbf{q}_{k},\mathbf{p}_{k+1}) =  
		- \mathbf{q}_{k+1} \mathbf{p}_{k+1} - \mathcal{H}^{-} (\mathbf{q}_{k+1}, \mathbf{p}_{k}),\notag\\
		\mathbf{q}_{k} &=  D_{1}\mathcal{R}(\mathbf{p}_{k},\mathbf{p}_{k+1}),\qquad 
		\mathbf{q}_{k+1} = D_{2}\mathcal{L}(\mathbf{q}_{k},\mathbf{q}_{k+1}),
		\label{eq:ragr-gen-func}
	\end{align*}
	is nothing but the Lagrangian function on the space of momenta satisfying the discrete Euler-Lagrange equations
	\begin{equation*}
		D_{2}\mathcal{R}(\mathbf{p}_{k-1},\mathbf{p}_{k}) + D_{1} \mathcal{R}(\mathbf{p}_{k},\mathbf{p}_{k+1}) = \mathbf{0}.
	\end{equation*}
		

\subsection{Elementary Divisors} 
\label{sub:elementary_divisors}
For elementary Schlesinger transformations 
$\left\{\begin{smallmatrix}
	\alpha&\beta\\\mu&\nu
\end{smallmatrix}\right\}$ the multiplier matrix $\mathbf{R}(z)$ has a very special simple form. In this section we 
describe some of the properties of such matrices that we need for our computations.
Consider a rational matrix function
\begin{equation*}
	\mathbf{R}(z) = \mathbf{I} + \frac{ \mathbf{G} }{ z - z_{0} }, \qquad\text{where $\mathbf{G} = \mathbf{f} \mathbf{g}^{\dag}$ 
	is a matrix of rank 1.}
\end{equation*}
Because of their role in representing general matrix functions as a product of factors of this form, such matrices 
are sometimes called elementary divisors \cite{Dzh:2009:FORMF}. Assuming that 
$\operatorname{tr}(\mathbf{G}) = \mathbf{g}^{\dag} \mathbf{f}\neq0$, which is the case we need, we can 
consider instead of $\mathbf{G}$ a corresponding rank-one projector 
$\mathbf{P} = \mathbf{f} (\mathbf{g}^{\dag} \mathbf{f})^{-1} \mathbf{g}^{\dag}$, $\mathbf{P}^{2} = \mathbf{P}$.
Then $\mathbf{G} = (\mathbf{g}^{\dag} \mathbf{f}) \mathbf{P} = (z_{0} - \zeta_{0})\mathbf{P}$, where $z_{0}$ and $\zeta_{0}$
are the zero and the pole of $\det \mathbf{R}(z)$  respectively. We also put $\mathbf{Q} = \mathbf{I} - \mathbf{P}$ to be the complementary
projector, $\mathbf{Q}^{2} = \mathbf{Q}$ and $\mathbf{PQ} = \mathbf{QP} = \mathbf{0}$.  

\begin{lemma}\label{lem:el-divs}
	Let 
	\begin{equation*}
	\mathbf{R}(z) = \mathbf{I} + \frac{ z_{0} - \zeta_{0}}{z - z_{0} } \frac{ \mathbf{f} \mathbf{g}^{\dag} }{ \mathbf{g}^{\dag} \mathbf{f} }
	 = \mathbf{I} + \frac{ z_{0} - \zeta_{0}}{z - z_{0} }\mathbf{P}.		
	\end{equation*}
	Then we have the following basic properties:
		\begin{equation*}
			\det \mathbf{R}(z) = \frac{ z - \zeta_{0} }{ z-z_{0} },\quad
			\mathbf{R}^{-1}(z) = 
				\mathbf{I} + \frac{\zeta_{0} - z_{0}}{z-\zeta_{0}} \frac{ \mathbf{f} \mathbf{g}^{\dag} }{ \mathbf{g}^{\dag} \mathbf{f} },\quad
				\mathbf{R}(z)\mathbf{f} = \left(\frac{z - \zeta_{0}}{z-z_{0}}\right) \mathbf{f},\quad
				\mathbf{g}^{\dag} \mathbf{R}(z) = \left(\frac{z - \zeta_{0}}{z-z_{0}}\right) \mathbf{g}^{\dag},		\label{eq:el-divs-basic-1}
		\end{equation*}
		\begin{equation*}		
				\mathbf{P}\mathbf{R}(z) = \mathbf{R}(z) \mathbf{P} = \frac{ z - \zeta_{0} }{ z - z_{0} }\mathbf{P},\quad
				\mathbf{P}\mathbf{R}^{-1}(z) = \mathbf{R}^{-1}(z) \mathbf{P} = \frac{ z - z_{0} }{ z - \zeta_{0} }\mathbf{P},\quad				
				\mathbf{R}(z)\mathbf{Q}=\mathbf{Q}\mathbf{R}(z) =\mathbf{Q};
		\label{eq:el-divs-basic-2}
		\end{equation*}
\end{lemma}

\begin{proof} This is a simple direct computation, see also \cite{Dzh:2009:FORMF}.
\end{proof}



\section{Schlesinger Transformations} 
\label{sec:schlesinger_transformations}
In this section we derive evolution equations for an elementary Schlesinger transformation 
$\left\{\begin{smallmatrix}
	\alpha&\beta\\\mu&\nu
\end{smallmatrix}\right\}$
in terms of the coefficient matrices $\mathbf{A}_{i}$ of the Fuchsian system and then  
lift them to get the discrete Schlesinger dynamic
on the decomposition space $\mathcal{B}\times \mathcal{C}$.
Take  $\mathbf{R}(z)$ to be 
of the form considered in Lemma~\ref{lem:el-divs}. Then we have the following result.

\begin{theorem}\label{thm:schlesinger} Consider  transformation (\ref{eq:sch-transf}) with 
	\begin{equation}
		\mathbf{R}(z) = \mathbf{I} + \frac{z_{0} - \zeta_{0}}{z - z_{0}} 
		\frac{\mathbf{f} \mathbf{g}^{\dag}}{\mathbf{g}^{\dag} \mathbf{f}}
		= \mathbf{I} + \frac{ z_{0} - \zeta_{0}}{z - z_{0} }\mathbf{P},
		\qquad z_{0}\neq \zeta_{0}.\label{eq:R-matrix-generic}
	\end{equation}
	Then 
	\begin{enumerate}[(i)]
		\item the poles $\{\bar{z}_{i}\}$ of $\bar{\mathbf{A}}(z)$ coincide with the poles $\{z_{i}\}$
		of $\mathbf{A}(z)$ if and only if, for some choice of indices 
		$\alpha$ and $\beta$, $\beta\neq \alpha$, we have $z_{0} = z_{\alpha}$, $\zeta_{0} = z_{\beta}$, and
		\begin{equation}
			\bar{\mathbf{A}}_{\alpha} \mathbf{P} = \mathbf{P} \mathbf{A}_{\alpha} - \mathbf{P}, \qquad 
			\mathbf{P}\bar{\mathbf{A}}_{\beta} = \mathbf{A}_{\beta} \mathbf{P} + \mathbf{P}.
			\label{eq:Ai-special-P}
		\end{equation}
		Then, for some indices	$\mu$ and $\nu$, 
		\begin{equation}
			 \quad 
			\mathbf{f} \sim \bar{\mathbf{b}}_{\alpha,\mu}\sim \mathbf{b}_{\beta,\nu},\quad
			\mathbf{g}^{\dag} \sim \mathbf{c}^{\mu\dag}_{\alpha} \sim \bar{\mathbf{c}}_{\beta}^{\nu\dag},\quad
			\bar{\theta}_{\alpha}^{\mu} = \theta_{\alpha}^{\mu} - 1,\quad \bar{\theta}_{\beta}^{\nu} = \theta_{\beta}^{\nu} + 1,
			\label{eq:sch-R-matrix-params}
		\end{equation}		
		where $\sim$ means \emph{proportional}, and
		so such $\mathbf{R}(z)$ defines the  elementary transformation $\left\{\begin{smallmatrix}
			\alpha&\beta\\\mu&\nu
		\end{smallmatrix}\right\}$.
		
		\item For $\mathbf{R}(z)$ satisfying (\ref{eq:sch-R-matrix-params}), 
		\begin{equation}
			\mathbf{R}(z) = \mathbf{I} + \frac{ z_{\alpha} - z_{\beta} }{ z - z_{\alpha} } 
			\frac{ \mathbf{b}_{\beta,\nu} \mathbf{c}_{\alpha}^{\mu\dag } }{ \mathbf{c}_{\alpha}^{\mu\dag } \mathbf{b}_{\beta,\nu}  },
			\label{eq:R-Sch}
		\end{equation}	the residue matrices of $\mathbf{A}(z)$ and 
		$\bar{\mathbf{A}}(z)$ are connected by the equations
		\begin{align}
		 \bar{\mathbf{A}}_{i} \mathbf{R}(z_{i}) &= \mathbf{R}(z_{i}) \mathbf{A}_{i}\quad\text{or equivalently}\quad 
			\mathbf{A}_{i} \mathbf{R}^{-1}(z_{i}) = \mathbf{R}^{-1}(z_{i}) \bar{\mathbf{A}}_{i},\quad i\neq \alpha,\beta,
			\label{eq:Ai-generic}\\
			\bar{\mathbf{A}}_{\beta} \mathbf{Q} &=  \mathbf{Q} \mathbf{A}_{\beta}
			\quad\text{and}\quad
			\mathbf{A}_{\alpha} \mathbf{Q} =  \mathbf{Q}\bar{\mathbf{A}}_{\alpha},
			\quad\text{ where } \mathbf{Q} = \mathbf{I} - \mathbf{P} = \mathbf{R}(z_{\beta}) = \mathbf{R}^{-1}(z_{\alpha}).
			\label{eq:Ai-special-Q}
		\end{align}
		\item Elementary Schlesinger transformation 
		$\left\{\begin{smallmatrix}
			\alpha&\beta\\\mu&\nu
		\end{smallmatrix}\right\}$
		corresponds to the following map on the coefficient matrices (discrete Schlesinger evolution equations):
		\begin{align}
			\bar{\mathbf{A}}_{i} &= \mathbf{R}(z_{i}) \mathbf{A}_{i} \mathbf{R}^{-1}(z_{i}),\qquad i\neq \alpha,\beta, \label{eq:Ai}\\
			\bar{\mathbf{A}}_{\alpha} &= \mathbf{A}_{\alpha} -[\mathbf{A}_{\alpha},\mathbf{P}]
			- \mathbf{P} + \sum_{i\neq \alpha}\left(\frac{ z_{\beta} - z_{\alpha} }{ z_{i} - z_{\alpha} }\right)\mathbf{P}\mathbf{A}_{i}\mathbf{Q},
			\label{eq:Aalpha}\\
			\bar{\mathbf{A}}_{\beta} &= \mathbf{A}_{\beta} + [\mathbf{A}_{\beta},\mathbf{P}] + \mathbf{P}
			+ \sum_{i\neq \beta} \left(\frac{ z_{\alpha} - z_{\beta} }{ z_{i} - z_{\beta} }\right)\mathbf{Q}\mathbf{A}_{i}\mathbf{P}.\label{eq:Abeta}
		\end{align}
	\end{enumerate}

\end{theorem}

\begin{proof}
	From the Schlesinger transformation equations
	\begin{equation}
			\bar{\mathbf{A}}(z)\mathbf{R}(z) = \mathbf{R}(z) \mathbf{A}(z) + \frac{d \mathbf{R}(z)}{dz} 
			\quad\text{ or }\quad 
			\mathbf{R}^{-1}(z)\bar{\mathbf{A}}(z) =  \mathbf{A}(z) \mathbf{R}^{-1}(z) - \frac{d \mathbf{R}^{-1}(z)}{dz}, 		
			\label{eq:sch-transf-A}
	\end{equation}
	and
	\begin{equation*}
		\frac{d \mathbf{R}(z)}{dz} \mathbf{R}^{-1}(z) = - \mathbf{R}(z) \frac{d \mathbf{R}^{-1}(z)}{dz} 
		= \left(\frac{1}{z - \zeta_{0}} - \frac{1}{z - z_{0}}\right)\frac{\mathbf{f} \mathbf{g}^{\dag}}{\mathbf{g}^{\dag} \mathbf{f}},
	\end{equation*}
	we see that
	\begin{equation*}
		\operatorname{tr}	\frac{d \mathbf{R}(z)}{dz} \mathbf{R}^{-1}(z) = \left(\frac{1}{z - \zeta_{0}} - \frac{1}{z - z_{0}}\right) 
		= \operatorname{tr} \bar{\mathbf{A}}(z) - \operatorname{tr} \mathbf{A}(z) 
		= \sum_{i=1}^{n} \frac{\operatorname{tr}(\bar{\mathbf{A}}_{i}) - \operatorname{tr}(\mathbf{A}_{i})}{z-z_{i}}.
	\end{equation*}
	Therefore,  $z_{0},\zeta_{0}\in \{z_{i}\}_{i=1}^{n}$. Let $z_{0} = z_{\alpha}$ and $\zeta_{0} = z_{\beta}$. Then
	$\operatorname{tr}(\bar{\mathbf{A}}_{\alpha}) - \operatorname{tr}(\mathbf{A}_{\alpha}) = -1$ and 
	$\operatorname{tr}(\bar{\mathbf{A}}_{\beta}) - \operatorname{tr}(\mathbf{A}_{\beta}) = 1$. 

	At $z_{\alpha}$ the first equation in (\ref{eq:sch-transf-A}) becomes
	\begin{align*}			
		\left(  \frac{\bar{\mathbf{A}}_{\alpha}}{z - z_{\alpha}} + \sum_{i\neq \alpha} \frac{\bar{\mathbf{A}}_{i}}{z - z_{i}} \right)
		\left(\mathbf{I} + \frac{z_{\alpha} - z_{\beta}}{z - z_{\alpha}} \mathbf{P}\right) 
		&=  \left(\mathbf{I} + \frac{z_{\alpha} - z_{\beta}}{z - z_{\alpha}} 
		\mathbf{P}\right) 
		\left( \frac{\mathbf{A}_{\alpha}}{z - z_{\alpha}} + \sum_{i\neq \alpha} 
		\frac{\mathbf{A}_{i}}{z - z_{i}}\right) - \frac{z_{\alpha} - z_{\beta}}{(z - z_{\alpha})^{2}}\mathbf{P}.\label{eq:sch-transf-alpha}
	\end{align*}
	The $(z-z_{\alpha})^{-2}$-terms give
	\begin{equation*}
	\bar{\mathbf{A}}_{\alpha} \mathbf{P}= \mathbf{P} \mathbf{A}_{\alpha} - \mathbf{P}, 
	\end{equation*}
	and so we see that $\mathbf{g}^{\dag}$ has to be an eigenvector of $\mathbf{A}_{\alpha}$ 
	and $\mathbf{f}$ has to be an eigenvector of 
	$\bar{\mathbf{A}}_{\alpha}$. Choose an index $\mu$ and let 
	$\mathbf{g}^{\dag} \sim \mathbf{c}_{\alpha}^{\mu\dag}$
	and $\mathbf{f} \sim \bar{\mathbf{b}}_{\alpha,\mu}$, 
	then
	$\theta_{\alpha}^{\mu} = \bar{\theta}_{\alpha}^{\mu} + 1$. Repeating this argument for 
	$z_{\beta}$ and the second equation in (\ref{eq:sch-transf-A}) 
	implies that $\mathbf{g}^{\dag} \sim \bar{\mathbf{c}}_{\beta}^{\nu\dag}$,
	$\mathbf{f} \sim \mathbf{b}_{\beta,\nu}$, and 
	$\theta_{\beta}^{\nu} = \bar{\theta}_{\beta}^{\nu} - 1$, for some choice of the index $\nu$, which
	completes the proof of (i).

	Since $\mathbf{R}(z)$ is regular (as a complex function) and invertible (as a matrix) at $z_{i}$ for $i\neq \alpha, \beta$, taking residues of 
	(\ref{eq:sch-transf-A}) at $z_{i}$ gives (\ref{eq:Ai-generic}), which is also equivalent to (\ref{eq:Ai}).
	Taking the residue of $\mathbf{R}(z)$ at $z_{\beta}$ of the first equation in (\ref{eq:sch-transf-A}) gives the first equation
	in 	(\ref{eq:Ai-special-Q}) and taking the residue at $z_{\alpha}$ of the second equation in (\ref{eq:sch-transf-A}) 
	gives the second equation in (\ref{eq:Ai-special-Q}). This competes the proof of (ii).

	Finally, taking the residue at infinity we get 
	$-\bar{\mathbf{A}}_{\infty} = \sum_{i=1}^{n} \bar{\mathbf{A}}_{i} = \sum_{i=1}^{n} \mathbf{A}_{i} = -\mathbf{A}_{\infty}$, 
	which becomes
	\begin{equation}
		\bar{\mathbf{A}}_{\alpha} + \bar{\mathbf{A}}_{\beta} = 
		\mathbf{A}_{\alpha} + \mathbf{A}_{\beta} 
		+ \sum_{i\neq \alpha, \beta} \left(\mathbf{A}_{i} - \bar{\mathbf{A}}_{i}\right)
		= \mathbf{A}_{\alpha} + \mathbf{A}_{\beta} 
		+ \sum_{i\neq \alpha, \beta} \left(\mathbf{A}_{i} - \mathbf{R}(z_{i}) \mathbf{A}_{i} \mathbf{R}^{-1}(z_{i})\right),\label{eq:A-inf}
	\end{equation}
	after using~(\ref{eq:Ai-generic}). It is easy to see that taking the residue at $z_{\alpha}$ of the first equation in (\ref{eq:sch-transf-A}) 
	or the residue at $z_{\beta}$ of the second equation in (\ref{eq:sch-transf-A}) results in equivalent equations.
	
	Right-multiplying (\ref{eq:A-inf}) by $\mathbf{Q}$ and using (\ref{eq:Ai-special-Q})
	\begin{equation*}
		\bar{\mathbf{A}}_{\alpha} \mathbf{Q}  = \mathbf{A}_{\alpha} \mathbf{Q} + 
		(\mathbf{A}_{\beta} \mathbf{Q} - \mathbf{Q} \mathbf{A}_{\beta}) + 
		\sum_{i\neq \alpha, \beta}\left(- \frac{ z_{\alpha} - z_{\beta} }{ z_{i} - z_{\alpha} }\right) \mathbf{P}\mathbf{A}_{i}\mathbf{Q}
		= \mathbf{A}_{\alpha} - \mathbf{A}_{\alpha}\mathbf{P}
		+ \sum_{i\neq \alpha}\left(\frac{z_{\beta} - z_{\alpha}}{ z_{i} - z_{\alpha} }\right) \mathbf{P}\mathbf{A}_{i}\mathbf{Q},
	\end{equation*}
	since $\mathbf{Q} \mathbf{A}_{\beta} = \mathbf{Q}\mathbf{A}_{\beta}\mathbf{Q}$ and
	\begin{equation*}
		\left(\mathbf{A}_{i} - \mathbf{R}(z_{i}) \mathbf{A}_{i} \mathbf{R}^{-1}(z_{i})\right) \mathbf{Q} = 
		\left( \mathbf{I} - \mathbf{R}(z_{i})\right)\mathbf{A}_{i}\mathbf{Q} = - \frac{ z_{\alpha} - z_{\beta} }{ z_{i} - z_{\alpha} } 
		\mathbf{P} \mathbf{A}_{i}\mathbf{Q}, \quad\text{since }\mathbf{R}^{-1}(z_{i}) \mathbf{Q} = \mathbf{Q}.
	\end{equation*} 
	Finally, using $\bar{\mathbf{A}}_{\alpha}\mathbf{Q} = \bar{\mathbf{A}}_{\alpha} - \bar{\mathbf{A}}_{\alpha}\mathbf{P} = 
	\bar{\mathbf{A}}_{\alpha} - \mathbf{P}\mathbf{A}_{\alpha} + \mathbf{P}$ and simplifying, we get (\ref{eq:Aalpha}). 
	Equation~(\ref{eq:Abeta}) is obtained in the same way, only we start by left-multiplying (\ref{eq:A-inf}) by $\mathbf{Q}$. 
	
\end{proof}

We now want to lift these equations to the decomposition space $\mathcal{B}\times \mathcal{C}$. Our strategy is the following.
Imposing the \emph{orthogonality conditions} 
\begin{equation}
	\mathbf{C}_{i}^{\dag} \mathbf{B}_{i} = \mathbf{\Theta}_{i},\quad\text{ and }\quad
	\bar{\mathbf{C}}_{i}^{\dag} \bar{\mathbf{B}}_{i} = \mathbf{\Theta}_{i}\label{eq:ortho-cond}
\end{equation}	
allows us to ``split'' equations~(\ref{eq:Ai}--\ref{eq:Abeta}) to get the map from 
$(\mathcal{B}\times \mathcal{C})_{(\mathbf{\Theta},\mathbf{A}_{\infty})}$ to
$(\bar{\mathcal{B}}\times \bar{\mathcal{C}})_{(\bar{\mathbf{\Theta}},\bar{\mathbf{A}}_{\infty})}$ that we 
then extend to the (dense open subset of the) whole of the decomposition space.

For convenience we list here some consequences of the orthogonality conditions that we often use:
	\begin{alignat*}{5}
	\mathbf{P}\mathbf{b}_{\beta,\nu} &= \mathbf{b}_{\beta,\nu},&\quad 
	\mathbf{Q}\mathbf{b}_{\beta,\nu} &= \mathbf{0}, &\quad
	\mathbf{P}\mathbf{b}_{\alpha,j} &= \mathbf{0},&\quad 
	\mathbf{Q}\mathbf{b}_{\alpha,j} &= \mathbf{b}_{\alpha,j}&\quad (j&\neq \mu),\\
	\mathbf{c}_{\alpha}^{\mu\dag}\mathbf{P} &= \mathbf{c}_{\alpha}^{\mu\dag},&\quad
	\mathbf{c}_{\alpha}^{\mu\dag}\mathbf{Q} &= \mathbf{0},&\quad
	\mathbf{c}_{\beta}^{j\dag}\mathbf{P} &= \mathbf{0},&\quad
	\mathbf{c}_{\beta}^{j\dag}\mathbf{Q} &= \mathbf{c}_{\beta}^{j\dag}&\quad (j&\neq \nu).
	\label{eq:ortho}
\end{alignat*}

We begin with the following easy Lemma.

\begin{lemma}\label{cor:schlesinger-lift} Let $\mathbf{A}(z)$ and $\bar{\mathbf{A}}(z)$ be connected by an elementary 
	Schlesinger transformation 
	$\left\{\begin{smallmatrix}	\alpha&\beta\\\mu&\nu \end{smallmatrix}\right\}$
	given by $\mathbf{R}(z)$ of the form (\ref{eq:R-matrix-generic}) satisfying (\ref{eq:sch-R-matrix-params})
	and let $\mathbf{A}_{i} = \mathbf{B}_{i}\mathbf{C}_{i}^{\dag}$ with $\mathbf{C}_{i}^{\dag}\mathbf{B}_{i} = \mathbf{\Theta}_{i}$,
	and similarly for $\bar{\mathbf{A}}_{i}$.
	Then the corresponding points 
	$(\mathbf{B},\mathbf{C}^{\dag})$ and 
	$(\bar{\mathbf{B}},\bar{\mathbf{C}}^{\dag})$ in the 
	decomposition space are related as follows:
	\begin{enumerate}[(i)]
		\item If $i\neq \alpha,\beta$, $\mathbf{R}(z_{i})$ is invertible and 
		\begin{equation}
			\bar{\mathbf{B}}_{i}\sim \mathbf{R}(z_{i}) \mathbf{B}_{i}\quad 
			(\text{equivalently, }\mathbf{B}_{i}\sim \mathbf{R}^{-1}(z_{i}) \bar{\mathbf{B}}_{i})
			\text{ and }\quad 
			\mathbf{C}_{i}^{\dag} \sim \bar{\mathbf{C}}_{i}^{\dag} \mathbf{R}(z_{i})\quad (
			\bar{\mathbf{C}}_{i}^{\dag} \sim \mathbf{C}_{i}^{\dag} \mathbf{R}^{-1}(z_{i})).\label{eq:BC-generic}
		\end{equation}
		Therefore 
		\begin{equation}
			\bar{\mathbf{B}}_{i} \bar{\mathbf{D}}_{i} = \mathbf{R}(z_{i}) \mathbf{B}_{i}, \qquad 
			\bar{\mathbf{\Delta}}_{i}\bar{\mathbf{C}}_{i}^{\dag} = \mathbf{C}_{i}^{\dag}\mathbf{R}^{-1}(z_{i}),
		\end{equation}
		where $\bar{\mathbf{D}}_{i}$ and $\bar{\mathbf{\Delta}}_{i}$ are \emph{diagonal} normalization matrices,
		$\bar{\mathbf{D}}_{i} \bar{\mathbf{\Delta}}_{i} = \mathbf{I}$.
		\item For the special indices $\alpha$ and $\beta$ we have 
		\begin{equation}
			\mathbf{C}_{\beta}^{\dag} \sim \bar{\mathbf{C}}_{\beta}^{\dag} \mathbf{Q},\quad
			\bar{\mathbf{B}}_{\beta}\sim \mathbf{Q} \mathbf{B}_{\beta},\quad
			 \text{ and }\quad
			\bar{\mathbf{C}}_{\alpha}^{\dag} \sim \mathbf{C}_{\alpha}^{\dag} \mathbf{Q}, \quad
			\mathbf{B}_{\alpha}\sim \mathbf{Q} \bar{\mathbf{B}}_{\alpha}.\label{eq:BC-special}
		\end{equation}
		It is still possible to write
		\begin{equation}
			\bar{\mathbf{B}}_{\beta} \bar{\mathbf{D}}_{\beta} = \mathbf{Q} \mathbf{B}_{\beta}, \qquad
			\bar{\mathbf{\Delta}}_{\alpha} \bar{\mathbf{C}}_{\alpha}^{\dag} = \mathbf{C}_{\alpha}^{\dag} \mathbf{Q},
		\end{equation}
		but since $\mathbf{Q}\mathbf{b}_{\beta,\nu} = \mathbf{c}_{\alpha}^{\mu\dag} \mathbf{Q} = \mathbf{0}$,
		 $\left(\bar{\mathbf{D}}_{\beta}\right)_{\nu}^{\nu} = \left(\bar{\mathbf{\Delta}}_{\alpha}\right)_{\mu}^{\mu} = 0$.
	\end{enumerate}
	
\end{lemma}

\begin{proof}
	Multiplying first equation in (\ref{eq:Ai-generic}) on the right by $\mathbf{B}_{i}$ gives
	$\bar{\mathbf{A}}_{i} \mathbf{R}(z_{i}) \mathbf{B}_{i} = \mathbf{R}(z_{i}) 
	\mathbf{A}_{i} \mathbf{B}_{i} = \mathbf{R}(z_{i}) \mathbf{B}_{i} \mathbf{\Theta}_{i}$,
	and so the matrix $\mathbf{R}(z_{i}) \mathbf{B}_{i}$ is the matrix of eigenvectors of 
	$\bar{\mathbf{A}}_{i}$, $\bar{\mathbf{B}}_{i} \sim \mathbf{R}(z_{i}) \mathbf{B}_{i}$ or 
	$\bar{\mathbf{B}}_{i} \bar{\mathbf{D}}_{i} = \mathbf{R}(z_{i}) \mathbf{B}_{i}$, where 
	$\bar{\mathbf{D}}_{i}$ is a diagonal matrix of non-zero proportionality coefficients.
	The equations for  	$\bar{\mathbf{C}}_{i}^{\dag}$ are proved similarly.
	From the normalization condition 
		\begin{equation*}
			\bar{\mathbf{\Delta}}_{i}\bar{\mathbf{C}}_{i}^{\dag} \bar{\mathbf{B}}_{i} \bar{\mathbf{D}}_{i} 
			= \bar{\mathbf{\Delta}}_{i} \bar{\mathbf{\Theta}}_{i} \bar{\mathbf{D}}_{i} =  
			\bar{\mathbf{\Delta}}_{i} \bar{\mathbf{D}}_{i} \bar{\mathbf{\Theta}}_{i} =  
			\mathbf{C}_{i}^{\dag} \mathbf{B}_{i} = \mathbf{\Theta}_{i} = \bar{\mathbf{\Theta}}_{i},
		\end{equation*}
	where all matrices are diagonal, we immediately get $\bar{\mathbf{D}}_{i} \bar{\mathbf{\Delta}}_{i} = \mathbf{I}$.	
	
	For the special indices $\alpha$ and $\beta$, away from the pairs $(\alpha,\mu)$ and $(\beta,\nu)$, the situation is exactly the same.
\end{proof}


\begin{theorem}\label{thm:evolution}
		The elementary Schlesinger transformation
		$\left\{\begin{smallmatrix}
			\alpha&\beta\\\mu&\nu
		\end{smallmatrix}\right\}$
		defines the map 
		\begin{equation*}
			(\mathcal{B}\times \mathcal{C})_{(\mathbf{\Theta},\mathbf{A}_{\infty})} \to 
			(\bar{\mathcal{B}}\times \bar{\mathcal{C}})_{(\bar{\mathbf{\Theta}},\bar{\mathbf{A}}_{\infty})}	
		\end{equation*}
		given by the following evolution equations (grouped for convenience).
		\begin{enumerate}[(i)]
			\item Transformation vectors:
			\begin{equation}
				\bar{\mathbf{b}}_{\alpha,\mu} = \frac{ 1 }{ c_{\alpha}^{\mu} } \mathbf{b}_{\beta,\nu},\qquad
				\bar{\mathbf{c}}_{\beta}^{\nu\dag} = c_{\beta}^{\nu} \mathbf{c}_{\alpha}^{\mu\dag}.
				\label{eq:bb-cb-generators}				
			\end{equation}
			\item Generic indices:
			\begin{equation}
				\bar{\mathbf{b}}_{i,j} = 
				\frac{ 1 }{ c_{i}^{j} } \mathbf{R}(z_{i}) \mathbf{b}_{i,j},\,(i\neq \alpha\text{ and if } i= \beta, j\neq \nu);\quad 
				\bar{\mathbf{c}}_{i}^{j\dag} = c_{i}^{j}  \mathbf{c}_{i}^{j\dag} \mathbf{R}^{-1}(z_{i}),\, (i\neq \beta\text{ and if } i=\alpha, j\neq \mu).
				\label{eq:bb-cd-generic}
			\end{equation}
			\item Special indices:
			\begin{align}
				\bar{\mathbf{b}}_{\alpha,j} &= \frac{ 1 }{ c_{\alpha}^{j} }\left(\mathbf{I} - 
				 \frac{ \mathbf{P} \left(\sum_{i\neq \alpha} 
				\frac{ z_{\beta} - z_{\alpha}}{ z_{i} - z_{\alpha} }\mathbf{A}_{i}
				\right) }{ \theta_{\alpha}^{\mu} - \theta_{\alpha}^{j} - 1 }\right)\mathbf{b}_{\alpha,j},\qquad j\neq \mu;
				\label{eq:bb-ak}\\
				\bar{\mathbf{c}}_{\beta}^{j\dag} &= c_{\beta}^{j}\mathbf{c}_{\beta}^{j\dag} \left(
				\mathbf{I} - 
				\frac{ \left(\sum_{i\neq \beta}
				\frac{ z_{\alpha} - z_{\beta} }{ z_{i} - z_{\beta} } \mathbf{A}_{i}\right) \mathbf{P}}{ 
				\theta_{\beta}^{\nu} - \theta_{\beta}^{j} + 1 } 
				\right),\qquad j\neq \nu;
				\label{eq:cb-bk}\\
				\bar{\mathbf{b}}_{\beta,\nu} &= \frac{ 1 }{ c_{\beta}^{\nu}  }\left( (\theta_{\beta}^{\nu} + 1) \mathbf{I} + 
				\mathbf{Q}	
				\left(\mathbf{I} +  \sum_{j\neq \nu}\frac{ \mathbf{b}_{\beta,j} \mathbf{c}_{\beta}^{j\dag}}{ 
				\theta_{\beta}^{\nu} - \theta_{\beta}^{j} + 1 }\right) \left(
				\sum_{i\neq \beta} \frac{ z_{\alpha} - z_{\beta}  }{ 
				z_i - z_{\beta}  } \mathbf{A}_{i}\right)\right)
				\frac{ \mathbf{b}_{\beta,\nu} }{  \mathbf{c}_{\alpha}^{\mu\dag} \mathbf{b}_{\beta,\nu}  };
				\label{eq:bb-bn}\\
				\bar{\mathbf{c}}_{\alpha}^{\mu\dag} &= c_{\alpha}^{\mu} 
				\frac{ \mathbf{c}_{\alpha}^{\mu\dag} }{ \mathbf{c}_{\alpha}^{\mu\dag} \mathbf{b}_{\beta,\nu}  }\left(
				 (\theta_{\alpha}^{\mu} -  1) \mathbf{I}
				+\left(\sum_{i\neq \alpha}\frac{ z_{\beta} - z_{\alpha}
				 }{ z_{i} - z_{\alpha}  } \mathbf{A}_{i}\right)
				\left( \mathbf{I} + \sum_{j\neq \mu} 
				\frac{ \mathbf{b}_{\alpha,j} \mathbf{c}_{\alpha}^{j\dag} }{ \theta_{\alpha}^{\mu} - \theta_{\alpha}^{j} -1 }
				\right) \mathbf{Q}
				\right).\label{eq:cb-am}
			\end{align}
		\end{enumerate}
		
\end{theorem}

\begin{proof} Part (i) is just a restatement of~(\ref{eq:sch-R-matrix-params}) and part (ii) follows immediately from 
	Lemma~\ref{cor:schlesinger-lift} if we put 
	$\bar{\mathbf{D}}_{i} = \bar{\mathbf{\Delta}}^{-1}_{i} = \operatorname{diag}\{c_{i}^{1},\dots c_{i}^{r_{i}} \}$.
	Note, however, that for special indices $\alpha$ and $\beta$, since
	 $\left(\bar{\mathbf{D}}_{\beta}\right)_{\nu}^{\nu} = \left(\bar{\mathbf{\Delta}}_{\alpha}\right)_{\mu}^{\mu} = 0$,
	we have no information about $\bar{\mathbf{b}}_{\beta,\nu}$ and $\bar{\mathbf{c}}_{\alpha}^{\mu\dag}$.
	
	The difficult part of the Theorem is to establish equations in (iii).  
	Right-multiplying~(\ref{eq:Aalpha}) by $\mathbf{Q}\mathbf{b}_{\alpha,j} = \mathbf{b}_{\alpha,j}$, $j\neq \mu$,
	we get
	\begin{equation*}
		\bar{\mathbf{A}}_{\alpha} \mathbf{Q}\mathbf{b}_{\alpha,j}  = \theta_{\alpha}^{j} \mathbf{b}_{\alpha,j}
		+ \sum_{i\neq \alpha}\left(\frac{z_{\beta} - z_{\alpha}}{ z_{i} - z_{\alpha} }\right) \mathbf{P}\mathbf{A}_{i}\mathbf{b}_{\alpha,j}.
	\end{equation*}
	At the same time, in view of (\ref{eq:bb-cd-generic}) and orthogonality
	\begin{equation*}
		\bar{\mathbf{A}}_{\alpha} = \sum_{j\neq \mu} c_{\alpha}^{j} \bar{\mathbf{b}}_{\alpha,j} \mathbf{c}_{\alpha}^{j\dag} \mathbf{Q}
		+ \bar{\mathbf{b}}_{\alpha,\mu}\bar{\mathbf{c}}_{\alpha}^{\mu\dag},\quad 
		\bar{\mathbf{A}}_{\alpha} \mathbf{Q}\mathbf{b}_{\alpha,j} = 
		c_{\alpha}^{j} \bar{\mathbf{b}}_{\alpha,j}\theta_{\alpha}^{j} + \bar{\mathbf{b}}_{\alpha,\mu}\bar{\mathbf{c}}_{\alpha}^{\mu\dag} \mathbf{b}_{\alpha,j}.
	\end{equation*}
	Thus,
	\begin{equation*}
		c_{\alpha}^{j} \bar{\mathbf{b}}_{\alpha,j}\theta_{\alpha}^{j} + \bar{\mathbf{b}}_{\alpha,\mu}\bar{\mathbf{c}}_{\alpha}^{\mu\dag} \mathbf{b}_{\alpha,j}
		= \theta_{\alpha}^{j} \mathbf{b}_{\alpha,j} + 
		\sum_{i\neq \alpha} \frac{ z_{\beta} - z_{\alpha} }{ z_{i} - z_{\alpha} } \mathbf{P}\mathbf{A}_{i}\mathbf{b}_{\alpha,j}.			
	\end{equation*}
To evaluate $\bar{\mathbf{c}}_{\alpha}^{\mu\dag} \mathbf{b}_{\alpha,j}$ we now left-multiply by 
$\bar{\mathbf{c}}_{\alpha}^{\mu\dag}$ and use the fact that 
$\bar{\mathbf{c}}_{\alpha}^{\mu\dag} \mathbf{P} =  (\bar{\theta}_{\alpha}^{\mu} \mathbf{c}_{\alpha}^{\mu\dag})/ 
(\mathbf{c}_{\alpha}^{\mu\dag}\bar{\mathbf{b}}_{\alpha,\mu})$:
\begin{align*}
	\bar{\theta}_{\alpha}^{\mu} \bar{\mathbf{c}}_{\alpha}^{\mu\dag} \mathbf{b}_{\alpha,j} &= 
	\theta_{\alpha}^{j} \bar{\mathbf{c}}_{\alpha}^{\mu\dag} \mathbf{b}_{\alpha,j}	+ 
	\sum_{i\neq \alpha}\frac{ z_{\beta} - z_{\alpha} }{ z_{i} - z_{\alpha} } \cdot 
	\frac{ \bar{\theta}_{\alpha}^{\mu} \mathbf{c}_{\alpha}^{\mu} \mathbf{A}_{i} \mathbf{b}_{\alpha,j} }{ 
	\mathbf{c}_{\alpha}^{\mu\dag}\bar{\mathbf{b}}_{\alpha,\mu} }.
	\intertext{Hence}
	\bar{\mathbf{c}}_{\alpha}^{\mu\dag} \mathbf{b}_{\alpha,j} &= 
	\frac{ \bar{\theta}_{\alpha}^{\mu} \mathbf{c}_{\alpha}^{\mu} }{ \mathbf{c}_{\alpha}^{\mu\dag}\bar{\mathbf{b}}_{\alpha,\mu}  }
	\left(\sum_{i\neq \alpha} \frac{ z_{\beta} - z_{\alpha} }{ z_{i} - z_{\alpha} } \mathbf{A}_{i}\right) 
	\frac{ \mathbf{b}_{\alpha,j} }{ \bar{\theta}_{\alpha}^{\mu} - \theta_{\alpha}^{j} }.
\end{align*}
Then 
\begin{equation*}
	\bar{\mathbf{b}}_{\alpha,j} c_{\alpha}^{j} \theta_{\alpha}^{j} = 
	\theta_{\alpha}^{j} \mathbf{b}_{\alpha,j} - \bar{\theta}_{\alpha,\mu} \mathbf{P} 
	\left(\sum_{i\neq \alpha} \frac{ z_{\beta} - z_{\alpha} }{ z_{i} - z_{\alpha} } \mathbf{A}_{i}\right) 
	\frac{ \mathbf{b}_{\alpha,j} }{ \bar{\theta}_{\alpha}^{\mu} - \theta_{\alpha}^{j} } + 
	\mathbf{P}\left(\sum_{i\neq \alpha} \frac{ z_{\beta} - z_{\alpha} }{ z_{i} - z_{\alpha} } \mathbf{A}_{i}\right)  \mathbf{b}_{\alpha,j},
\end{equation*}
which, after easy simplification, becomes (\ref{eq:bb-ak}). Equation (\ref{eq:cb-bk}) is proved in the exactly same way.

Next, observe that in view of~(\ref{eq:bb-cb-generators}), (\ref{eq:bb-cd-generic}), and (\ref{eq:cb-bk}) we can write
\begin{align}
	\bar{\mathbf{A}}_{\beta} &= \sum_{j\neq v}\bar{\mathbf{b}}_{\beta,j} \bar{\mathbf{c}}_{\beta}^{j\dag} + 
	\bar{\mathbf{b}}_{\beta,\nu} \bar{\mathbf{c}}_{\beta}^{\nu\dag}
	= \sum_{j\neq \nu} \mathbf{Q}\mathbf{b}_{\beta,j}\mathbf{c}_{\beta}^{j\dag}
	\left( \mathbf{I}  - \frac{ \left(\sum_{i\neq \beta}
	\frac{ z_{\alpha} - z_{\beta} }{ z_{i} - z_{\beta} } \mathbf{A}_{i}\right) \mathbf{P}}{ 
	\theta_{\beta}^{\nu} - \theta_{\beta}^{j} + 1 } 	\right) 
	+ c_{\beta}^{\nu} \bar{\mathbf{b}}_{\beta,\nu} \mathbf{c}_{\alpha}^{\mu\dag}.\label{eq:Abeta-1}
	\intertext{Thus, we can compute $\bar{\mathbf{A}}_{\beta}\mathbf{b}_{\beta,\nu}$ either using~(\ref{eq:Abeta-1}) or (\ref{eq:Abeta}):}
	\bar{\mathbf{A}}_{\beta}\mathbf{b}_{\beta,\nu} &= c_{\beta}^{\nu}(\mathbf{c}_{\alpha}^{\mu\dag}\mathbf{b}_{\beta,\nu})  \bar{\mathbf{b}}_{\beta,\nu} 
	- \sum_{j\neq \nu} \frac{\mathbf{Q}\mathbf{b}_{\beta,j}\mathbf{c}_{\beta}^{j\dag}}{\theta_{\beta}^{\nu} - \theta_{\beta}^{j} + 1} 
	\left( \left(\sum_{i\neq \beta}
	\frac{ z_{\alpha} - z_{\beta} }{ z_{i} - z_{\beta} } \mathbf{A}_{i}\right)\mathbf{b}_{\beta,\nu} \right) \notag\\
	&= (\theta_{\beta}^{\nu} + 1)\mathbf{b}_{\beta,\nu} + 
	\mathbf{Q}\sum_{i\neq \beta} \frac{ z_{\alpha} - z_{\beta} }{ z_{i} - z_{\beta} } \mathbf{A}_{i}\mathbf{b}_{\beta,\nu}.\notag
\end{align}
Solving for $\bar{\mathbf{b}}_{\beta,\nu}$ and simplifying gives (\ref{eq:bb-bn}). Equation~(\ref{eq:cb-am}) is proved similarly.
\end{proof}


\section{Hamiltonian Formulation} 
\label{sec:hamiltonian_formulation}
The main objective of this part is to give the Hamiltonian formulation of the discrete dynamic described in the 
previous section. First, pick points $z_{1},\dots,z_{n}$, $z_{i}\neq z_{j}$, and  characteristic indices
$\theta_{i}^{j}$, $1\leq i\leq n$, $1\leq j\leq r_{i}$, $\theta_{i}^{j}\neq \theta_{i}^{k}$ 
for $i\neq j$ (this is an important assumption that corresponds to Assumption~\ref{assume:diagonal} but is slightly more restrictive)
and put $\bar{\theta}_{\alpha}^{\mu} = \theta_{\alpha}^{\mu}$, $\bar{\theta}_{\beta}^{\nu} = \theta_{\beta}^{\nu} + 1$ and
$\bar{\theta}_{i}^{j} = \theta_{i}^{j}$ otherwise. These are  parameters of our dynamic. 

As before, put $\mathbf{\Theta}_{i} = \operatorname{diag}\{\theta_{i}^{1},\dots,\theta_{i}^{r_{1}} \}$
and $\bar{\mathbf{\Theta}}_{i} = \operatorname{diag}\{\bar{\theta}_{i}^{1},\dots,\bar{\theta}_{i}^{r_{1}} \}$ and let
\begin{equation}
	\mathbf{P} = \frac{ \mathbf{b}_{\beta,\nu} \bar{\mathbf{c}}_{\beta}^{\nu\dag} }{ \bar{\mathbf{c}}_{\beta}^{\nu\dag} \mathbf{b}_{\beta,\nu} },
	\qquad \mathbf{R}(z) = \mathbf{I} + \frac{ z_{\alpha} - z_{\beta} }{ z- z_{\alpha}  }\mathbf{P},\qquad \mathbf{Q} = \mathbf{I} - \mathbf{P}. 
	\label{eq:R-matrix-Ham}
\end{equation}

Our goal then is to find a function $\mathcal{H}^{+} = \mathcal{H}^{+}(\mathbf{B},\bar{\mathbf{C}}^{\dag})$ on $\mathcal{B}\times \bar{\mathcal{C}}$
so that, if we use equations (\ref{eq:hplus-gen-eqns}) to define $\bar{\mathbf{b}}_{i,j}$ and $\mathbf{c}_{i}^{j\dag}$  (recall Remark~\ref{rem:coords}),
\begin{equation*}
	\bar{\mathbf{b}}_{i,j} = \frac{ \partial \mathcal{H}^{+} }{ \partial \bar{\mathbf{c}}_{i}^{j\dag}}(\mathbf{B},\bar{\mathbf{C}}^{\dag})\quad
	\text{ and }\quad 
	\mathbf{c}_{i}^{j\dag} = \frac{ \partial \mathcal{H}^{+} }{ \partial \mathbf{b}_{i,j}}(\mathbf{B},\bar{\mathbf{C}}^{\dag}),
\end{equation*}
and construct corresponding matrices $\mathbf{B}_{i}$, $\bar{\mathbf{B}}_{i}$, $\mathbf{C}_{i}^{\dag}$, $\bar{\mathbf{C}}_{i}^{\dag}$,
$\mathbf{A}_{i} = \mathbf{B}_{i} \mathbf{C}_{i}^{\dag}$, and $\bar{\mathbf{A}}_{i} = \bar{\mathbf{B}}_{i} \bar{\mathbf{C}}_{i}^{\dag}$, then the following 
is true, \emph{cf}.~equations~(\ref{eq:sch-R-matrix-params}--\ref{eq:Ai-special-Q}) and (\ref{eq:A-inf}):
\begin{align}
	\bar{\mathbf{b}}_{\alpha,\mu}&\sim \mathbf{b}_{\beta,\nu} \quad\text{ and }\quad\mathbf{c}_{\alpha}^{\mu\dag}\sim \bar{\mathbf{c}}_{\beta}^{\nu\dag};
	\label{eq:transf-Ham}\\
	\mathbf{C}_{i}^{\dag} \mathbf{B}_{i} &= \mathbf{\Theta}_{i} \quad\text{ and }\quad\bar{\mathbf{C}}_{i}^{\dag} \bar{\mathbf{B}}_{i} = \bar{\mathbf{\Theta}}_{i};
	\label{eq-orto-Ham}\\
	\bar{\mathbf{A}}_{i} \mathbf{R}(z_{i}) &= \mathbf{R}(z_{i})\mathbf{A}_{i}\text{ for }i\neq \alpha,\beta,\qquad 
	\bar{\mathbf{A}}_{\beta} \mathbf{Q} = \mathbf{Q} \mathbf{A}_{\beta}, \quad\text{ and }\quad\mathbf{Q}\bar{\mathbf{A}}_{\alpha} = \mathbf{A}_{\alpha} \mathbf{Q};
	\label{eq:Schlesinger-Ham}\\
	\sum_{i} \bar{\mathbf{A}}_{i} 
	&= \sum_{i} \mathbf{A}_{i}.\label{eq:resinf-Ham}
\end{align}

It is convenient to separate special indices $(\alpha,\mu)$ and $(\beta,\nu)$ from generic indices $(i,j)$. Thus, we put
\begin{equation}
	\mathbf{A}_{\alpha}' = \sum_{j\neq \mu} \mathbf{b}_{\alpha,j} \mathbf{c}_{\alpha}^{j\dag} = 
	\mathbf{B}_{\alpha}' (\mathbf{C}_{\alpha}^{\dag})'\quad
	\Big(\text{thus, } \mathbf{A}_{\alpha} = \sum_{j} \mathbf{b}_{\alpha,j} \mathbf{c}_{\alpha}^{j\dag} = \mathbf{A}_{\alpha}' +
	\mathbf{b}_{\alpha,\mu}\mathbf{c}_{\alpha}^{\mu\dag}\Big),
\end{equation}	
and so on. Then, since in view of (\ref{eq:R-matrix-Ham}) and (\ref{eq:transf-Ham})
\begin{equation}
	\mathbf{Q} \mathbf{b}_{\beta,\nu} = \mathbf{Q}\bar{\mathbf{b}}_{\alpha,\mu} = \mathbf{0},\qquad 
	\mathbf{c}_{\alpha}^{\mu\dag}\mathbf{Q} = \bar{\mathbf{c}}_{\beta}^{\nu\dag}\mathbf{Q} = \mathbf{0},
\end{equation}
equations for $i=\alpha,\beta$ in (\ref{eq:Schlesinger-Ham}) become $\bar{\mathbf{A}}_{\beta}' \mathbf{Q} = \mathbf{Q} \mathbf{A}_{\beta}'$, and 
$\mathbf{Q}\bar{\mathbf{A}}_{\alpha}' = \mathbf{A}_{\alpha}' \mathbf{Q}$.

Next, observe that the orthogonality conditions (\ref{eq-orto-Ham}) impose some \emph{a priori} conditions on the domain of $\mathcal{H}^{+}$ that 
cut out a certain locus $(\mathcal{B}\times\bar{\mathcal{C}})^{\operatorname{ort}}\subset (\mathcal{B}\times\bar{\mathcal{C}})$. 
Indeed, from (\ref{eq:transf-Ham}) and the condition 
$\bar{\mathbf{C}}_{\alpha}^{\dag} \bar{\mathbf{B}}_{\alpha} = \bar{\mathbf{\Theta}}_{\alpha}$ we get 
$(\bar{\mathbf{C}}_{\alpha}^{\dag})' \mathbf{b}_{\beta,\nu} = \mathbf{0}$. Similarly, $\bar{\mathbf{c}}_{\beta}^{\nu} \mathbf{B}_{\alpha}' = \mathbf{0}$.
For $i\neq \alpha,\beta$ we get
\begin{equation*}
	\mathbf{\Theta}_{i} (\bar{\mathbf{C}}_{i}^{\dag} \mathbf{R}(z_{i}) \mathbf{B}_{i})
	= \bar{\mathbf{C}}_{i}^{\dag} \bar{\mathbf{A}}_{i} \mathbf{R}(z_{i}) \mathbf{B}_{i} = 
	\bar{\mathbf{C}}_{i}^{\dag} \mathbf{R}(z_{i}) \mathbf{A}_{i} \mathbf{B}_{i} = 
	(\bar{\mathbf{C}}_{i}^{\dag} \mathbf{R}(z_{i}) \mathbf{B}_{i}) \mathbf{\Theta}_{i},
\end{equation*}
and in view of our assumption $\theta_{i}^{j}\neq \theta_{i}^{k}$ for $j\neq k$, 
$\bar{\mathbf{C}}_{i}^{\dag} \mathbf{R}(z_{i}) \mathbf{B}_{i}$ must be diagonal. 
We also need to require that it is invertible (which is an open condition). Similarly, 
$((\bar{\mathbf{C}}_{\beta}^{\dag})' \mathbf{Q} \mathbf{B}_{\beta}')$ must be diagonal (and we also require it is invertible).
We will later see that we need $(\bar{\mathbf{C}}_{\alpha}^{\dag})' \mathbf{B}_{\alpha}'$ to be diagonal (and invertible) as well. 
Thus, we put
\begin{equation}
	(\mathcal{B}\times\bar{\mathcal{C}})^{\operatorname{ort}} = \{
	(\bar{\mathbf{C}}_{\alpha}^{\dag})' \mathbf{b}_{\beta,\nu} = \mathbf{0},\,
	\bar{\mathbf{c}}_{\beta}^{\nu} \mathbf{B}_{\alpha}' = \mathbf{0},\,
	(\bar{\mathbf{C}}_{\alpha}^{\dag})' \mathbf{B}_{\alpha}'\,,
	\bar{\mathbf{C}}_{i}^{\dag} \mathbf{R}(z_{i}) \mathbf{B}_{i}\text{ are diagonal and invertible, }i\neq \alpha
	\}.\label{eq:Ham-ortho-locus}
\end{equation}

We can illustrate the relationship between evolution dynamic and discrete Hamiltonian by the diagram on Figure~\ref{fig:discr-ham}:
\begin{figure}[h]
	\centering
		\includegraphics[width=3.5in]{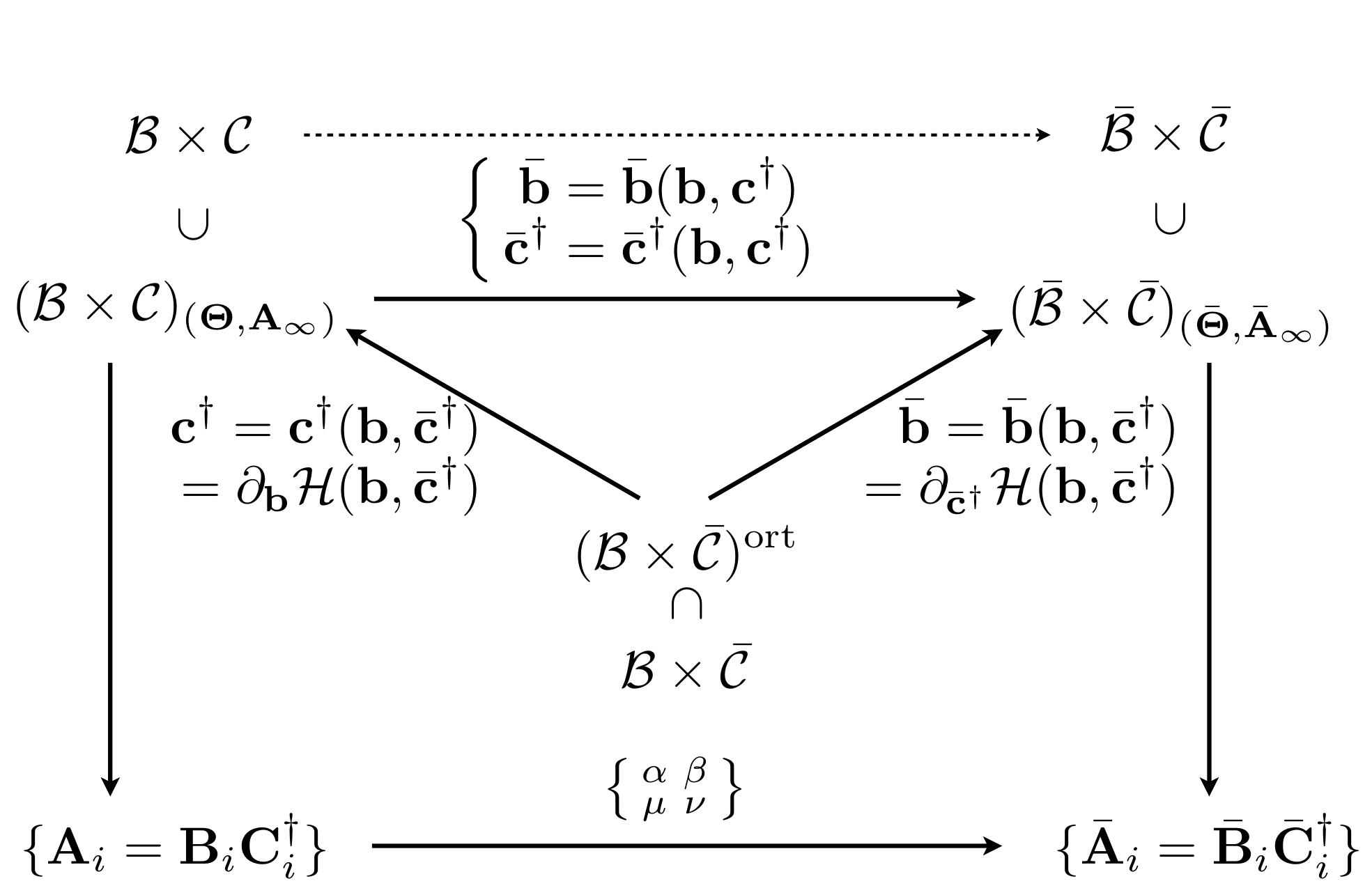}
	\caption{Discrete Hamiltonian Dynamic}
	\label{fig:discr-ham}
\end{figure}

\begin{theorem}\label{thm:eq-Ham-solve} If we restrict to $(\mathcal{B}\times\bar{\mathcal{C}})^{\operatorname{ort}}$, 
	equations~(\ref{eq:transf-Ham}--\ref{eq:resinf-Ham}) can then be solved for 
	$\bar{\mathbf{b}}_{i,j} = \bar{\mathbf{b}}_{i,j}(\mathbf{B},\bar{\mathbf{C}}^{\dag})$ and
	$\mathbf{c}_{i}^{j\dag} = \mathbf{c}_{i}^{j\dag}(\mathbf{B},\bar{\mathbf{C}}^{\dag})$ to give the following expressions.
	\begin{enumerate}[(a)]
		\item Generic indices $i\neq \alpha$, $(i,j)\neq (\beta,\nu)$ (recall that $\mathbf{R}(z_{\beta}) = \mathbf{Q}$):
		\begin{equation}
			\bar{\mathbf{b}}_{i,j} = \bar{\theta}_{i}^{j} \frac{ \mathbf{R}(z_{i}) \mathbf{b}_{i,j} }{ 
			\bar{\mathbf{c}}_{i}^{j\dag} \mathbf{R}(z_{i}) \mathbf{b}_{i,j} },
			\qquad
			\mathbf{c}_{i}^{j\dag} = \theta_{i}^{j} \frac{ \bar{\mathbf{c}}_{i}^{j\dag} \mathbf{R}(z_{i}) }{ 
			\bar{\mathbf{c}}_{i}^{j\dag} \mathbf{R}(z_{i}) \mathbf{b}_{i,j} }.\label{eq:Ham-generic}			
		\end{equation}
		\item Special index $\alpha$:
		\begin{alignat}{2}
			\bar{\mathbf{b}}_{\alpha,\mu} &= \bar{\theta}_{\alpha}^{\mu} \frac{ \mathbf{b}_{\beta,\nu} }{ 
			\bar{\mathbf{c}}_{\alpha}^{\mu\dag} \mathbf{b}_{\beta,\nu} },
			&\qquad
			\mathbf{c}_{\alpha}^{\mu\dag} &= \theta_{\alpha}^{\mu} \frac{ \bar{\mathbf{c}}_{\beta}^{\nu\dag} }{ 
			\bar{\mathbf{c}}_{\beta}^{\nu\dag} \mathbf{b}_{\alpha,\mu} },\label{eq:Ham-am}\\
			\intertext{and for $j\neq \mu$,}
			\bar{\mathbf{b}}_{\alpha,j} &= \bar{\theta}_{\alpha}^{j} \frac{ \mathbf{R}_{\alpha}\mathbf{b}_{\alpha,j} }{ 
			\bar{\mathbf{c}}_{\alpha}^{j\dag} \mathbf{R}_{\alpha} \mathbf{b}_{\alpha,j} },
			&\qquad
			\mathbf{c}_{\alpha}^{j\dag} &= \theta_{\alpha}^{j} \frac{ \bar{\mathbf{c}}_{\alpha}^{j\dag} \mathbf{R}_{\alpha}}{ 
			\bar{\mathbf{c}}_{\alpha}^{j\dag} \mathbf{R}_{\alpha} \mathbf{b}_{\alpha,j} },\quad\text{ where }
			\mathbf{R}_{\alpha} = \mathbf{I} - \frac{ \mathbf{b}_{\alpha,\mu} \bar{\mathbf{c}}_{\beta}^{\nu\dag} }{ 
			\bar{\mathbf{c}}_{\beta}^{\nu\dag} \mathbf{b}_{\alpha,\mu} } - 
			\frac{ \mathbf{b}_{\beta,\nu} \bar{\mathbf{c}}_{\alpha}^{\mu\dag} }{ \bar{\mathbf{c}}_{\alpha}^{\mu\dag} \mathbf{b}_{\beta,\nu} }.
			\label{eq:Ham-aj}
		\end{alignat}
		\item Special index $(\beta,\nu)$:
		\begin{align}
			\bar{\mathbf{b}}_{\beta,\nu} &= (\theta_{\beta}^{\nu} - \theta_{\alpha}^{\mu} + 1) 
			\frac{ \mathbf{b}_{\beta,\nu} }{ \bar{\mathbf{c}}_{\beta}^{\nu\dag} \mathbf{b}_{\beta,\nu}} 
			+ \theta_{\alpha}^{\mu} \frac{ \mathbf{b}_{\alpha,\mu} }{ \bar{\mathbf{c}}_{\beta}^{\nu\dag} \mathbf{b}_{\alpha,\mu}} 
			- \sum_{j\neq \mu} \theta_{\alpha}^{j} \frac{ \bar{\mathbf{c}}_{\alpha}^{j\dag} \mathbf{b}_{\alpha,\mu} }{ 
			\bar{\mathbf{c}}_{\beta}^{\nu\dag} \mathbf{b}_{\alpha,\mu} } \cdot 
			\frac{ \mathbf{b}_{\alpha,j} }{ \bar{\mathbf{c}}_{\alpha}^{j\dag} \mathbf{b}_{\alpha,j} }\label{eq:Ham-bbn}\\
		 	&	\qquad- \mathbf{Q} \bigg(
			\mathop{\sum\sum}_{\substack{i\neq \alpha,\ j \\ (i,j) \neq (\beta,\nu)}}   \left(\frac{ z_{\beta} - z_{\alpha} }{ z_{i} - z_{\alpha}}\right)
			\theta_{i}^{j} \frac{ \mathbf{b}_{i,j} \bar{\mathbf{c}}_{i}^{j\dag} }{ \bar{\mathbf{c}}_{i}^{j\dag}
			\mathbf{R}(z_{i}) \mathbf{b}_{i,j} }
			\bigg)\frac{ \mathbf{b}_{\beta,\nu} }{ \bar{\mathbf{c}}_{\beta}^{\nu\dag} \mathbf{b}_{\beta,\nu} };\notag\\
			\mathbf{c}_{\beta}^{\nu\dag} &= (\theta_{\beta}^{\nu} - \theta_{\alpha}^{\mu} + 1) 
			\frac{ \bar{\mathbf{c}}_{\beta}^{\nu\dag} }{ \bar{\mathbf{c}}_{\beta}^{\nu\dag} \mathbf{b}_{\beta,\nu} }
			+ \bar{\theta}_{\alpha}^{\mu} \frac{ \bar{\mathbf{c}}_{\alpha}^{\mu\dag} }{ \bar{\mathbf{c}}_{\alpha}^{\mu\dag} \mathbf{b}_{\beta,\nu} }
			- \sum_{j\neq \mu} \theta_{\alpha}^{j} \frac{ \bar{\mathbf{c}}_{\alpha}^{\mu\dag} }{ \bar{\mathbf{c}}_{\alpha}^{\mu\dag} \mathbf{b}_{\beta,\nu} }
			\cdot \frac{ \mathbf{b}_{\alpha,j} \bar{\mathbf{c}}_{\alpha}^{j\dag} }{ \bar{\mathbf{c}}_{\alpha}^{j\dag} \mathbf{b}_{\alpha,j} }
			\label{eq:Ham-cbn}\\
			&\qquad - \frac{ \bar{\mathbf{c}}_{\beta}^{\nu\dag} }{ \bar{\mathbf{c}}_{\beta}^{\nu\dag} \mathbf{b}_{\beta,\nu} }
			\bigg( 
				\mathop{\sum\sum}_{\substack{i\neq \alpha,\ j\\(i,j)\neq(\beta,\nu)}}\left(\frac{ z_{\beta} - z_{\alpha} }{ z_{i} - z_{\alpha} }\right)
				\theta_{i}^{j} \frac{ \mathbf{b}_{i,j} \bar{\mathbf{c}}_{i}^{j\dag} }{ \bar{\mathbf{c}}_{i}^{j\dag} 
				\mathbf{R}(z_{i})\mathbf{b}_{i,j} }
			\bigg)\mathbf{Q}.\notag
		\end{align}
	\end{enumerate}	
\end{theorem}
\begin{proof}
	To prove (a) for $i\neq \alpha,\beta$, we right-multiply the equation $\bar{\mathbf{A}}_{i} \mathbf{R}(z_{i}) = \mathbf{R}(z_{i})\mathbf{A}_{i}$
	by $\mathbf{B}_{i}$ to obtain $\bar{\mathbf{B}}_{i}(\bar{\mathbf{C}}_{i}^{\dag} \mathbf{R}(z_{i})\mathbf{B}_{i}) = \mathbf{R}(z_{i})\mathbf{B}_{i} \mathbf{\Theta}_{i}$
	and then use (\ref{eq:Ham-ortho-locus}) to get 
	$\bar{\mathbf{B}}_{i} = \mathbf{R}(z_{i})\mathbf{B}_{i} \mathbf{\Theta}_{i} (\bar{\mathbf{C}}_{i}^{\dag}\mathbf{R}(z_{i})\mathbf{B}_{i})^{-1}$, which is the 
	first equation in (\ref{eq:Ham-generic}). The second equation is obtained in the same way, only we start by-left multiplying by $\bar{\mathbf{C}}_{i}^{\dag}$.
	The exact same argument works for indices $(\beta,j)$ with $j\neq \nu$. 
	
	To prove (b), notice that equations~(\ref{eq:Ham-am}) follow immediately from (\ref{eq:transf-Ham}--\ref{eq-orto-Ham}). Equations~(\ref{eq:Ham-aj})
	are slightly less straightforward. First, given 
	$\mathbf{R}_{\alpha}$ as in~(\ref{eq:Ham-aj}), it is easy to see
	\begin{equation*}
		\mathbf{R}_{\alpha}\mathbf{Q} = \mathbf{I} - \frac{ \mathbf{b}_{\beta,\nu} \bar{\mathbf{c}}_{\alpha}^{\mu\dag} }{ 
		\bar{\mathbf{c}}_{\alpha}^{\mu\dag} \mathbf{b}_{\beta,\nu} }\text{ and }
		\mathbf{Q}\mathbf{R}_{\alpha} = \mathbf{I} - \frac{ \mathbf{b}_{\alpha,\mu} \bar{\mathbf{c}}_{\beta}^{\nu\dag} }{ 
		\bar{\mathbf{c}}_{\beta}^{\nu\dag} \mathbf{b}_{\alpha,\mu} }.
	\end{equation*}
	Then, since by (\ref{eq-orto-Ham}) we require 
	$\bar{\mathbf{c}}_{\alpha}^{\mu\dag} \bar{\mathbf{b}}_{\alpha,j} = \mathbf{c}_{\alpha}^{j\dag}\mathbf{b}_{\alpha,\mu} = 0$ for $j\neq \mu$,
	we get $\mathbf{R}_{\alpha}\mathbf{Q}\bar{\mathbf{A}}_{\alpha}' = \bar{\mathbf{A}}_{\alpha}'$, 
	$\mathbf{A}_{\alpha}' \mathbf{Q} \mathbf{R}_{\alpha} = \mathbf{A}_{\alpha}'$ and so equation
	$\mathbf{Q}\bar{\mathbf{A}}_{\alpha}' = \mathbf{A}_{\alpha}' \mathbf{Q}$ in (\ref{eq:Schlesinger-Ham}) becomes 
	$\bar{\mathbf{A}}_{\alpha}\mathbf{R}_{\alpha} = \mathbf{R}_{\alpha}\mathbf{A}_{\alpha}'$. The rest of the argument proceeds exactly as in part (a),
	with one minor modification that 
	$(\bar{\mathbf{C}}_{\alpha}^{\dag})' \mathbf{R}_{\alpha} \mathbf{B}_{\alpha}'$ becomes just $(\bar{\mathbf{C}}_{\alpha}^{\dag})'\mathbf{B}_{\alpha}'$
	in view of orthogonality.
	
	Given (a) and (b), equation~(\ref{eq:resinf-Ham}) can be written as
	\begin{align}
		\bar{\mathbf{b}}_{\beta,\nu} \bar{\mathbf{c}}_{\beta}^{\nu\dag} - \mathbf{b}_{\beta,\nu} \mathbf{c}_{\beta}^{\nu\dag} &= 
		\theta_{\alpha}^{\mu} \frac{ \mathbf{b}_{\alpha,\mu} \bar{\mathbf{c}}_{\beta}^{\nu\dag} }{ \bar{\mathbf{c}}_{\beta}^{\nu\dag} \mathbf{b}_{\alpha,\mu} }
		- \bar{\theta}_{\alpha}^{\mu} \frac{ \mathbf{b}_{\beta,\nu} \bar{\mathbf{c}}_{\alpha}^{\mu\dag} }{ 
		\bar{\mathbf{c}}_{\alpha}^{\mu\dag} \mathbf{b}_{\beta,\nu} }
		+ \sum_{j\neq \mu} \theta_{\alpha}^{j} \frac{ \mathbf{b}_{\alpha,j}  \bar{\mathbf{c}}_{\alpha}^{j\dag} \mathbf{R}_{\alpha} - 
		\mathbf{R}_{\alpha} \mathbf{b}_{\alpha,j} \bar{\mathbf{c}}_{\alpha}^{j\dag} }{ \bar{\mathbf{c}}_{\alpha}^{j\dag} \mathbf{b}_{\alpha,j} }
		\label{eq:Ham-star}\\
		&\qquad + \sum_{j\neq \nu} \theta_{\beta}^{j} \frac{ \mathbf{b}_{\beta,j}  \bar{\mathbf{c}}_{\beta}^{j\dag} \mathbf{Q} - 
		\mathbf{Q} \mathbf{b}_{\beta,j} \bar{\mathbf{c}}_{\beta}^{j\dag} }{ \bar{\mathbf{c}}_{\beta}^{j\dag} \mathbf{Q} \mathbf{b}_{\beta,j} }
		+ \sum_{i\neq \alpha,\beta}\sum_{j} \theta_{i}^{j} \frac{ \mathbf{b}_{i,j}  \bar{\mathbf{c}}_{i}^{j\dag} \mathbf{R}(z_{i}) - 
		\mathbf{R}(z_{i}) \mathbf{b}_{i,j} \bar{\mathbf{c}}_{i}^{j\dag} }{ \bar{\mathbf{c}}_{i}^{j\dag} \mathbf{b}_{i,j} }.\notag
	\end{align}
	To extract $\bar{\mathbf{b}}_{\beta,\nu}$ we multiply (\ref{eq:Ham-star}) on the left by $\mathbf{Q}$ 
	(recall that $\mathbf{Q}\mathbf{b}_{\beta,\nu} = \mathbf{0}$)
	and on the right by $\mathbf{b}_{\alpha,\mu}$ and use (\ref{eq-orto-Ham}) and (\ref{eq:Ham-ortho-locus}):
	\begin{align*}
		\mathbf{Q}(\bar{\mathbf{b}}_{\beta,\nu} \bar{\mathbf{c}}_{\beta}^{\nu\dag}) \mathbf{b}_{\alpha,\mu} &= 
		\left(\bar{\mathbf{b}}_{\beta,\nu} - (\bar{\mathbf{c}}_{\beta}^{\nu\dag} \bar{\mathbf{b}}_{\beta,\nu})  
		\frac{ \mathbf{b}_{\beta,\nu} }{ \bar{\mathbf{c}}_{\beta}^{\nu\dag} \mathbf{b}_{\beta,\nu} }\right) 
		\bar{\mathbf{c}}_{\beta}^{\nu\dag} \mathbf{b}_{\alpha,\mu} = 
		\left(\bar{\mathbf{b}}_{\beta,\nu} - \bar{\theta}_{\beta}^{\nu}  
		\frac{ \mathbf{b}_{\beta,\nu} }{ \bar{\mathbf{c}}_{\beta}^{\nu\dag} \mathbf{b}_{\beta,\nu} }\right) 
		\bar{\mathbf{c}}_{\beta}^{\nu\dag} \mathbf{b}_{\alpha,\mu}\\
		&= \theta_{\alpha}^{\mu} \mathbf{Q}\mathbf{b}_{\alpha,\mu} + 
		\sum_{j\neq \mu} \theta_{\alpha}^{j} \left(- \frac{ \bar{\mathbf{c}}_{\alpha}^{j\dag} \mathbf{b}_{\alpha,\mu} }{ 
		\bar{\mathbf{c}}_{\alpha}^{j\dag} \mathbf{b}_{\alpha,j} }\right) \mathbf{b}_{\alpha,j} 
		+ \sum_{j\neq \nu} \theta_{\beta}^{j} \left(-
		\frac{ (\bar{\mathbf{c}}_{\beta}^{j\dag} \mathbf{b}_{\beta,\nu})(\bar{\mathbf{c}}_{\beta}^{\nu\dag} \mathbf{b}_{\alpha,\mu}) }{ 
		(\bar{\mathbf{c}}_{\beta}^{j\dag} \mathbf{Q} \mathbf{b}_{\beta,j})(\bar{\mathbf{c}}_{\beta}^{\nu\dag} \mathbf{b}_{\beta,\nu}) }\right) 
		\mathbf{Q} \mathbf{b}_{\beta,j}\\
		&\qquad + \sum_{i\neq \alpha,\beta}\sum_{j} \left(\frac{ z_{\alpha} - z_{\beta} }{ z_{i} - z_{\alpha} }\right)
		\frac{ (\bar{\mathbf{c}}_{i}^{j\dag} \mathbf{b}_{\beta,\nu}) (\bar{\mathbf{c}}_{\beta}^{\nu\dag} \mathbf{b}_{\alpha,\mu}) }{ 
		(\bar{\mathbf{c}}_{i}^{j\dag} \mathbf{R}(z_{i}) \mathbf{b}_{i,j})(\bar{\mathbf{c}}_{\beta}^{\nu\dag} \mathbf{b}_{\beta,\nu}) } \mathbf{Q} b_{i,j}.
	\end{align*}	
	After straightforward simplification this gives~(\ref{eq:Ham-bbn}). Equation~(\ref{eq:Ham-cbn}) is obtained in the same way,
	except that we start by multiplying (\ref{eq:Ham-star}) on the right by $\mathbf{Q}$ 
	and on the left by $\bar{\mathbf{c}}_{\alpha}^{\mu\dag}$.
\end{proof}

We now show that these equations are Hamiltonian.

\begin{theorem}\label{thm:Hamiltonian}
	Let 
	\begin{align}
		\mathcal{H}^{+}(\mathbf{B},\bar{\mathbf{C}}^{\dag}; \mathbf{\Theta})
		&= (\theta_{\beta}^{\nu} - \theta_{\alpha}^{\mu} + 1)\log(\bar{\mathbf{c}}_{\beta}^{\nu\dag} \mathbf{b}_{\beta,\nu})
		+ \theta_{\alpha}^{\mu} \log(\bar{\mathbf{c}}_{\beta}^{\nu\dag} \mathbf{b}_{\alpha,\mu}) + 
		(\theta_{\alpha}^{\mu} - 1) \log(\bar{\mathbf{c}}_{\alpha}^{\mu\dag} \mathbf{b}_{\beta,\nu}) \notag\\
		&\qquad + 
		\sum_{j\neq \mu} \theta_{\alpha}^{j} 
		\log(\bar{\mathbf{c}}_{\alpha}^{j\dag} \mathbf{R}_{\alpha} \mathbf{b}_{\alpha,j}) +
		\sum_{j\neq \nu} \theta_{\beta}^{j} 
		\log(\bar{\mathbf{c}}_{\beta}^{j\dag} \mathbf{Q} \mathbf{b}_{\beta,j}) + 
		\sum_{i\neq \alpha, \beta}\sum_{j} \theta_{i}^{j} 
		\log(\bar{\mathbf{c}}_{i}^{j\dag} \mathbf{R}(z_{i}) \mathbf{b}_{i,j}).\label{eq:Hamiltonian}
	\end{align}		
	Then, if we restrict to  $(\mathcal{B}\times\bar{\mathcal{C}})^{\operatorname{ort}}$, equations (\ref{eq:Ham-generic}--\ref{eq:Ham-cbn})
	are given by
	\begin{equation*}
		\bar{\mathbf{b}}_{i,j} = \frac{ \partial \mathcal{H}^{+} }{ \partial \bar{\mathbf{c}}_{i}^{j\dag}}(\mathbf{B},\bar{\mathbf{C}}^{\dag})\quad
		\text{ and }\quad 
		\mathbf{c}_{i}^{j\dag} = \frac{ \partial \mathcal{H}^{+} }{ \partial \mathbf{b}_{i,j}}(\mathbf{B},\bar{\mathbf{C}}^{\dag}).
	\end{equation*} 
\end{theorem}	
\begin{proof}
	This statement is immediate for all generic indices. The non-trivial part is to check it for derivatives of $\mathcal{H}^{+}$ w.r.t.~ 
	the vectors $\mathbf{b}_{\alpha,\mu}$, $\mathbf{b}_{\beta,\nu}$, $\bar{\mathbf{c}}_{\alpha}^{\mu\dag}$, and  
	$\bar{\mathbf{c}}_{\beta}^{\nu\dag}$ that appear in $\mathbf{R}(z_{i})$, $\mathbf{R}_{\alpha}$, and $\mathbf{Q}$. 
	Using (\ref{eq:Ham-ortho-locus}), we get
	\begin{align*}
		\frac{ \partial \mathcal{H}^{+} }{ \partial \bar{\mathbf{c}}_{\alpha}^{\mu\dag}} &= 
		(\theta_{\alpha}^{\mu} - 1) \frac{ \mathbf{b}_{\beta,\nu} }{ \bar{\mathbf{c}}_{\alpha}^{\mu\dag} \mathbf{b}_{\beta,\nu} }
		+ \sum_{j\neq \mu} \theta_{\alpha}^{j} \frac{ (\bar{\mathbf{c}}_{\alpha}^{j\dag} \mathbf{b}_{\beta,\nu}) 
		\partial_{\bar{\mathbf{c}}_{\alpha}^{j\dag}} \left( - \frac{ \bar{\mathbf{c}}_{\alpha}^{\mu\dag} \mathbf{b}_{\alpha,j} }{ 
		\bar{\mathbf{c}}_{\alpha}^{\mu\dag} \mathbf{b}_{\beta,j} }\right) }{ 
		\bar{\mathbf{c}}_{\alpha}^{j\dag} \mathbf{R}_{\alpha} \mathbf{b}_{\alpha,j} } = 
		(\theta_{\alpha}^{\mu} - 1) \frac{ \mathbf{b}_{\beta,\nu} }{ \bar{\mathbf{c}}_{\alpha}^{\mu\dag} \mathbf{b}_{\beta,\nu} }
		= \bar{\mathbf{b}}_{\alpha,\mu},\\
		\frac{ \partial \mathcal{H}^{+} }{ \partial \bar{\mathbf{c}}_{\beta}^{\nu\dag}} &= 
		(\theta_{\beta}^{\nu} - \theta_{\alpha}^{\mu} + 1) \frac{ \mathbf{b}_{\beta,\nu} }{ \bar{\mathbf{c}}_{\beta}^{\nu\dag} \mathbf{b}_{\beta,\nu} }
		+ \theta_{\alpha}^{\mu} \frac{ \mathbf{b}_{\alpha,\mu} }{ \bar{\mathbf{c}}_{\beta}^{\nu\dag} \mathbf{b}_{\alpha,\mu} }
		 - \sum_{j\neq \mu} \theta_{\alpha}^{j} \frac{ \bar{\mathbf{c}}_{\alpha}^{j\dag} \mathbf{b}_{\beta,\nu} }{ 
		\bar{\mathbf{c}}_{\alpha}^{j\dag} \mathbf{b}_{\alpha,j} }\partial_{\bar{\mathbf{c}}_{\beta}^{\nu\dag}} 
		\left(\frac{ \bar{\mathbf{c}}_{\beta}^{\nu\dag} \mathbf{b}_{\alpha,j} }{ \bar{\mathbf{c}}_{\beta}^{\nu\dag} \mathbf{b}_{\beta,\nu} }\right)
		\\
		&\qquad - \sum_{j\neq \nu}\theta_{\beta}^{j} \frac{ \bar{\mathbf{c}}_{\beta}^{j\dag} \mathbf{b}_{\beta,\nu} }{ 
		\bar{\mathbf{c}}_{\beta}^{j\dag} \mathbf{Q} \mathbf{b}_{\beta,j} } \partial_{\bar{\mathbf{c}}_{\beta,\nu}} 
		\left(\frac{ \bar{\mathbf{c}}_{\beta}^{\nu\dag} \mathbf{b}_{\beta,j} }{ \bar{\mathbf{c}}_{\beta}^{\nu\dag} \mathbf{b}_{\beta,\nu} }\right)
		+ \sum_{i\neq \alpha,\beta} \sum_{j}\theta_{i}^{j} 
		\frac{ \left(\frac{ z_{\alpha} - z_{\beta} }{ z_{i} - z_{\alpha} }\right)  \bar{\mathbf{c}}_{i}^{j\dag} \mathbf{b}_{\beta,\nu} }{ 
		\bar{\mathbf{c}}_{i}^{j\dag} \mathbf{R}(z_{i}) \mathbf{b}_{i,j} } \partial_{\bar{\mathbf{c}}_{\beta,\nu}} 
		\left(\frac{ \bar{\mathbf{c}}_{\beta}^{\nu\dag} \mathbf{b}_{i,j} }{ \bar{\mathbf{c}}_{\beta}^{\nu\dag} \mathbf{b}_{\beta,\nu} }\right)
		= \bar{\mathbf{b}}_{\beta,\nu},
	\end{align*}
	since
	\begin{equation*}
		\partial_{\bar{\mathbf{c}}_{\beta,\nu}} 
		\left(\frac{ \bar{\mathbf{c}}_{\beta}^{\nu\dag} \mathbf{b}_{i,j} }{ \bar{\mathbf{c}}_{\beta}^{\nu\dag} \mathbf{b}_{\beta,\nu} }\right)
		= \frac{ (\bar{\mathbf{c}}_{\beta}^{\nu\dag} \mathbf{b}_{\beta,\nu}) \mathbf{b}_{i,j} - 
		(\bar{\mathbf{c}}_{\beta}^{\nu\dag} \mathbf{b}_{i,j}) \mathbf{b}_{\beta,\nu}}{ 
		(\bar{\mathbf{c}}_{\beta}^{\nu\dag} \mathbf{b}_{\beta,\nu})^{2} }  = 
		\frac{ \mathbf{Q} \mathbf{b}_{i,j} }{ (\bar{\mathbf{c}}_{\beta}^{\nu\dag} \mathbf{b}_{\beta,\nu}) }.
	\end{equation*}
	The other two equations are established in the same way.
\end{proof}

Finally, we show that the evolution equations~(\ref{eq:bb-cb-generators}--\ref{eq:cb-am}) can be obtained directly from the Hamiltonian.

\begin{lemma}\label{lem:eq-Ham-evol}
	Equations~(\ref{eq:Ham-generic}--\ref{eq:Ham-cbn}) give the same dynamic
	$(\mathcal{B}\times \mathcal{C})_{(\mathbf{\Theta},\mathbf{A}_{\infty})} \to 
	(\bar{\mathcal{B}}\times \bar{\mathcal{C}})_{(\bar{\mathbf{\Theta}},\bar{\mathbf{A}}_{\infty})}$ as
	in Theorem~\ref{thm:evolution}.
\end{lemma}	
\begin{proof}
	Equation~(\ref{eq:bb-cb-generators}) is just a restatement of~(\ref{eq:Ham-am}). It then follows that $\mathbf{R}(z)$ is given by 
	(\ref{eq:R-Sch}). As usual, the statement is immediate for all indices $i\neq \alpha,\beta$ (note that the 
	relationship between the normalization coefficients for $\bar{\mathbf{b}}_{i,j}$ and $\bar{\mathbf{c}}_{i}^{j\dag}$
	follows from the requirements~(\ref{eq:ortho-cond}) for the domain and the range of the map).	
	Equation~(\ref{eq:Ham-generic}) 
	gives the expression for $\bar{\mathbf{b}}_{\beta,j}$ in (\ref{eq:bb-cd-generic}) for $j\neq \nu$, and 
	right-multiplying $\mathbf{c}_{\alpha}^{j\dag}\sim \bar{\mathbf{c}}_{\alpha}^{j\dag} \mathbf{R}_{\alpha}$ in (\ref{eq:Ham-aj})
	by $\mathbf{Q}$ gives $\bar{\mathbf{c}}_{\alpha}^{j\dag}\sim \mathbf{c}_{\alpha}^{j\dag}\mathbf{Q}$, which is just
	the expression for $\bar{\mathbf{c}}_{\alpha}^{j\dag}$ in
	(\ref{eq:bb-cd-generic}) for $j\neq \mu$. So it remains to establish equations~(\ref{eq:bb-ak}--\ref{eq:cb-am}).

	First, we see that (\ref{eq:Ham-bbn}) now becomes
	\begin{align}
		c_{\beta}^{\nu}\bar{\mathbf{b}}_{\beta,\nu} 
		&= 
		(\theta_{\beta}^{\nu} + 1) \frac{ \mathbf{b}_{\beta,\nu} }{ \mathbf{c}_{\alpha}^{\mu\dag} \mathbf{b}_{\beta,\nu}  }
		+ \mathbf{Q}
		\bigg(  
			\sum_{i\neq \beta} \left(\frac{ z_{\alpha} - z_{\beta} }{ z_{i} - z_{\beta} }\mathbf{A}_{i}\right) 
		\bigg)\frac{ \mathbf{b}_{\beta,\nu} }{ \mathbf{c}_{\alpha}^{\mu\dag} \mathbf{b}_{\beta,\nu}  } - \sum_{j\neq \nu} 
		\left(\frac{ \bar{\mathbf{c}}_{\beta}^{j\dag} \mathbf{b}_{\beta,\nu} }{ 
		\mathbf{c}_{\alpha}^{\mu\dag} \mathbf{b}_{\beta,\nu} }\right) \bar{\mathbf{b}}_{\beta,j}.\label{eq:bbn-temp}\\
		\intertext{Left-multiplying it by $\bar{\mathbf{c}}_{\beta}^{k\dag}$, $k\neq \mu$, gives}
		0&= (\theta_{\beta}^{\nu} + 1 - \theta_{\beta}^{k}) \left(\frac{ \bar{\mathbf{c}}_{\beta}^{k\dag} \mathbf{b}_{\beta,\nu} }{ 
		\mathbf{c}_{\alpha}^{\mu\dag} \mathbf{b}_{\beta,\nu} }\right) + 
		c_{\beta}^{k} \mathbf{c}_{\beta}^{k\dag}\bigg(  
			\sum_{i\neq \beta} \left(\frac{ z_{\alpha} - z_{\beta} }{ z_{i} - z_{\beta} }\mathbf{A}_{i} \right)
		\bigg)\frac{ \mathbf{b}_{\beta,\nu} }{ \mathbf{c}_{\alpha}^{\mu\dag} \mathbf{b}_{\beta,\nu}  }.\label{eq:bbn-coeff}
	\end{align}
	Solving (\ref{eq:bbn-coeff}) for $(\bar{\mathbf{c}}_{\beta}^{k\dag} \mathbf{b}_{\beta,\nu})/(\bar{\mathbf{c}}_{\alpha}^{\mu\dag} \mathbf{b}_{\beta,\nu})$
	and substituting it into (\ref{eq:bbn-temp}) gives, after simplifying, equation~(\ref{eq:bb-bn}). Similarly,~(\ref{eq:Ham-cbn}) becomes
	\begin{align}
		\bar{\theta}_{\alpha}^{\mu} \frac{ \bar{\mathbf{c}}_{\alpha}^{\mu\dag} }{ \bar{\mathbf{c}}_{\alpha}^{\mu\dag} \mathbf{b}_{\beta,\nu} }
		- \sum_{j\neq \mu}  \frac{ \bar{\mathbf{c}}_{\alpha}^{\mu\dag} \mathbf{b}_{\alpha,j} }{ \bar{\mathbf{c}}_{\alpha}^{\mu\dag} \mathbf{b}_{\beta,\nu} }
		\cdot \mathbf{c}_{\alpha}^{j\dag}\mathbf{Q}
		&= 
		\frac{ \mathbf{c}_{\alpha}^{\mu\dag} }{ \mathbf{c}_{\alpha}^{\mu\dag} \mathbf{b}_{\beta,\nu} }
		\bigg( (\theta_{\alpha}^{\mu} - 1)\mathbf{I} + 
			\sum_{i\neq \alpha} \left(\frac{ z_{\beta} - z_{\alpha} }{ z_{i} - z_{\alpha} }\right)
			\mathbf{A}_{i} \mathbf{Q}
		\bigg).\label{eq:cam-temp}
		\intertext{Right-multiplying it by $\mathbf{b}_{\beta,k}$, $k\neq \nu$ gives}
		(\bar{\theta}_{\alpha}^{\mu} - \theta_{\alpha}^{k}) 
		\frac{ \bar{\mathbf{c}}_{\alpha}^{\mu\dag} \mathbf{b}_{\alpha,k} }{ \bar{\mathbf{c}}_{\alpha}^{\mu\dag} \mathbf{b}_{\beta,\nu} }
		&= \frac{ \mathbf{c}_{\alpha}^{\mu\dag} }{ \mathbf{c}_{\alpha}^{\mu\dag} \mathbf{b}_{\beta,\nu}  }
		\bigg( 
			\sum_{i\neq \alpha} \left(\frac{ z_{\beta} - z_{\alpha} }{ z_{i} - z_{\alpha} }\right)
			\mathbf{A}_{i} \mathbf{b}_{\alpha,k},
		\bigg).\label{eq:cam-coeff}
	\end{align}
	Solving (\ref{eq:cam-coeff}) for $(\bar{\mathbf{c}}_{\alpha}^{\mu\dag} \mathbf{b}_{\alpha,k})/(\bar{\mathbf{c}}_{\alpha}^{\mu\dag} \mathbf{b}_{\beta,\nu})$,
	substituting it into (\ref{eq:cam-temp}), and simplifying gives~(\ref{eq:cb-am}). Substituting the expression for $\bar{\mathbf{c}}_{\alpha}^{\mu\dag}$
	into $\mathbf{R}_{\alpha}$ in $\bar{\mathbf{b}}_{\alpha,j} \sim \mathbf{R}_{\alpha}\mathbf{b}_{\alpha,j}$ and simplifying gives~(\ref{eq:bb-ak}),
	where we again use~(\ref{eq:ortho-cond}) to adjust the normalization coefficient. Finally, right-multiplying 
	$\mathbf{c}_{\beta}^{j\dag} \sim \bar{\mathbf{c}}_{\beta}^{j\dag} \mathbf{Q}$ by 
	$\mathbf{I} - \bar{\mathbf{b}}_{\beta,\nu} (\mathbf{c}_{\alpha}^{\mu\dag} \bar{\mathbf{b}}_{\beta,\nu})^{-1} \mathbf{c}_{\alpha}^{\mu\dag}$
	gives
	\begin{equation*}
		\bar{\mathbf{c}}_{\beta}^{\nu\dag}\sim \mathbf{c}_{\beta}^{j\dag}\left(\mathbf{I} - 
		\frac{ \bar{\mathbf{b}}_{\beta,\nu}  \mathbf{c}_{\alpha}^{\mu\dag} }{ \mathbf{c}_{\alpha}^{\mu\dag} \bar{\mathbf{b}}_{\beta,\nu} }\right).
	\end{equation*}
	Using the expression for $\bar{\mathbf{b}}_{\beta,\nu}$ and simplifying gives~(\ref{eq:cb-bk}) and completes the proof of the Lemma.
\end{proof}	

\section{Examples} 
\label{sec:examples}

In this section we consider two explicit examples of reductions from
discrete Schlesinger transformations to difference Painlev\'e equations.
 We also emphasize the role played by the geometry
of Painlev\'e equations in not only determining the type of the equation,
but also in studying the relationship between different explicit
forms of equations of the same type.

Recall that, according to the geometric theory of Painlev\'e equations 
developed in
\cite{Sak:2001:RSAWARSGPE}, both continuous and discrete 
Painlev\'e equations can be classified by the type of 
their space of initial conditions, introduced
by Okamoto \cite{Oka:1979:SLFAAEDSOAPCFDPP} in the continuous case. This space can be obtained 
by successively blowing up the projective plane $\mathbb{P}^{2}$ at 
nine (possibly infinitely-close) points that lie on a unique
(possibly singular) cubic curve. The type of the resulting surface
$X$ is then given by the configurations of irreducible components $D_{i}$ of its \emph{unique}
anti-canonical divisor $D$ or, equivalently, by the root 
sub-lattice $R$ spanned 
by $D_{i}$ in the orthogonal complement $D^{\perp}$ of $D$ in the Picard lattice
$\operatorname{Pic}(X)$. The orthogonal complement $R^{\perp}$ of $R$ in $D^{\perp}$
then corresponds to the symmetries of the equation --- its Weyl group acts as
automorphisms, called the \emph{Cremona isometries}, of $\operatorname{Pic}(X)$ 
preserving $R\subset D^{\perp}$ and the translational
part of the Weyl group corresponds to discrete Painlev\'e equations.

We also want to stress the following point, see also \cite{Sak:2007:PDPEATLF}. Given a discrete
Painlev\'e equation, we can determine its type by constructing its space of initial 
conditions. However, there is no converse process, although there are some particularly nice 
forms (that we call \emph{standard}) of equations of each type, see 
\cite{Sak:2007:PDPEATLF, GraRamOht:2003:AUDOTAQVADIEATST, Mur:2004:NEFDPE}.
Further, there can be many 
\emph{non-equivalent equations} (i.e., equations that can not be transformed into each other 
by some change of coordinates)
of the same type --- such equations can correspond to different translation directions in the
affine Weyl group of symmetries of $X$. And even when the two different equations are equivalent,
finding the actual change of coordinates that will transform one equation into the other can be 
a very challenging problem. In each case, understanding the geometry and the blowing-down structure 
of the space of initial conditions can be of great help in determining the relationship between
two different equations of the same type,  as we illustrate by two examples in this section. 

For the first example, we consider the Fuchsian equation 
of matrix size $m=2$ with $n=3$ finite poles and rank-one condition 
$r_1=r_2=r_3=1$ for the residue matrices at those poles. 
It is well-known that continuous isomonodromic deformations
of this system are described by the classical continuous 
Painlev\'e equation $P_{\text{VI}}$, the type of the space of 
initial conditions is $D_{4}^{(1)}$, and 
the discrete Painlev\'e V equation d-$P_{\text{V}} = $ d-${P}(D_{4}^{(1)})$  appears as B\"acklund
transformations of $P_{\text{VI}}$. Thus, elementary Schlesinger 
transformations of this system should be described by d-$P_{\text{V}}$, and we  
explicitly show that this is indeed the case. We consider this example using two different 
parametrizations. One parametrization is given by the symplectic coordinates for $P_{\text{VI}}$.
Unfortunately, in these coordinates the space of initial conditions is not minimal and we need to 
do blowing-downs in addition to blow-ups. Next, we find an isomorphism between the resulting 
surface $X$ and the Okamoto surface $\tilde{X}$ of the standard form of d-$P_{V}$ by finding an appropriate 
blowing-down structure in $Pic(X)$ that matches the blowing-down structure of $\tilde{X}$. This 
isomorphism can be used to explicitly find the change of coordinates that will transform one 
equation into the other, see the
recent preprint \cite{CarTak:2012:ANOMORSOFBDS} by A.~Carstea and T.~Takenawa 
for a more detailed introduction into this technique. We then outline how to compute the same example in a different 
parameterization that seems to be more natural for the discrete case.

The second example is new and is more interesting (and also more complicated). In \cite{Sak:2007:PDPEATLF}
one of the authors (H.S.) posed a problem of writing down discrete
Painlev\'e equations of type $A_0^{(1)**}$, $A_1^{(1)*}$ and
$A_2^{(1)*}$ as Schlesinger transformations of some Fuchsian equations.
One difficulty here is that these equations do not correspond to any 
continuous Painlev\'e differential equations. This 
problem was solved theoretically by Boalch in \cite{Boa:2009:QADPE}, where it was shown
that the moduli spaces of monodromy groups of some Fuchsian equations are
algebraically isomorphic to the spaces of initial conditions for these discrete
Painlev\'e equations. In particular, Boalch showed that the Fuchsian system with matrix size
$m=3$ with $n=2$ finite poles (and so there are no continuous deformation 
parameters) and rank-two residue matrices, $r_1=r_2=2$, has as its symmetry
the affine Weyl group of type $E_6^{(1)}$, and so the type of the corresponding 
surface is either $A_{2}^{(1)}$ (for the \emph{multiplicative} or \emph{$q$-difference} case)
or $A_{2}^{(1)*}$ (for the \emph{additive} or \emph{difference} case). Since we consider the additive
case, we expect that 
discrete Schlesinger transformations of this system should have type d-$P(A_{2}^{(1)*})$. However,
an elementary discrete Schlesinger transformation, when written in coordinates, becomes very 
complicated and in fact it can not be reduced to a standard form of d-$P(A_{2}^{(1)*})$, 
as written in \cite{Sak:2007:PDPEATLF}. Using 
the geometry of the space of initial conditions, we see that this is an example of a situation 
when the standard form of the equation corresponds to a 
direction that is a combination of two different directions given by 
different elementary Schlesinger transformations, and then we explicitly show how to combine
them into a single Schlesinger transformation that corresponds to the standard form of d-$P(A_{2}^{(1)*})$,
this is a new result. 

\subsection{Discrete Painlev\'e--V equation or d-$P\big(D_4^{(1)}\big)$} 
\label{sub:discrete_painlev'e_v_equation}
\subsubsection{Elementary Schlesinger transformations} 
\label{ssub:schlesinger_transformations_and_d_p__v_}
For this example we consider  a Fuchsian system of the spectral type $11,11,11,11$, i.e., we have $n=3$ (finite) poles and 
the matrix size is $m=2$. In view of Assumption~\ref{assume:rank-reduce}, it is possible to make the rank of the residue 
matrices at the finite poles equal to one.
Further, using the M\"obius transformations on $\mathbb{P}^{1}$,
we can map $(z_{0}=\infty, z_{1}, z_{3})$ to $(\infty,1,0)$. 
We put $z_{2} = t$ and from now on we use $(1,t,0)$ instead of the indices 
$(1,2,3)$, so $\theta_{2} = \theta_{t}$, etc. Then
\begin{equation*}
	\mathbf{A}(z) = \frac{ \mathbf{A}_{1} }{ z-1 } + \frac{ \mathbf{A}_{t} }{ z-t } + \frac{ \mathbf{A}_{0} }{ z },
	\qquad 
	\mathbf{A}_{\infty} = -(\mathbf{A}_{0} + \mathbf{A}_{1} + \mathbf{A}_{t}) = \begin{bmatrix}
		\kappa_{1} & 0 \\ 0 & \kappa_{2}
	\end{bmatrix},
\end{equation*}
and we have the following \emph{Riemann scheme} and the corresponding \emph{Fuchs relation}:
\begin{equation*}
	\left\{
	\begin{tabular}{cccc}
		$z = 0$ 		& $	z = 1 $		& 	$z = t $		& 	$z = \infty$	\\
		$\theta_{0}$	& $\theta_{1}$	& $\theta_{t}$ 	& 	$\kappa_{1}$ \\
		$0$			& 	$0$			& 	$0$ 			& 	$\kappa_{2}$ 
	\end{tabular}
	\right\},\qquad 
	\theta_{0}  + \theta_{1} + \theta_{t} + \kappa_{1} + \kappa_{2} = 0.
\end{equation*}

In the rank-one case there is no need for the second index, and so we put $\mathbf{b}_{i} = \mathbf{b}_{i,1}$,
$\mathbf{c}_{i}^{\dag} = \mathbf{c}_{i}^{1\dag}$. Then the spectral decomposition is simply 
$\mathbf{A}_{i} = \mathbf{b}_{i}\mathbf{c}_{i}^{\dag}$ with 
$\operatorname{tr} (\mathbf{A}_{i}) = \mathbf{c}_{i}^{\dag} \mathbf{b}_{i} = \theta_{i}$.

Normalizing the 
eigenvectors $\mathbf{b}_{i}$ and $\mathbf{c}_{i}^{\dag}$ , we can write
\begin{equation*}
	\mathbf{A}_{i} = a_{i} \begin{bmatrix}
		1 \\ b_{i}
	\end{bmatrix} \begin{bmatrix}
		c_{i} & 1
	\end{bmatrix} = a_{i} \begin{bmatrix} 1 \\\beta_{i}-w \end{bmatrix} \begin{bmatrix} \gamma_{i} + w & 
	1 \end{bmatrix},\qquad \operatorname{tr}(\mathbf{A}_{i}) = \theta_{i} = a_{i} (c_{i} + b_{i}) = a_{i}(\beta_{i} + \gamma_{i}).
\end{equation*}
Here $w = - b_{0}$ and so $\beta_{0} = 0$ and $\gamma_{0} = \theta_{0}/a_{0}$. 
The reason for introducing $w$ is to simplify the relations between 
parameters $a_{i}$, $b_{i}$, and $c_{i}$ that follow from the trace conditions and the equation 
$\mathbf{A}_{\infty} = -\operatorname{diag}\{\kappa_{1},\kappa_{2}\}$. These relations, written in terms of 
$a_{i}$, $\beta_{i}$, and $\gamma_{i}$, are 
\begin{align}
		a_{0} + a_{1} + a_{t} &= 0 \label{eq:dpv-rel12}\\
	a_{1} \beta_{1}\gamma_{1} + a_{t} \beta_{t} \gamma_{t} + w(\kappa_{1} - \kappa_{2}) &=0 \label{eq:dpv-rel21}\\
	a_{1} \beta_{1} + a_{t} \beta_{t}  + \kappa_{2} &=0 \label{eq:dpv-rel22}\\
	\theta_{0} + a_{1} \gamma_{1} + a_{t} \gamma_{t}  + \kappa_{1} &=0, \label{eq:dpv-rel11}
\end{align}
where the last equation also follows from (\ref{eq:dpv-rel22}) and the Fuchs relation. Thus, 
everything is
parameterized by $a_{1}$, $\beta_{1}$, $a_{t}$, and $\beta_{t}$ subject to one additional constraint 
$a_{1} \beta_{1} + a_{t} \beta_{t} + \kappa_{2} = 0$. One additional degree of freedom is related to 
the global gauge action by constant non-degenerate diagonal matrices that allows us to normalize one of the
$a_{i}$s, but it is convenient to keep this freedom for computations. Finally, note that the normalization
parameter $a_{i}$ should be considered as a part of either $\mathbf{b}_{i}$ or $\mathbf{c}_{i}^{\dag}$,
where the actual choice depends on the context.

Same coordinates were used in \cite{Sak:2010:IDA4PTE} to describe the isomonodromic dynamic in the continuous
case and the reduction to $P_{\text{VI}}$. With $a_{i}$ attached to $\mathbf{c}_{i}^{\dag}$, the standard symplectic 
form on the decomposition space reduces as follows, see \textbf{Proposition~4} in \cite{Sak:2010:IDA4PTE} for a more general statement:
\begin{align*}
	\omega &= \sum_{i=1}^{3} \operatorname{tr}\left(d \mathbf{C}_{i}^{\dag} \wedge d \mathbf{B}_{i}\right)
	= d a_{0} \wedge d(-w) + d a_{1}  \wedge d(\beta_{1} - w) + d a_{t} \wedge d(\beta_{t} - w)  \\
	&= d(a_{0} + a_{1} + a_{t}) \wedge d(-w)  +  d a_{1} \wedge d \beta_{1} + d a_{t} \wedge d \beta_t  
	= d a_{1} \wedge d \beta_{1} + d a_{t} \wedge d \beta_{t} \\
	&=\frac{ da_{1} }{ a_{1} }\wedge d(a_{1} \beta_{1}) + \frac{ d a_{t} }{ a_{t} }\wedge d(a_{t} \beta_{t})
	= \left(\frac{ da_{t} }{ a_{t} } - \frac{ da_{1} }{ a_{1} }\right)\wedge d(a_{t} \beta_{t}) 
	= d\left(\frac{ a_{t} }{ a_{1} }\right)\wedge d \left(a_{1} \beta_{t}\right),
\end{align*}
where we used (\ref{eq:dpv-rel22}) to eliminate $a_{1} \beta_{1}$. It is convenient to introduce new symplectic
coordinates $p = (a_{1} \beta_{t})/t$ and $q = - (t a_{t})/a_{1}$. 
This gives, using (\ref{eq:dpv-rel12}--\ref{eq:dpv-rel11}),
\begin{alignat*}{3}
	a_{1} \beta_{1} &= pq - \kappa_{2}, &\qquad a_{1} \beta_{t} &= tp, 									&\qquad a_{1} \gamma_{1} &= \theta_{1} + \kappa_{2} - pq,\\
	a_{t} \beta_{t} &= - pq, 			&\qquad a_{t} \beta_{1} &= \frac{ q }{ t } (\kappa_{2} - pq), 	&\qquad a_{t} \gamma_{t} &= \theta_{t} + pq.
\end{alignat*}
For the continuous deformations, the only deformation parameter is $t$ and the corresponding Hamiltonian, given by
equation~(\ref{eq:cont-Ham}), under this reduction becomes
\begin{align*}
	\mathcal{H}_{t} &= \frac{ \operatorname{tr}(\mathbf{A}_{t} \mathbf{A}_{0}) }{ t } + \frac{ \operatorname{tr}(\mathbf{A}_{t} \mathbf{A}_{1}) }{ t-1 }
	= \frac{ (\theta_{t} + pq)(\theta_{0} + p(q-t)) }{ t } + 
	\frac{ (\theta_{1} + \kappa_{2} - p(q - t)) (t \theta_{t} + \kappa_{2} q -pq(q-t)) }{ t(t-1) }.
\end{align*}
Since 
\begin{align*}
	\omega &= d\left(- \frac{ q }{ t }\right) \wedge d(tp) = dp \wedge dq - d \left( \frac{ pq }{ t }\right)\wedge dt,\\
	\omega- d\mathcal{H}_{t} \wedge dt &= dp\wedge dq - d\left( \mathcal{H}_{t} + \frac{ pq }{ t }\right)\wedge dt = 
	dp\wedge dq - d \mathcal{H}_{VI}\wedge dt,
\end{align*}
where $ \mathcal{H}_{VI} = \mathcal{H}_{t} + pq/t$ is the Hamiltonian function for $P_{\text{VI}}$.

Unfortunately, such a reduction procedure does not work in the discrete case and we need to work directly with the evolution equations.
We give a brief outline of such computations and show the reduction to d-$P_{\text{V}}$. 

From equations (\ref{eq:Ai}--\ref{eq:Abeta}), that are more convenient here,
we almost immediately get the following relations:
\begin{equation}
	\bar{w} - w = \bar{\beta}_{1} - \beta_{t} = \gamma_{1} - \bar{\gamma}_{t} 
	\text{ (and so $\bar{\beta}_{1} + \bar{\gamma}_{t}  = \beta_{t} + \gamma_{1}$)},\qquad 
	\beta_{t} \bar{\gamma}_{t} = t \bar{\beta}_{1} \gamma_{1}\,.	\label{eq:dpv-evol}	
\end{equation}
Multiplying the last equation in (\ref{eq:dpv-evol}) by $a_{1}\bar{a_{t}}$, we get
\begin{equation}
		(\theta_{1} + \kappa_{2} - p q) \bar{q} (\bar{\kappa}_{2} - \bar{p}\bar{q})
		=  tp (\bar{\theta}_{t} + \bar{p}\bar{q}).\label{eq:dpv-eq1-pq}
\end{equation}
Combining the first equation in (\ref{eq:dpv-evol}) with (\ref{eq:dpv-rel21}) gives
\begin{equation*}
	a_{1}\beta_{1}(\gamma_{1} + \beta_{t}) + \beta_{t}(a_{t}\gamma_{t} - a_{1} \beta_{1} + \kappa_{1} - \kappa_{2}) =
	\bar{a}_{1} \bar{\beta}_{t}(\bar{\gamma}_{t} + \bar{\beta}_{1}) + \bar{\beta}_{1}(\bar{a}_{1} \bar{\gamma}_{1} - 
	\bar{a}_{t}\bar{\beta}_{t} + \kappa_{1}- \kappa_{2}).
\end{equation*}
Dividing by $\gamma_{1} + \beta_{t} = \bar{\gamma}_{t} + \bar{\beta}_{1}$ and doing some re-arranging (and using
(\ref{eq:dpv-eq1-pq}) in the process), we get
\begin{equation}
	pq + \bar{p}\bar{q} - \kappa_{2} = \frac{ t(\theta_{t} + \kappa_{1}) }{ q - t - \frac{ \theta_{1} + \kappa_{2} }{ p } }
	+ \frac{ \bar{\theta}_{1} + \kappa_{1} }{ q - 1 - \frac{ \theta_{1} + \kappa_{2} }{ p } }.
	\label{eq:dpv-eq2-pq}
\end{equation}

We now claim that after a certain change of variables equations (\ref{eq:dpv-eq1-pq}) and (\ref{eq:dpv-eq2-pq}) 
transform into the standard form for d-$P_{\text{V}}$. Although this can be seen directly without too much 
difficulty, we prefer to use the geometry of the Okamoto space of initial conditions for this equation to explain
how to find this coordinate transformation, illustrating how such geometric machinery can be used.


\subsubsection{Geometry of d-$P(D_{4}^{(1)})$} 
\label{ssub:geometry_of_d_p__v_}
Let us briefly review the geometric description of the difference Painlev\'e V equation from the point of view of the theory of 
rational surfaces \cite{Sak:2001:RSAWARSGPE}. We think about it
as a birational map $\varphi: \mathbb{P}^{1}\times \mathbb{P}^{1} \dashrightarrow \mathbb{P}^{1}\times \mathbb{P}^{1}$ 
with parameters $a_{0},\dots, a_{4},s$ or, equivalently, as an automorphism of a field
$\mathbb{C}(a_{0},\dots,a_{4};s; f,g)$,
\begin{equation*}
	\left(\begin{matrix}
		a_{1}\quad a_{2}\quad a_{3}\\
		\quad a_{4}\quad a_{0}\quad
	\end{matrix}; s; f,g\right) \mapsto 
	\left(\begin{matrix}
		\bar{a}_{1}\quad \bar{a}_{2}\quad \bar{a}_{3}\\
		\quad \bar{a}_{4}\quad \bar{a}_{0}\quad
	\end{matrix}; \bar{s}; \bar{f},\bar{g}\right),\qquad\qquad
	\begin{aligned}
		\bar{a}_{1} &= a_{1} + \delta,\quad & \bar{a}_{2} &= a_{2} - \delta,\quad & \bar{a}_{3} &= a_{3},\\
		\bar{a}_{4} &= a_{4},\quad & \bar{a}_{0} &= a_{0} + \delta,\quad & \bar{s} &= s,
	\end{aligned}
\end{equation*} 
where $\delta = a_{0} + a_{1} + 2 a_{2} + a_{3} + a_{4}$  
and where $\bar{f}$ and $\bar{g}$ are given by the following system that is usually called 
the discrete Painlev\'e-V equation
\begin{equation}
	\left\{
	\begin{aligned}
		f + \bar{f} &= a_{3} + \frac{ a_{1} }{ g+1 } + \frac{ a_{0} }{ s g + 1 }\\
		g \bar{g} &= \frac{ (\bar{f} + \bar{a}_{2}) (\bar{f} + \bar{a}_{2} + \bar{a}_{4}) }{ s \bar{f} (\bar{f} - \bar{a}_{3}) }
	\end{aligned}
	\right.;\label{eq:dpv-standard}
\end{equation}
we refer to it as the standard form of d-$P_{\text{V}}$.
We lift this map to an \emph{isomorphism} $\hat{\varphi}$ between rational surfaces $X_{\mathbf{a}} = X_{(a_{0},\dots a_{4})}$ 
and $X_{\bar{\mathbf{a}}}$
obtained from $\mathbb{P}^{1}\times \mathbb{P}^{1}$
via a sequence of blow-ups that resolve the indeterminacies of the map $\varphi$,
\begin{equation*}
  \raisebox{0.6in}{\xymatrix@C+2em{
  X_{\mathbf{a}} \ar[r]_{\hat{\varphi}}^{\simeq} \ar@{-->}[d] & X_{\bar{\mathbf{a}}} \ar@{-->}[d] \\
  \mathbb{P}^{1}\times \mathbb{P}^{1} \ar@{-->}[r]_-{\varphi} & \mathbb{P}^{1}\times \mathbb{P}^{1}
  }}.
\end{equation*}
Decomposing $\varphi$ as a sequence of two maps, 
$(f,g)\overset{\varphi_{1}}{\longmapsto} (\bar{f},g)\overset{\varphi_{2}}{\longmapsto} (\bar{f},\bar{g})$
it is easy to see that the indeterminate points of $\varphi_{1}$ are $(\infty,-1)$ and $(\infty,-1/s)$ and the 
indeterminate points of $\varphi_{2}$ are $(-\bar{a}_{2},0)$, $(-\bar{a}_{2} - \bar{a}_{4},0)$, $(0,\infty)$,
and $(\bar{a}_{3},\infty)$. Applying $\varphi_{1}^{-1}$ we obtain that the indeterminate points of 
$\varphi$ are $p_{1}(-a_{2}-a_{4},0)$, $p_{2}(-a_{2},0)$, $p_{3}(0,\infty)$, $p_{4}(a_{3},\infty)$,
$p_{5}(\infty,-1)$, and $p_{6}(\infty,-1/s)$. In exactly the same way we find that the indeterminate points
of $\varphi^{-1}$ are $\bar{p}_{1}(-\bar{a}_{2}-\bar{a}_{4},0),\dots ,\bar{p}_{6}(\infty,-1/\bar{s})$. 

As usual, we denote a local coordinate description of a blow-up at a point $(x_{0},y_{0})$ by
$(x,y)\leftarrow (u,v)\cup(U,V)$ where $x-x_{0} = u = UV$ and $y-y_{0} = uv = V$, then the exceptional
divisor is given by $u = V = 0$. For example, extending the map $\varphi$ at the blow-up at $p_{1}(-a_{2}-a_{4},0)$, 
we use $u_{1} = f + a_{2} + a_{4}$ and $u_{1} v_{1} = g$. Then $\bar{f} = \bar{f}(u_{1},v_{1})$ and 
$\bar{g} = g(u_{1}, v_{1})$ become well-defined and we see that the exceptional divisor $E_{1}$ is mapped 
to the line $\bar{f} = -\bar{a}_{2}$. On the level of divisor classes on $\operatorname{Pic}(X)$ we get the map
${E}_{1}\mapsto H_{f} - {E}_{2}$, where $H_{f}$ is the class of the total 
transform of the line $f=\text{const}$.

Proceeding along the same lines and noticing that we need to blow-up at two additional points 
$p_{7}(0,a_{1})$ on $E_{5}$ and $p_{8}(0,a_{0}/s)$ on $E_{7}$ (here we give the coordinates of the 
points in the $(u_{i}, v_{i})$ charts), we get the following blow-up diagram describing the construction of the 
surface $X_{\mathbf{a}}$ corresponding to our equation:
\begin{center}
	\includegraphics[width=6.2in]{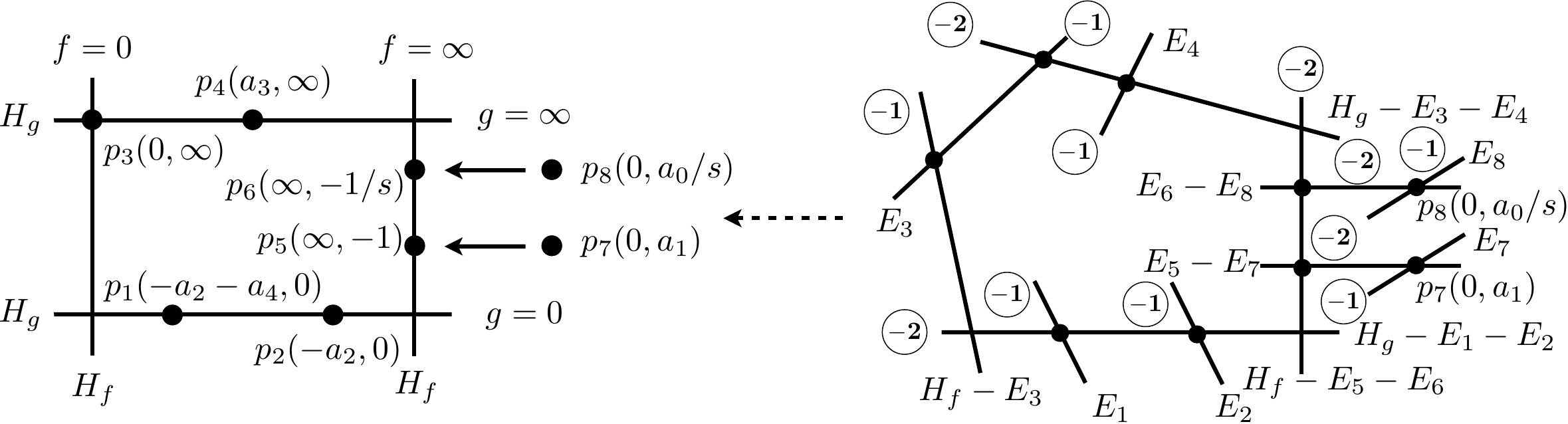}
\end{center}

The Picard lattice of $X$ is generated by the total transforms 
$H_{f}$ and $H_{g}$ of the coordinate lines and the classes of the exceptional divisors $E_{i}$,
\begin{equation*}
	\operatorname{Pic}(X) = \mathbb{Z} H_{f} \oplus \mathbb{Z} H_{g} \oplus \bigoplus_{i=1}^{8} \mathbb{Z}E_{i}.
\end{equation*}
The anti-canonical divisor $-K_{X}$ uniquely decomposes as a positive linear combination of $-2$-curves $D_{i}$,
\begin{equation*}
	-K_{X} = 2 H_{f} + 2 H_{g} - \sum_{i=1}^{8} E_{i} = D_{0} + D_{1} + 2 D_{2} + D_{3} + D_{4}, 
\end{equation*}
where
\begin{equation*}
	D_{0} = H_{g} - E_{1} - E_{2},\quad D_{1} = H_{g} - E_{3} - E_{4},\quad D_{2} = H_{f} - E_{5} - E_{6},\quad 
	D_{3} = E_{5} - E_{7},\quad
	D_{4} = E_{6} - E_{8}.
\end{equation*}
The configuration of these $-2$-curves (on which the blow-up points are located) is described by the affine Dynkin diagram
of type $D_{4}^{(1)}$,
\begin{center}
	\includegraphics[width=4in]{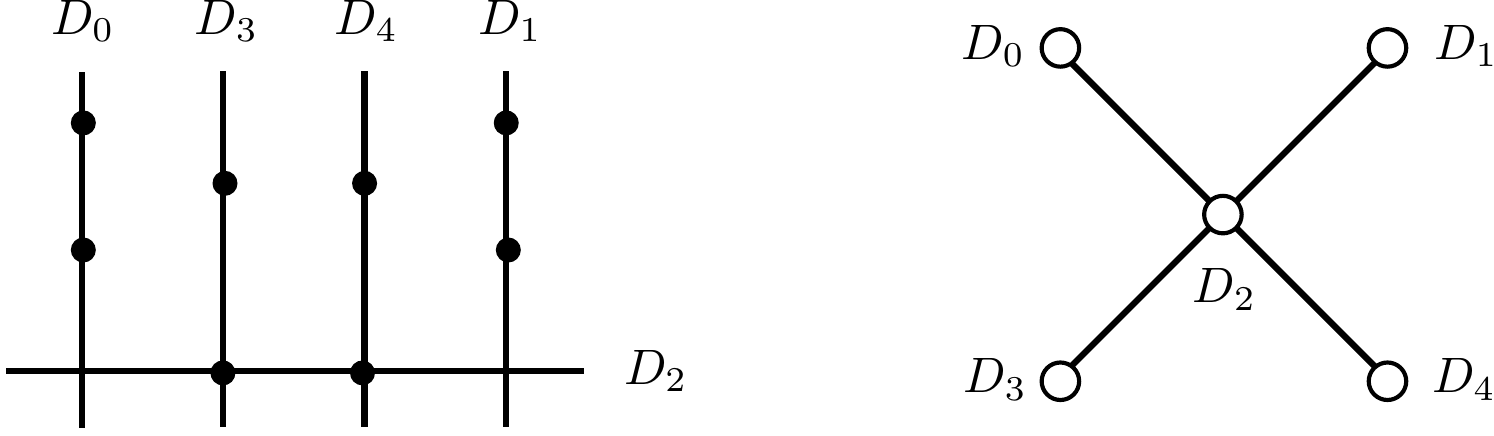},
\end{center}
which is why we say that the difference Painlev\'e V equation has the type d-$P(D_{4}^{(1)})$. The corresponding divisors 
$D_{i}$ generate a 
sub-lattice $R=\operatorname{Span}_{\mathbb{Z}}\{D_{0},\dots,D_{4}\}\subset (-K_{X})^{\perp}\simeq E_{8}^{(1)}$ in 
$\operatorname{Pic}(X)$. Symmetries of $X$ are given by the affine Weyl group of \emph{Cremona isometries} 
generated by the \emph{complementary} root sub-lattice $R^{\perp}$ in $(-K_{X})^{\perp}$ acting as automorphisms of
$\operatorname{Pic}(X)$. In this case, $R^{\perp}=\operatorname{Span}_{\mathbb{Z}}\{\alpha_{0},\dots,\alpha_{4}\}$, where
\begin{equation*}
	\begin{aligned}
		\alpha_{0}&= E_{1} - E_{2} \\  
		\alpha_{1}&= E_{3} - E_{4} \\  
		\alpha_{2}&= H_{f} - E_{1} - E_{3} \\
		\alpha_{3}&= H_{g} - E_{5} - E_{7} \\
		\alpha_{4}&= H_{g} - E_{6} - E_{8} \\
	\end{aligned}\qquad\qquad\qquad
		\raisebox{-0.5in}{\includegraphics[height=1.1in]{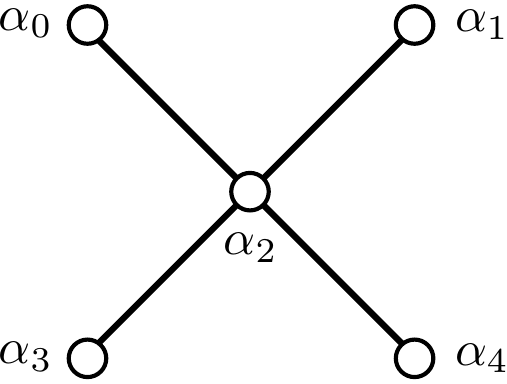}},
\end{equation*}
which is again of the type $D_{4}^{(1)}$. Finally, we can compute the action of $\varphi_{*}$ on $\operatorname{Pic}(X)$ to be
\begin{alignat*}{2}
	 H_{f}&\mapsto 5 H_{f} + 2 H_{g} -2 E_{1} - 2 E_{2} - 2 E_{3} - 2 E_{4} - E_{5} - E_{6} - E_{7} - E_{8},& \qquad 
		 E_{1}&\mapsto H_{f}-E_{2},\\
	 H_{g}&\mapsto 2 H_{f} + H_{g} - E_{1} - E_{2} - E_{3} - E_{4},& \qquad 
	 E_{2}&\mapsto H_{f}-E_{1},\\
	 E_{5}&\mapsto 2 H_{f} + H_{g}-E_1-E_2-E_3-E_4-E_8, & \qquad 	 E_{3}&\mapsto H_{f}-E_{4},\\
	 E_{6}&\mapsto 2 H_{f} + H_{g}-E_1-E_2-E_3-E_4-E_7, & \qquad 	 E_{4}&\mapsto H_{f}-E_{3},\\
	 E_{7}&\mapsto 2 H_{f} + H_{g}-E_1-E_2-E_3-E_4-E_6,\\
	 E_{8}&\mapsto 2 H_{f} + H_{g}-E_1-E_2-E_3-E_4-E_5,
\end{alignat*}
and so the induced action $\varphi_{*}$ on the sub-lattice $R^{\perp}$	is given by \emph{translation}:
\begin{equation*}
	\varphi_{*}: (\alpha_{0}, \alpha_{1}, \alpha_{2}, \alpha_{3}, \alpha_{4}) \mapsto	
	(\alpha_{0}, \alpha_{1}, \alpha_{2}, \alpha_{3}, \alpha_{4}) + 
	(0,0,1,-1,-1)\delta,	\qquad\text{where }\delta = - K_{X}.
\end{equation*}


\subsubsection{Reduction to the standard form} 
\label{ssub:reduction_to_the_standard_form}
We now demonstrate how to reduce equations (\ref{eq:dpv-eq1-pq}, \ref{eq:dpv-eq2-pq}) to equations 
(\ref{eq:dpv-standard}). First, we resolve the indeterminacies of 
$\psi: (p,q)\to (\bar{p},\bar{q})$ to get
an isomorphism $\hat{\psi}: \tilde{X}_{\mathfrak{a}}\to \tilde{X}_{\bar{\mathfrak{a}}}$. The surface
$\tilde{X}=\tilde{X}_{\mathfrak{a}}$, with $\mathbf{\mathfrak{a}}$ denoting the collection of parameters of the map,
$\mathbf{\mathfrak{a}}=\{\theta_{0}, \theta_{1}, \theta_{t}, \kappa_{1}, \kappa_{2} \}$, is
constructed by the following blow-up diagram:
\begin{center}
	\includegraphics[width=6.2in]{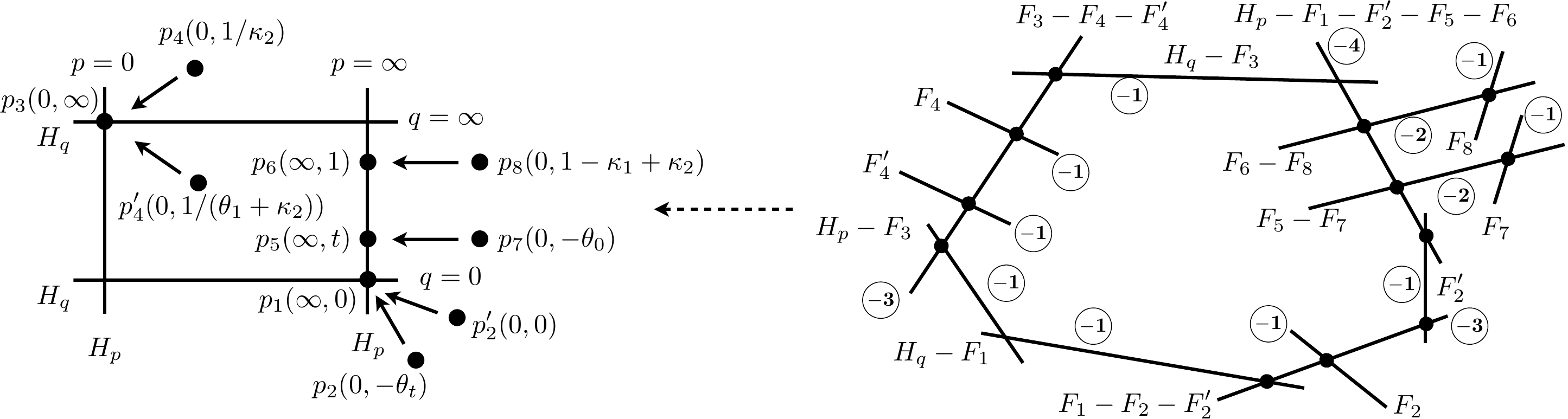}.
\end{center}
Here the coordinates of the blow-up points are given w.r.t.~the following local coordinate systems:
\begin{align*}
	p_{1}\left(\frac{ 1 }{ p } = 0, q = 0 \right) &\leftarrow p_{2}\left(\frac{ 1 }{ p } = 0, pq = - \theta_{t}\right)
	\text{ and } p_{2}'\left(\frac{ 1 }{ pq } = 0, q = 0\right),\\
	p_{3}\left( p = 0, \frac{ 1 }{ q } = 0 \right) &\leftarrow p_{4}\left(p = 0, \frac{ 1 }{ pq } = \frac{ 1 }{ \kappa_{2} }\right)
	\text{ and } p_{4}'\left(p = 0, \frac{ 1 }{ pq } = \frac{ 1 }{ \theta_{1} + \kappa_{2} }\right),\\
	p_{5}\left( \frac{ 1 }{ p } = 0, q = t \right) &\leftarrow p_{7}\left(\frac{ 1 }{ p } = 0 , p(q - t) = - \theta_{0}\right),\\
	p_{6}\left( \frac{ 1 }{ p } = 0, q = 1 \right) &\leftarrow p_{8}\left(\frac{ 1 }{ p } = 0 , p(q - 1) = 1 - \kappa_{1} + \kappa_{2})\right).
\end{align*}

Note that the resulting configuration of exceptional divisors looks quite different from the previous case and moreover, 
$\operatorname{rank}(\operatorname{Pic}(\tilde{X})) = 12$. The reason for this is that the surface
$\tilde{X}$ is not \emph{relatively minimal} for $\hat{\psi}$, see also \cite{CarTak:2012:ANOMORSOFBDS} 
--- there are two pairs of $-1$-curves that are
exchanged by $\hat{\psi}$: $F_{2}\leftrightarrow F_{4}'$ and $F_{2}'\leftrightarrow H_{q} - F_{3}$, and so blowing 
down any of these pairs will still result in an isomorphism between the corresponding surfaces. We blow down $F_{2}'$ and 
$H_{q} - F_{3}$, which eliminates $-4$ and $-3$-curves and gives us the diagram on the left. This diagram 
now looks very similar to the configuration 
we had for the standard form of d-$P_{V}$ (on the right), 
\begin{center}
	\begin{tabular}{cc}
		\quad\includegraphics[width=3in]{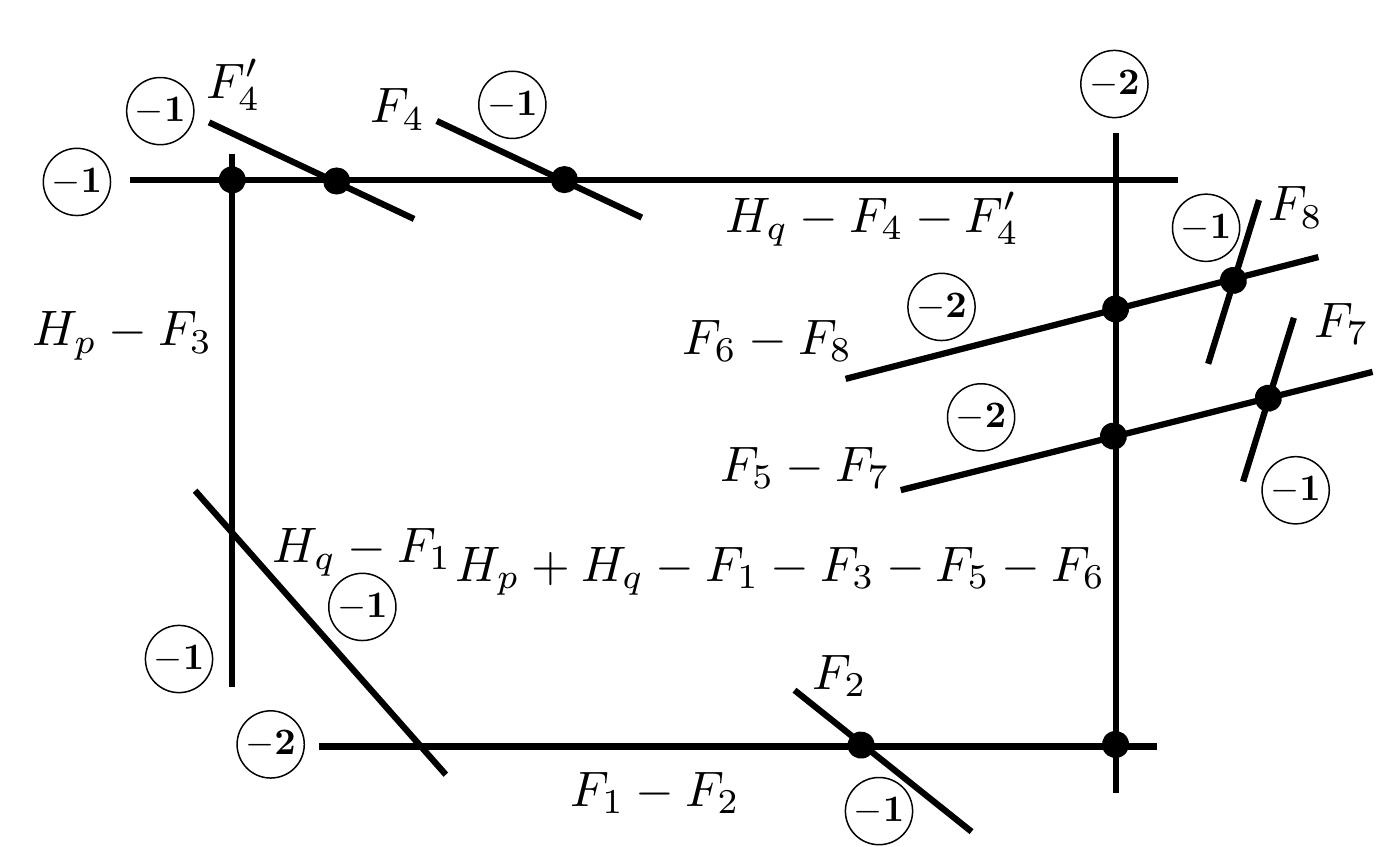}\qquad &
			\qquad\raisebox{0.17in}{\includegraphics[width=2.6in]{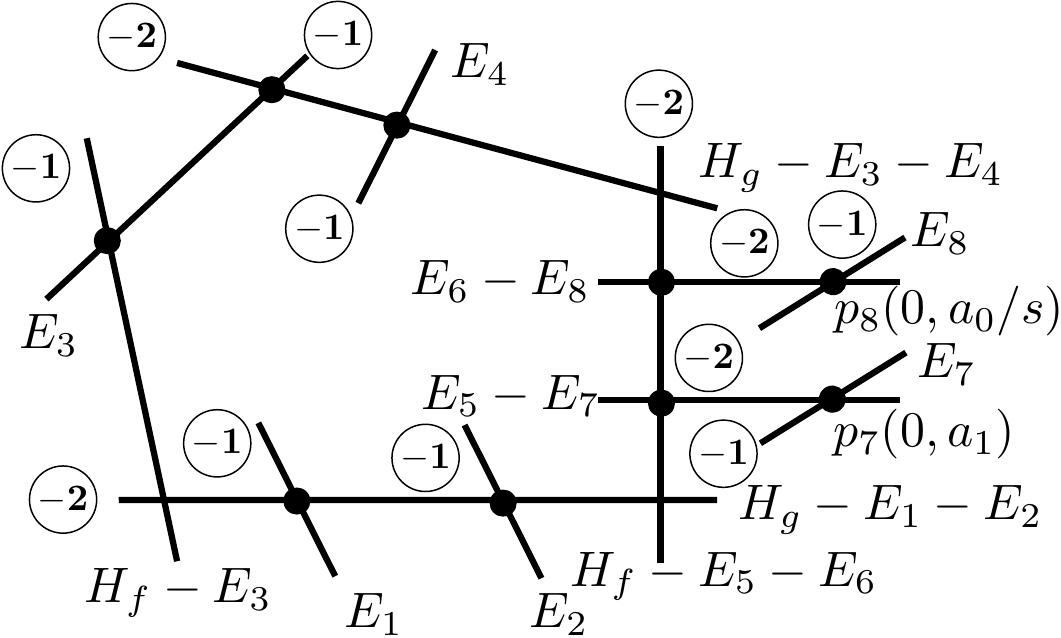}}\quad 
	\end{tabular},
\end{center}
and in particular we immediately see that $\tilde{X}$ is again of type $D_{4}^{(1)}$. We now need to find the blowing-down structure for
$\tilde{X}$ to match the blowing-down structure for $X$. 
Thus, in $\operatorname{Pic}(\tilde{X})$ we put $\mathcal{E}_{9} = F_{2}'$, $\mathcal{E}_{10} = H_{q} - F_{3}$ and we are looking for
\emph{rational} $0$-curves $\mathcal{H}_{f}$, $\mathcal{H}_{g}$, and $-1$-curves $\mathcal{E}_{i}$, $i=1,\dots,8$ such that 
\begin{equation*}
	\mathcal{H}_{f}\bullet \mathcal{H}_{g} = 1,\qquad \mathcal{E}_{i}^{2} = -1, \qquad
	\mathcal{H}_{f}^{2} = \mathcal{H}_{g}^{2} = \mathcal{H}_{f} \bullet \mathcal{E}_{i} = 
	\mathcal{H}_{f} \bullet \mathcal{E}_{i} = \mathcal{E}_{i} \bullet \mathcal{E}_{j} = 0,\quad 1\leq i\neq j\leq 10. 
\end{equation*}
Using the above diagrams, the genus formula $2g(C) -2 = C^{2} + C\bullet K_{\tilde{X}}$, and the distinguished $-2$-curve $D_{2}$,
we see that it makes sense to take $\mathcal{H}_{f} = H_{p} + H_{q} - F_{1} - F_{3}$ and
$\mathcal{E}_{i} = F_{i}$, $i=5,\dots8$. Then the possible choices for $\mathcal{H}_{g}$ are $H_{q}$, 
$H_{p} + H_{q} - F_{2} - F_{3}$, $H_{p} + H_{q} - F_{4} - F_{3}$, and $H_{p} + H_{q} - F_{4}' - F_{3}$. Attempting to represent
$F_{1}-F_{2} = \mathcal{H}_{g} - \mathcal{E}_{1} - \mathcal{E}_{2}$ eliminates the first two possibilities, 
and the remaining two are equivalent. To summarize, we have the following blowing-down structure:
\begin{alignat*}{3}
	\mathcal{H}_{f} &= H_{p} + H_{q} - F_{1} - F_{3}, &\qquad  \mathcal{E}_{3} &= H_{p} - F_{3}, 
	&\qquad \mathcal{E}_{i} &= F_{i},\qquad i=2,4,\dots,8,\\
	\mathcal{H}_{g} &= H_{p} + H_{q} - F_{3} - F'_{4}, &\qquad \mathcal{E}_{9} &= F_{2}',\\
	\mathcal{E}_{1} &= H_{p} + H_{q} - F_{1} - F _{3} - F'_{4} , &\qquad \mathcal{E}_{10} &= H_{q} - F_{3}.
\end{alignat*}
To find the expressions for coordinates $f$ and $g$ we proceed as follows. The linear system $|\mathcal{H}_{f}|$ consists of curves of bi-degree $(1,1)$
passing through $p_{1}$ an $p_{3}$, and so $|\mathcal{H}_{f}|= \{c_{1} pq + c_{2} =0\}$. Thus, $[pq:1]$ is a projective coordinate on $|H_{f}|$, and
so we can take $f= pq$. Similarly, $\left[ pq -(\theta_{1}+\kappa_{2}) : p \right]$ is a projective coordinate on $|\mathcal{H}_{g}|$. We take 
$g = -\frac{ 1 }{ t }\left(q - \frac{ \theta_{1} + \kappa_{2} }{ p }\right)$, where the additional factor $-1/t$ is needed to ensure that the 
$(f,g)$-coordinates of $p_{5}$ are $(\infty,-1)$, as in the standard case. 

We can now match the parameters of our elementary Schlesinger transformations with the parameters $s$ and $a_{i}$ in the standard 
form (\ref{eq:dpv-standard}) of d-$P_{\text{V}}$. 
For example, $\mathcal{E}_{1}$ blows down to $(f,g)=(\theta_{1}+\kappa_{2},0) = (-a_{2}-a_{4},0)$. Doing this for other
divisors $\mathcal{E}_{i}$ results in the following
\begin{equation*}
	a_{0} = \theta_{1} - 1 + k_{1}, \quad a_{1} = - \theta_{t}- k_{1},\quad a_{2} = \theta_{t},\quad a_{3}= \kappa_{2},\quad
	a_{4} = - \theta_{1} - \theta_{t} - \kappa_{2},\quad \delta = - 1.
\end{equation*}
Thus, not only we can now immediately see that equations (\ref{eq:dpv-eq1-pq}, \ref{eq:dpv-eq2-pq}) transform into the standard form
(\ref{eq:dpv-standard}) of d-$P_{\text{V}}$, but also that under this identification of parameters, 
\begin{equation*}
	\bar{\theta}_{1} = -\bar{a}_{2} -\bar{a}_{3} - \bar{a}_{4} = - a_{2} + \delta - a_{3} - a_{4} = \theta_{1} - 1,\qquad
	\bar{\theta}_{t} = - \bar{a}_{0} - \bar{a}_{1} - \bar{a}_{2} - \bar{a}_{3} - \bar{a}_{4} - 1 = 
	\theta_{t}+1,
\end{equation*}
and the other parameters remain unchanged. Thus, we can \emph{identify the action of d-$P_{\text{V}}$ on the Riemann Scheme} of the Fuchsian system
under this correspondence. We use the same observation in the next example.

\subsubsection{Different Parameterization} 
\label{ssub:different_parameterization}
A more direct derivation of d-$P_{\text{V}}$ can be obtained if we make a different coordinate choice (which seems to be more natural
for Schlesinger transformations and is used in the next example as well).

We use the global gauge group action to map vectors $\mathbf{b}_{t}$ and $\mathbf{b}_{1}$ to the standard basis vectors and then 
use the trivial transformations to make components of $\mathbf{b}_{0}$ equal to $1$ to get the following parameterization:
\begin{equation*}
	\mathbf{b}_{t} = \begin{bmatrix} 1\\ 0	\end{bmatrix},\quad
	\mathbf{c}_{t} = \begin{bmatrix} \theta_{t}& \upalpha	\end{bmatrix},\quad
	\mathbf{b}_{1} = \begin{bmatrix} 0\\ 1	\end{bmatrix},\quad
	\mathbf{c}_{1} = \begin{bmatrix} \upbeta& \theta_{1}	\end{bmatrix},\quad
	\mathbf{b}_{0} = \begin{bmatrix} 1\\ 1	\end{bmatrix},\quad
	\mathbf{c}_{0} = \begin{bmatrix} \upgamma& \theta_{0}-\upgamma	\end{bmatrix},\quad
\end{equation*}
where three parameters $\upalpha$, $\upbeta$, and $\upgamma$ are constrained by the condition 
that the eigenvalues of $\mathbf{A}_{\infty}$ are $\kappa_{1}$, $\kappa_{2}$. Indeed, with this parameterization,
\begin{equation*}
- \mathbf{A}_{\infty} = \begin{bmatrix}
	\theta_{t} + \upgamma & \theta_{0} + \upalpha - \upgamma \\
	\upbeta + \upgamma & \theta_{0} + \theta_{1} - \upgamma
\end{bmatrix},\qquad 
\begin{aligned}
	\operatorname{tr} \mathbf{A}_{\infty} &= \kappa_{1} + \kappa_{2} = - (\theta_{0} + \theta_{1} + \theta_{t}),\\
	\det \mathbf{A}_{\infty} &= \kappa_{1} \kappa_{2} = (\theta_{0} + \theta_{1} - \upgamma)(\theta_{t} + \upgamma)
	- (\theta_{0} + \upalpha - \upgamma)(\upbeta + \upgamma).	
\end{aligned}	
\end{equation*}
The trace condition is just the Fuchs relation, and the determinant condition allow us to express, e.g., 
$\upalpha$ in terms of $\upbeta$ and $\upgamma$. Thus, if we put $\upbeta = x$ and $\upgamma = y$, 
$\upalpha(x,y)$ is given by the expression
\begin{equation*}
	\upalpha(x,y) = \frac{ x(y - \theta_{0}) + y(\theta_{1} - \theta_{t}) + (\theta_{1} + \kappa_{1} + \theta_{0})(\theta_{t} + \kappa_{1})}{ x+y }.
\end{equation*}

Using the evolution equations~(\ref{eq:bb-cb-generators}---\ref{eq:cb-am}) for the elementary 
Schlesinger transformation $\left\{\begin{smallmatrix} 1&t\\1&1\end{smallmatrix}\right\}$ then gives us the following map 
$\chi:(x,y)\to (\bar{x},\bar{y})$:
\begin{equation}
	\left\{ 
	\begin{aligned}
		\bar{x} &= \frac{ 	
		  ((x +\theta_{1}) y - 
		       \theta_{0} x) ((t - 1) ((x + \theta_{1}) y + 
		          \theta_{1} x)  + 
		       x (t x - \theta_{t})) ( t(x + y) - y)   
		}{ t x  ((t(x + y) - y) (\theta_{1} + x) -   x (1 + \theta_{t})) }\\
		&\qquad \qquad + \frac{  tx (x+y)((\theta_{1}-1)(\theta_{t}+1) - \kappa_{1}\kappa_{2}) 
	+ x y \kappa_{1}\kappa_{2} }{ t  ((t(x + y) - y) (\theta_{1} + x) -   x (1 + \theta_{t})) }\\[10pt]
		\bar{y} &= \frac{ (t(x + y) - y) (\theta_{0} x - (x+\theta_{1})y ) + x y (\theta_{t} + 1) }{ x(t(x + y) - y) }		
	\end{aligned}
	\right.\label{eq:dpv-sch-simple}
\end{equation}

The Okamoto surface $X$ for this map is given by the blow-up diagram
\begin{center}
	\includegraphics[width=5.5in]{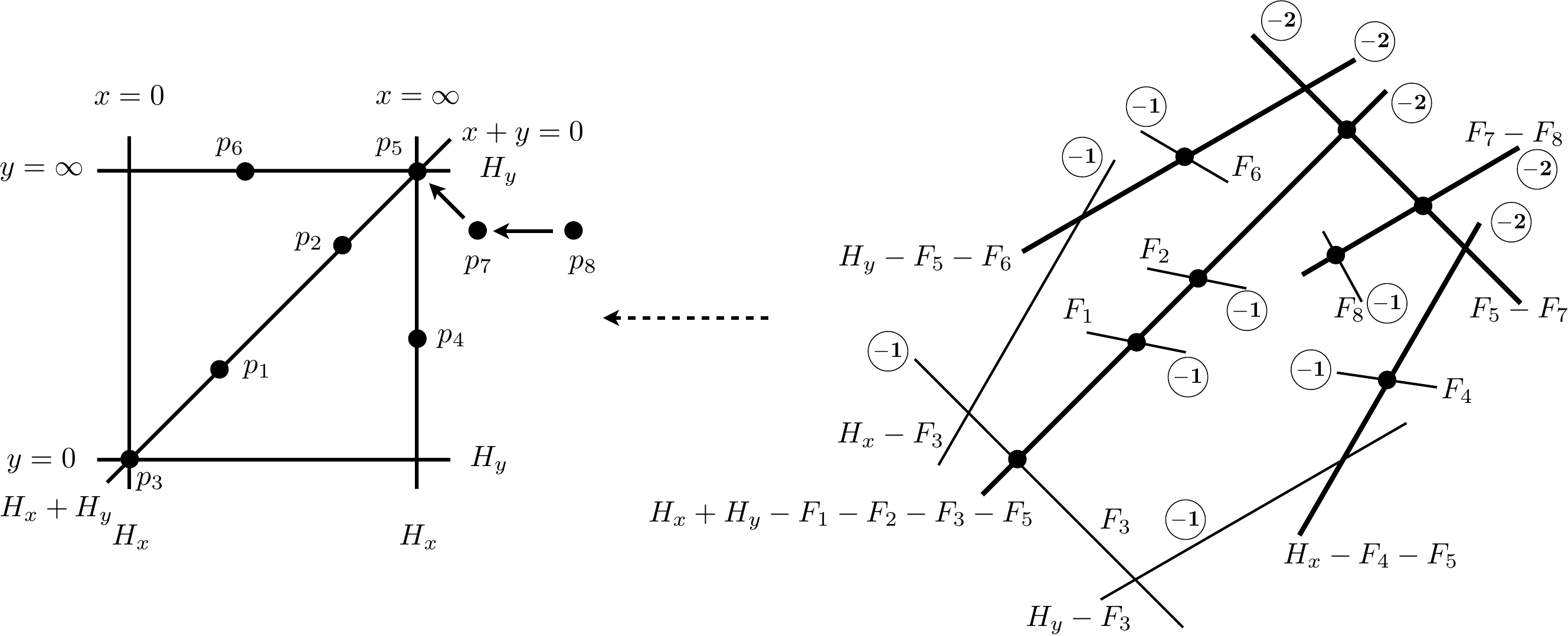},
\end{center}
where the coordinates of the blow-up points are 
\begin{align*}
	&p_{1}(x=\theta_{t} + \kappa_{1},y= - \theta_{t} - \kappa_{1}),\quad p_{2}(x=\theta_{t} + \kappa_{2},y= - \theta_{t} - \kappa_{2}),\quad
	p_{3}(x=0,y=0),\\ 
	&p_{4}\left(\frac{ 1 }{ x } = 0,y=\theta_{0}\right),\quad p_{6}\left(x = -\theta_{1},\frac{ 1 }{ y } = 0,\right),\\
	&p_{5}\left(\frac{ 1 }{ x }=0,\frac{ 1 }{ y }=0\right)\leftarrow p_{7}\left(\frac{ 1 }{ x } = 0,\frac{ x }{ y } = \frac{ 1-t }{ t }\right)
	\leftarrow p_{8}\left(\frac{ 1 }{ x } = 0, t(x+y)-y  = \theta_{t} + 1 \right).
\end{align*}

We compute the action of $\chi_{*}$ on $\operatorname{Pic}(X)$ to be
\begin{alignat*}{2}
	 H_{x}&\mapsto 2 H_{x} + 3 H_{y} -2 F_{1} - 2 F_{2} - F_{4} - F_{5} - F_{6} - F_{7},& \qquad 
		 F_{1}&\mapsto H_{y}-F_{2},\\
	 H_{y}&\mapsto 2 H_{x} + 4 H_{y} - 2 F_{1} - 2 F_{2} - 2 F_{4} - F_{5} - F_{6} - F_{7} - F_{8},& \qquad 
	 F_{2}&\mapsto H_{y}-F_{1},\\
	 F_{3}&\mapsto 2 H_{x} + 3 H_{y} - 2 F_{1} - 2 F_{2} - F_{4} - F_{5} - F_{6} - F_{7} - F_{8}, & \qquad 	 F_{4}&\mapsto F_{3},\\
	 F_{5}&\mapsto H_{x} + 2 H_{y} -F_{1} - F_{2} - F_{4} - F_{6} - F_{7}, & \qquad 	 
	 F_{6}&\mapsto H_{x} + 2 H_{y} -F_{1} - F_{2} - F_{4} - F_{5} - F_{7},\\
	 F_{7}&\mapsto H_{x} + 2 H_{y} -F_{1} - F_{2} - F_{4} - F_{5} - F_{6},& \qquad 
	 F_{8}&\mapsto H_{x} + H_{y} -F_{1} - F_{2} - F_{4}.
\end{alignat*}

To match this picture with the standard one, we look at the $-2$-curves. The distinguished $-2$-curve 
$D_{2} = H_{f} - E_{5} - E_{6}$ should correspond to $F_{5} - F_{7}$. Next, if we let 
$D_{0} = H_{g} - E_{1} - E_{2}$ correspond to $H_{x} + H_{y} - F_{1} - F_{2} - F_{3} - F_{5}$, 
$D_{1} = H_{g} - E_{3} - E_{4}$ should correspond to $H_{x} - F_{4} - F_{5}$ to match the dynamic 
$D_{0}\leftrightarrow D_{1}$. Let 
$D_{3} = E_{5} - E_{7}$ correspond to $H_{y} - F_{5} - F_{6}$, then  $D_{4} = E_{6} - E_{7}$ 
corresponds to the remaining $-2$-curve $F_{7} - F_{8}$. A change of basis giving such identification
can be taken as 
\begin{equation*}
	\mathcal{H}_{f} = H_{y}, \quad  \mathcal{H}_{g} = H_{x} + H_{y} - F_{3} - F_{5}, \quad 
	\mathcal{E}_{3} = H_{y} - F_{3},\quad \mathcal{E}_{5} = H_{y} - F_{5},\quad
	\mathcal{E}_{6} = F_{7},\quad \mathcal{E}_{7} = F_{6},
\end{equation*}
and $\mathcal{E}_{i} = F_{i}$ for $i=1,2,4,8$.
Under this identification the action of $\chi_{*}$ on the symmetry sub-lattice $R^{\perp}$ coincides
with the action of $\varphi_{*}$.

Let us now find the new coordinates $f$ and $g$. Since $|\mathcal{H}_{f}| = |H_{y}|$, we can take $f = y$. The linear system
$|\mathcal{H}_{g}|$ consists of curves of bi-degree $(1,1)$ passing through $p_{3}(0,0)$ and $p_{5}(\infty,\infty)$
and so $|\mathcal{H}_{g}|= \{c_{1} x + c_{2} y =0\}$. Thus, $[x:y]$ is a projective coordinate on $|H_{g}|$, and so we can provisionally 
put $g = x/y$, which gives $g(p_{1}) = g(p_{2}) = -1$, $g(p_{3}) = g(p_{4}) = \infty$, $g(p_{5}) = 0$, and we want them to be 
$0$, $\infty$, and $-1$ respectively. This is easy to fix by the M\"obius transformation $g\mapsto - (g + 1)$, and so our final change of coordinates 
is $f = y$ and $g = -(x+y)/y$. It is then easy to see that under this change of coordinates 
equations~(\ref{eq:dpv-sch-simple}) transforms into the standard form~(\ref{eq:dpv-standard}). The matching of the standard
parameters with the Fuchsian system parameters is given by  
\begin{equation*}
	a_{0} = -(\theta_{t} + 1), \quad a_{1} = \theta_{1},\quad a_{2} = \theta_{t} + \kappa_{2},\quad a_{3}= \theta_{0},\quad
	a_{4} = \kappa_{1} - \kappa_{2},\quad s = t,\quad \delta = - 1,
\end{equation*}
and under this identification of parameters, 
\begin{equation*}
	\bar{a}_{0} = -(\bar{\theta}_{t} + 1) = a_{0} + \delta, \quad \bar{a}_{1} = \bar{\theta}_{1} = a_{1} + \delta,
	\quad \bar{a}_{2} = \bar{\theta}_{t} + \bar{\kappa}_{2} = a_{2} - \delta,\quad \bar{a}_{3}= \theta_{0} = a_{3},\quad
	\bar{a}_{4} = \bar{\kappa}_{1} - \bar{\kappa}_{2} = a_{4},
\end{equation*}
as it should be.


\subsection{Discrete Painlev\'e equation d-$P(A_2^{(1)*})$} 
\label{sub:discrete_painlev'e_d_p_a_2}
\subsubsection{Schlesinger Transformations} 
\label{ssub:schlesinger_transformations}

We now consider Fuchsian systems of the spectral type $111,111,111$,
which corresponds to $n=2$ (finite) poles and the matrix size $m=3$. As before, we make
$\operatorname{rank}(\mathbf{A}_{i}) = 2$ using scalar gauge transformations and map the
finite poles to $z_{0} = 0$ and $z_{1}= 1$ by a  M\"obius transformation. Note that this 
example does not have any continuous deformation parameters but it still admits non-trivial 
Schlesinger transformations. So
\begin{equation*}
	\mathbf{A}(z) = \frac{ \mathbf{A}_{0} }{ z } + \frac{ \mathbf{A}_{1} }{ z-1 },
	\qquad 
	\mathbf{A}_{i} = \mathbf{B}_{i} \mathbf{C}_{i}^{\dag} = 
	\begin{bmatrix}
	\mathbf{b}_{i,1} & \mathbf{b}_{i,2}	
	\end{bmatrix} \begin{bmatrix}
		\mathbf{c}_{i}^{1\dag}\\[2pt] \mathbf{c}_{i}^{2\dag}
	\end{bmatrix},
\end{equation*}
and the corresponding \emph{Riemann scheme} and the \emph{Fuchs relation} are
\begin{equation*}
	\left\{
	\begin{tabular}{cccc}
		$z = 0$ 		& $	z = 1 $		& 	$z = \infty$	\\
		$\theta_{0}^{1}$	& $\theta_{1}^{1}$	& 	$\kappa_{1}$ \\
		$\theta_{0}^{2}$	& $\theta_{1}^{2}$	& 	$\kappa_{2}$ \\
		$ 0 $			& 	$ 0$		& 	$\kappa_{3}$ 
	\end{tabular}
	\right\},\qquad 
	\theta_{0}^{1} + \theta_{0}^{2} + \theta_{1}^{1} + \theta_{1}^{2}  + \sum_{j=1}^{3} \kappa_{j}= 0.
\end{equation*}

As in the previous example, we use the global similarity transformations to 
map the vectors $\mathbf{b}_{0,1}$, $\mathbf{b}_{0,2}$, and $\mathbf{b}_{1,1}$ to the standard basis 
vectors and then use the trivial transformations to adjust the scale on the coordinate axes so that all components of
$\mathbf{b}_{1,2}$ are equal to $1$. Then, using the condition $\mathbf{C}_{i}^{\dag} \mathbf{B}_{i} = \mathbf{\Theta}_{i}$,
we get the following parameterization:
\begin{equation*}
	\mathbf{B}_{0} = \begin{bmatrix}
		1 & 0 \\ 0 & 1 \\ 0 & 0
	\end{bmatrix}, \qquad \mathbf{C}_{0}^{\dag} = \begin{bmatrix}
		\theta_{0}^{1} & 0 & \upalpha \\ 0 & \theta_{0}^{2} & \upbeta
	\end{bmatrix},\qquad 
	\mathbf{B}_{1} = \begin{bmatrix}
		0 & 1 \\ 0 & 1 \\ 1 & 1
	\end{bmatrix}, \qquad 
	\mathbf{C}_{1}^{\dag} = \begin{bmatrix}
		-\upgamma - \theta_{1}^{1} & \upgamma & \theta_{1}^{1} \\ \theta_{1}^{2} - \updelta & \updelta & 0
	\end{bmatrix}.
\end{equation*}
Requiring that the eigenvalues of $\mathbf{A}_{\infty}$ are $\kappa_{1}$, $\kappa_{2}$,
and $\kappa_{3}$ results in the equations $\operatorname{tr} (\mathbf{A}_{\infty}) = \kappa_{1} + \kappa_{2} + \kappa_{3}$
(which is just the Fuchs relation), 
$|\mathbf{A}_{\infty}|_{11} + |\mathbf{A}_{\infty}|_{22} + |\mathbf{A}_{\infty}|_{33} = 
\kappa_{2}\kappa_{3} + \kappa_{3}\kappa_{1} + \kappa_{1}\kappa_{2}$ (where $|\mathbf{A}|_{ij}$ denotes the 
$(ij)$-minor of $\mathbf{A}$) and $\operatorname{det}(\mathbf{A}_{\infty}) = \kappa_{1} \kappa_{2} \kappa_{3}$.
The last two equations are linear in $\upalpha$, $\upbeta$, $\upgamma$, and $\updelta$ so we can write them as a 
linear system on $\upalpha$ and $\upbeta$:
\begin{gather}
	(\upgamma + \updelta + \theta_{1}^{1} - \theta_{1}^{2})\upalpha - (\upgamma + \updelta )\upbeta = 
	\kappa_{2}\kappa_{3} + \kappa_{3}\kappa_{1} + \kappa_{1}\kappa_{2} + (\theta_{0}^{2} - \theta_{0}^{1}) \updelta 
	- (\theta_{0}^{2} + \theta_{1}^{1}) (\theta_{0}^{1} + \theta_{1}^{2}) - \theta_{0}^{2} \theta_{1}^{1}),\label{eq:dpa-subs}	\\
	-(\theta_{0}^{2} (\upgamma + \updelta + \theta_{1}^{1} - \theta_{1}^{2}) + 
	\theta_{1}^{2}\upgamma + \theta_{1}^{1} \updelta )	\upalpha +  
	(\theta_{0}^{1}(\upgamma + \updelta) + \theta_{1}^{2}\upgamma + \theta_{1}^{1} \updelta) \upbeta = \kappa_{1} \kappa_{2} \kappa_{3}
	+ \theta_{1}^{1}((\theta_{0}^{1} - \theta_{0}^{2})\updelta  + \theta_{0}^{2} (\theta_{0}^{1} + \theta_{1}^{2})). \notag
\end{gather}
At this point we can choose parameterization variables $x$ and $y$ to simplify the structure of the substitution 
rule. Noticing that the coefficients of the matrix of the above linear system are written in terms of the expressions
$\upgamma + \updelta$, $\upgamma + \updelta + \theta_{1}^{1} - \theta_{1}^{2}$, and 
$\theta_{1}^{2}\upgamma + \theta_{1}^{1} \updelta$, we make the following choice of coordinates:
\begin{equation}
	x =  \frac{ (\upgamma + \updelta)(\theta_{0}^{1} - \theta_{0}^{2}) }{ \theta_{1}^{1} - \theta_{1}^{2} },
	\qquad 
	y = \frac{ \theta_{1}^{2} \upgamma + \theta_{1}^{1} \updelta }{ \upgamma + \updelta +  \theta_{1}^{1} - \theta_{1}^{2}}.
\end{equation}
The reasons for this particular choice are first to simplify the entries of the matrix of the above linear system, and 
second to simplify its determinant, which is the denominator for the expressions $\upalpha(x,y)$ and $\upbeta(x,y)$ --- its zeroes
contribute to the indeterminate points for $\upalpha(x,y)$ and $\upbeta(x,y)$ and through that to the indeterminate points 
of the dynamic as well. We then get 
\begin{align*}
	\upalpha(x,y) &= \frac{ 1 }{ (x + y) (\theta_{1}^{1} - \theta_{1}^{2}) } 
	\left(y r_{1} + \frac{ x(\theta_{0}^{2} r_{1} + r_{2}) }{ x + \theta_{0}^{1} - \theta_{0}^{2}  }\right),\qquad 
	\upbeta(x,y) = \frac{ 1 }{ (x + y) (\theta_{1}^{1} - \theta_{1}^{2}) } \left(
	(y + \theta_{0}^{2}) r_{1} + r_{2}\right),
	\intertext{where}
	r_{1} &= r_{1}(x,y) = \kappa_{1} \kappa_{2} +  \kappa_{2} \kappa_{3} +  \kappa_{3} \kappa_{1} - 
	 (y - \theta_{1}^{2})(x - \theta_{0}^{2}) - \theta_{0}^{1} (y + \theta_{0}^{2}) - 
	\theta_{1}^{1} (\theta_{0}^{1} + \theta_{0}^{2} + \theta_{1}^{2}),\\
	r_{2} &= r_{2}(x,y) =\kappa_{1} \kappa_{2} \kappa_{3} + 
	\theta_{1}^{1}((y - \theta_{1}^{2})(x - \theta_{0}^{2})  + \theta_{0}^{1} (y + \theta_{0}^{2})) 
\end{align*}
are the right-hand-sides of the linear system (\ref{eq:dpa-subs}) (written now in variables $x$ and $y$).

Using the evolution equations~(\ref{eq:bb-cb-generators}---\ref{eq:cb-am}) for the elementary 
Schlesinger transformation $\left\{\begin{smallmatrix} 0&1\\1&1\end{smallmatrix}\right\}$ then gives us the following map 
$\varphi:(x,y)\to (\bar{x},\bar{y})$: 
\begin{equation*}
	\left\{
	\begin{aligned}
		\bar{x} &= \frac{(\upalpha - \upbeta)  (\upalpha x (\theta_{1}^{1} - \theta_{1}^{2}) + 
		(1 + \theta_{0}^{2})(x (y - \theta_{1}^{2}) + y (\theta_{0}^{1} - \theta_{0}^{2})))  }{ 
		(\upalpha - \upbeta) (x (y - \theta_{1}^{2}) + (\theta_{0}^{1} - \theta_{0}^{2})y) - 
		\upalpha (\theta_{1}^{1} + 1)(\theta_{0}^{1} - \theta_{0}^{2}) }\\
		\bar{y} &= \frac{ (\upalpha - \upbeta)(y(x + \theta_{0}^{1} - \theta_{0}^{2}) - \theta_{1}^{2} x)
		 }{ \upalpha (\theta_{0}^{1} - \theta_{0}^{2}) }		
		\label{eq:dpa-21*-sch}
	\end{aligned}
	\right.\quad,		
\end{equation*}
where we still need to substitute $\upalpha=\upalpha(x,y)$ and $\upbeta = \upbeta(x,y)$, but this can be easily done 
using computer algebra.
%
We claim that the resulting dynamical system has the type $A_{2}^{(1)*}$, as expected. Indeed, the Okamoto 
surface for the map $\psi: (x,y)\to (\bar{x},\bar{y})$ is given by the blow-up diagram
\begin{center}
	\includegraphics[width=5.5in]{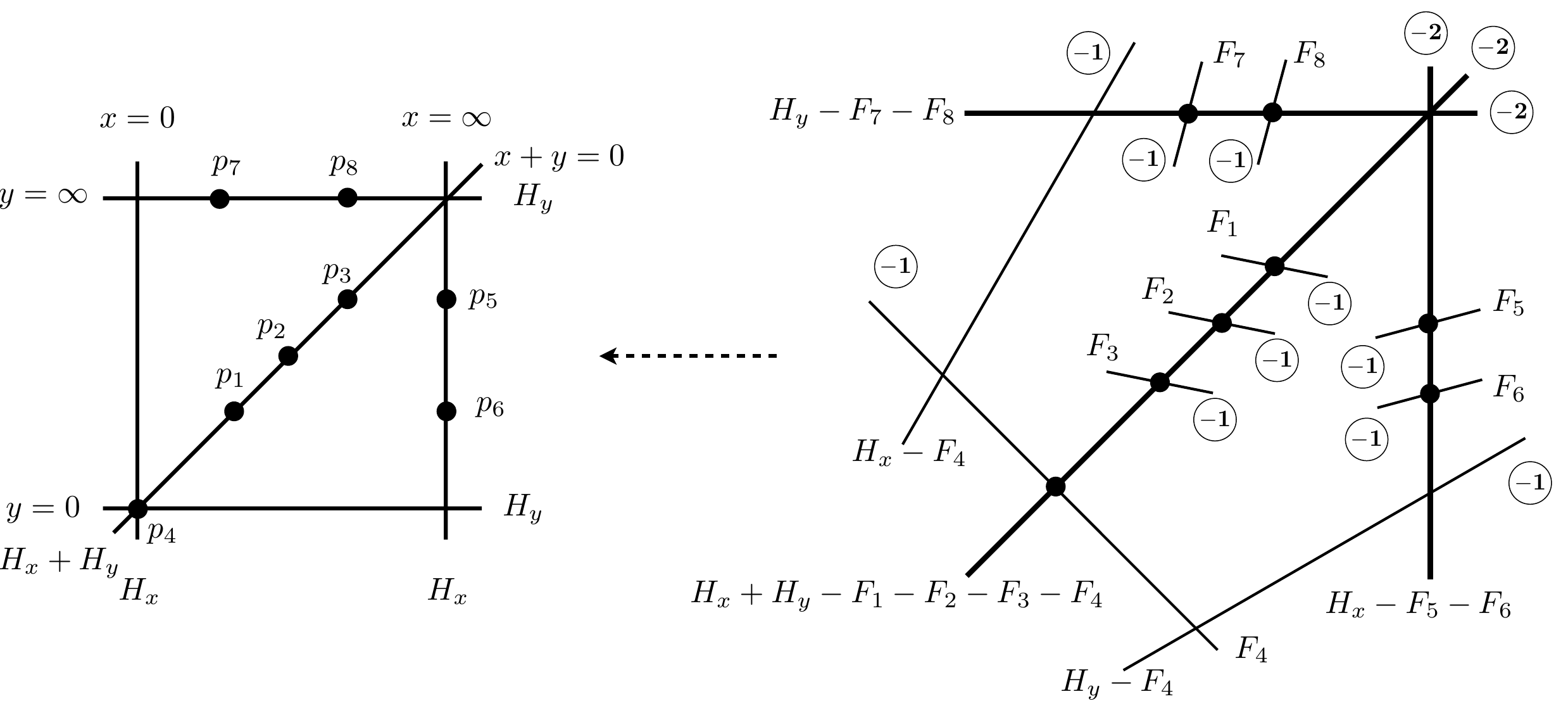}\quad,
\end{center}
where the coordinates of the blow-up points are 
\begin{alignat*}{4}
	&p_{1}(\theta_{0}^{2} + \kappa_{1}, - \theta_{0}^{2} - \kappa_{1}), & \quad 
	&p_{3}(\theta_{0}^{2} + \kappa_{3}, - \theta_{0}^{2} - \kappa_{3}), &\quad
	&p_{5}\left(\infty,\theta_{1}^{1}\right),&\quad 
	&p_{7}\left(\theta_{0}^{2} - \theta_{0}^{1},\infty\right),\\
	&p_{2}(\theta_{0}^{2} + \kappa_{2}, - \theta_{0}^{2} - \kappa_{2}), &\quad
	&p_{4}(0,0),&\quad
	&p_{6}\left(\infty,\theta_{1}^{2}\right),&\quad 
	&p_{8}\left(\theta_{0}^{2}+1,\infty\right).
\end{alignat*}
The anti-canonical divisor $-K_{X}=2 H_{f} + 2 H_{g} - \sum_{i=1}^{8} E_{i}$ 
uniquely decomposes as a positive linear combination of $-2$-curves $D_{i}$,
\begin{equation*}
-K_{X}=D_{0} + D_{1} + D_{2},
\quad\text{ where }	D_{0} = H_{x} + H_{y} - F_{1} - F_{2} - F_{3} - F_{4},
\quad D_{1} = H_{x} - F_{5} - F_{6}, \quad D_{2} = H_{y} - F_{7} - F_{8}. 
\end{equation*}
So we see that the configuration of components $D_{i}$ is indeed described by the 
Dynkin diagram of type $A_{2}^{(1)}$. To this diagram
correspond two different types of surfaces, the generic one corresponds to the multiplicative
system of type $A_{2}^{(1)}$, and the degenerate configuration, where all three components $D_{i}$ 
intersect at one point,
corresponds to the additive system denoted by $A_{2}^{(1)*}$, this is our case (on the picture 
below we use the equivalent description of $\mathbf{X}$ as a blow-up of $\mathbb{P}^{2}$ instead of 
$\mathbb{P}^{1} \times \mathbb{P}^{1}$, since all lines are then equivalent):
\begin{center}
	\begin{tabular}{ccc}
		\includegraphics[height=1.25in]{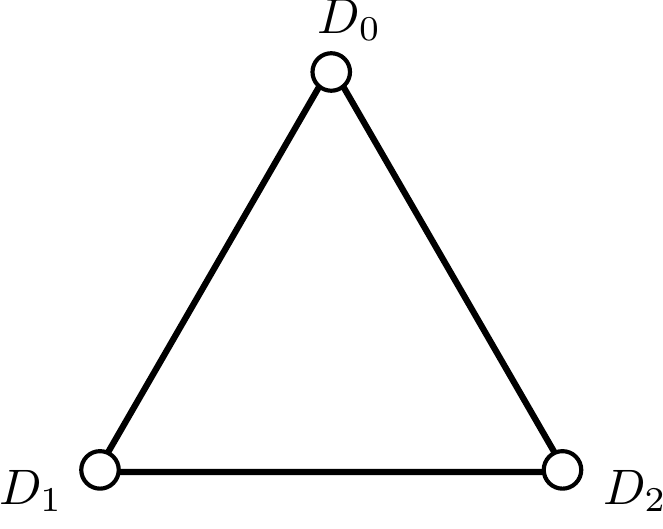} & 
		\includegraphics[height=1.2in]{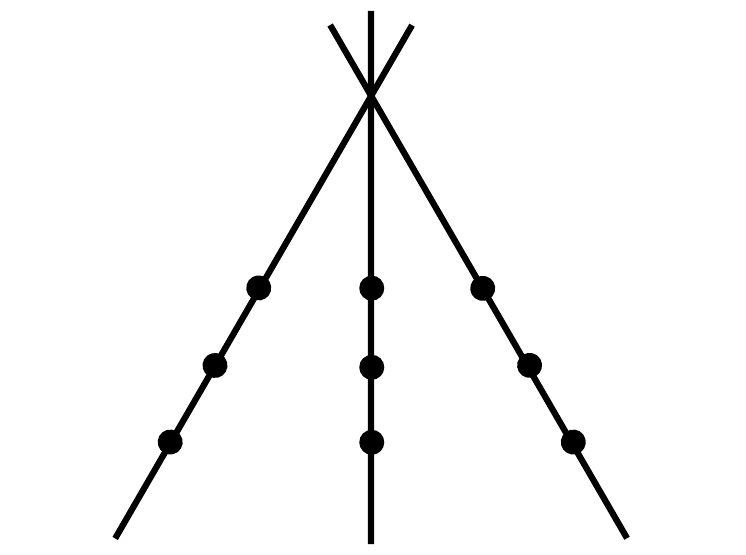} & 
		\includegraphics[height=1.2in]{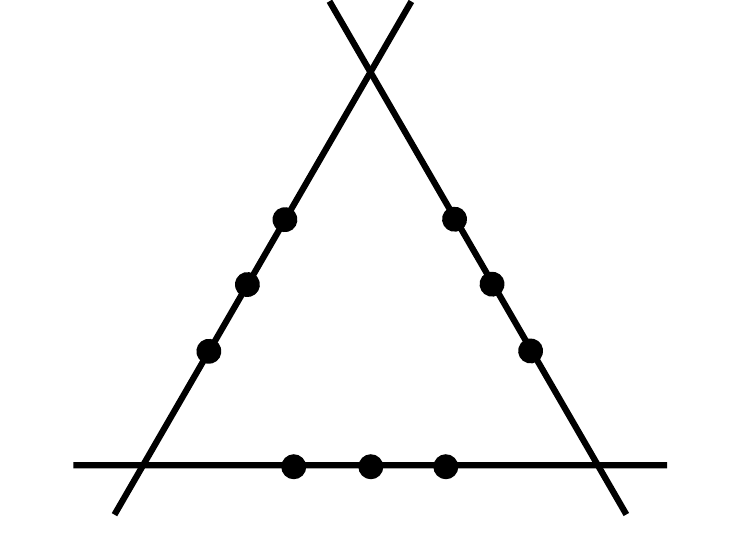} \\
		Dynkin diagram $A_{2}^{(1)}$ & $A_{2}^{(1)*}$-surface & $A_{2}^{(1)}$-surface.
	\end{tabular}.
\end{center}
Thus, we have showed that an elementary Schlesinger transformation of this Fuchsian system
reduces to a discrete Painlev\'e equation of type d-$P(A_{2}^{(1)*})$.
We now want to compare the Schlesinger dynamic with the standard Painlev\'e dynamic of this 
type.

\subsubsection{Comparison with the standard case} 
\label{ssub:geometry_of_d_p_a__2_1_}
We take as the model equation for the d-$P(A_{2}^{(1)*})$ dynamic the following 
equation written by Grammaticos, Ramani, and Ohta, \cite{GraRamOht:2003:AUDOTAQVADIEATST}, see also
Murata \cite{Mur:2004:NEFDPE} and Sakai \cite{Sak:2007:PDPEATLF}.
We consider d-$P(A_{2}^{(1)*})$ to be a
birational map $\varphi: \mathbb{P}^{1}\times \mathbb{P}^{1} \dashrightarrow \mathbb{P}^{1}\times \mathbb{P}^{1}$ 
with parameters $b_{1},\dots, b_{8}$ 
\begin{equation}
	\left(\begin{matrix}
		b_{1} & b_{2} & b_{3} & b_{4}\\
		b_{5} & b_{6} & b_{7} & b_{8}
	\end{matrix}; f,g\right) \mapsto 
	\left(\begin{matrix}
		\bar{b}_{1} & \bar{b}_{2} & \bar{b}_{3} & \bar{b}_{4}\\
		\bar{b}_{5} & \bar{b}_{6} & \bar{b}_{7} & \bar{b}_{8}
	\end{matrix}; \bar{f},\bar{g}\right),
	\quad
	\begin{aligned}
		\bar{b}_{1} &= b_{1} ,\quad & 
		\bar{b}_{3} &= b_{3}, \quad & 
		\bar{b}_{5} &= b_{5} + \delta ,\quad & 
		\bar{b}_{7} &= b_{7} - \delta ,\\
		\bar{b}_{2} &= b_{2} \quad & 
		\bar{b}_{4} &= b_{4} \quad & 
		\bar{b}_{6} &= b_{6} + \delta, \quad & 
		\bar{b}_{8} &= b_{8} - \delta,
	\end{aligned}
\end{equation} 
$\delta = b_{1} + \cdots + b_{8}$, and   $\bar{f}$ and $\bar{g}$ are given by the equation
\begin{equation}
	\left\{
	\begin{aligned}
		(f + g)(\bar{f}+g) & =\frac{(g+b_1)(g+b_2)(g+b_3)(g+b_4)}{(g-b_5)(g-b_6)}\\
		(\bar{f}+g)(\bar{f}+\bar{g})& =\frac{(\bar{f}-b_1)(\bar{f}-b_2)(\bar{f}-b_3)(\bar{f}-b_4)}{(\bar{f}+b_7-\delta)(\bar{f}+b_8-\delta)}
	\end{aligned}
	\right..\label{eq:dpA2-st}	
\end{equation}
Note  that equation~(\ref{eq:dpA2-st}) is the asymmetric version of equation~(1.3) in \cite{GraRamOht:2003:AUDOTAQVADIEATST}.
We see that the resulting blow-up diagram  matches the one that we obtained for the Schlesinger transformation dynamic. 
\begin{center}
	\includegraphics[width=5in]{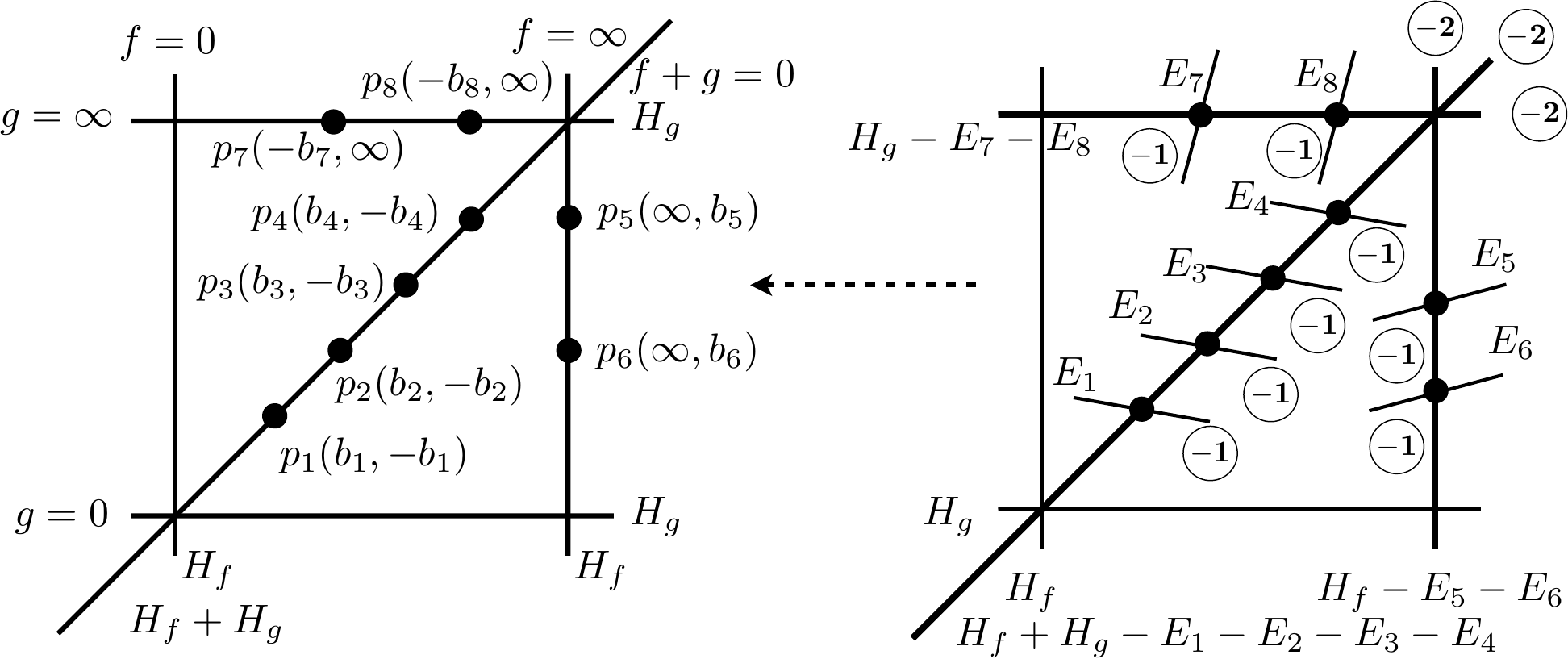}
\end{center}
Thus, we can put $\mathcal{H}_{f} = H_{x}$ and $\mathcal{H}_{g} = H_{y}$ and let $f=x$, $g=y$. We also put 
$\mathcal{E}_{i} = F_{i}$ and use this to identify Schlesinger parameters with parameters $b_{i}$:
\begin{equation*}
	b_{1} = \theta_{0}^{2} + \kappa_{1},\quad 
	b_{2} = \theta_{0}^{2} + \kappa_{2},\quad
	b_{3} = \theta_{0}^{2} + \kappa_{3},\quad
	b_{4} = 0, \quad
	b_{5} = \theta_{1}^{1},\quad
	b_{6} = \theta_{1}^{2},\quad
	b_{7} = \theta_{0}^{1} - \theta_{0}^{2},\quad
	b_{8} = - \theta_{0}^{2} - 1.
\end{equation*}
The dynamic itself, however, is different. There are two ways to see that. First, in view of the above identification 
of parameters we see that the dynamic on the Riemann Schemes is different.
Indeed, since $\delta = b_{1} + \cdots + b_{8} = -1$, the induced action of the standard Painlev\'e dynamic on 
the  Riemann scheme is
\begin{align*}
	\left\{
	\begin{tabular}{cccc}
		$z = 0$ 		& $	z = 1 $		& 	$z = \infty$	\\
		$\theta_{0}^{1}$	& $\theta_{1}^{1}$	& 	$\kappa_{1}$ \\
		$\theta_{0}^{2}$	& $\theta_{1}^{2}$	& 	$\kappa_{2}$ \\
		$ 0 $			& 	$ 0$		& 	$\kappa_{3}$ 
	\end{tabular}
	\right\}		&	\overset{\text{d-$P(A_2^{(1)*})$}}{\strut\longmapsto}
	\left\{
	\begin{tabular}{cccc}
		$z = 0$ 		& $	z = 1 $		& 	$z = \infty$	\\
		$\theta_{0}^{1}$	& $\theta_{1}^{1}-1$	& 	$\kappa_{1}+1$ \\
		$\theta_{0}^{2}-1$	& $\theta_{1}^{2}-1$	& 	$\kappa_{2}+1$ \\
		$ 0 $			& 	$ 0$		& 	$\kappa_{3}+1$ 
	\end{tabular}
	\right\},
	\intertext{whereas our elementary Schlesinger transformation acts as }
	\left\{
	\begin{tabular}{cccc}
		$z = 0$ 		& $	z = 1 $		& 	$z = \infty$	\\
		$\theta_{0}^{1}$	& $\theta_{1}^{1}$		& 	$\kappa_{1}$ \\
		$\theta_{0}^{2}$	& $\theta_{1}^{2}$		& 	$\kappa_{2}$ \\
		$0$			& 	$0$					& 	$\kappa_{3}$ 
	\end{tabular}
	\right\} &\overset{\left\{\begin{smallmatrix} 0&1\\1&1\end{smallmatrix}\right\}}{\strut\longmapsto}
	\left\{
	\begin{tabular}{cccc}
		$z = 0$ 		& $	z = 1 $		& 	$z = \infty$	\\
		$\theta_{0}^{1} - 1$	& $\theta_{1}^{1} + 1$		& 	$\kappa_{1}$ \\
		$\theta_{0}^{2}$	& $\theta_{1}^{2}$		& 	$\kappa_{2}$ \\
		$0$			& 	$0$			 		& 	$\kappa_{3}$ 
	\end{tabular}
	\right\}. 
\end{align*}

Second, we can explicitly compute the induced dynamic for both $\varphi_{*}$ and $\psi_{*}$ on the Picard lattice. 
Then we find out that the action on the components of the anti-canonical divisor is the same and is
a permutation 
$(D_{0}D_{1}D_{2})$, but the resulting translations on the symmetry sub-lattice are different. 
Indeed, take the symmetry sub-lattice $R^{\perp}=\operatorname{Span}_{\mathbb{Z}}\{\alpha_{0},\dots,\alpha_{6}\}$, where
\begin{equation*}
	\begin{aligned}
		\alpha_{0}&= E_{3} - E_{4},& \quad \alpha_{1}&= E_{2} - E_{3}, \\  
		\alpha_{2}&= E_{1} - E_{2},& \quad  \alpha_{3}&= H_{f} - E_{1} - E_{7},\\
		\alpha_{4}&= E_{7} - E_{8},& \quad  \alpha_{5}&= H_{g} - E_{1} - E_{5}, \\
		\alpha_{6}&= E_{5} - E_{6}
	\end{aligned}\qquad\qquad
		\raisebox{-0.5in}{\includegraphics[height=1.1in]{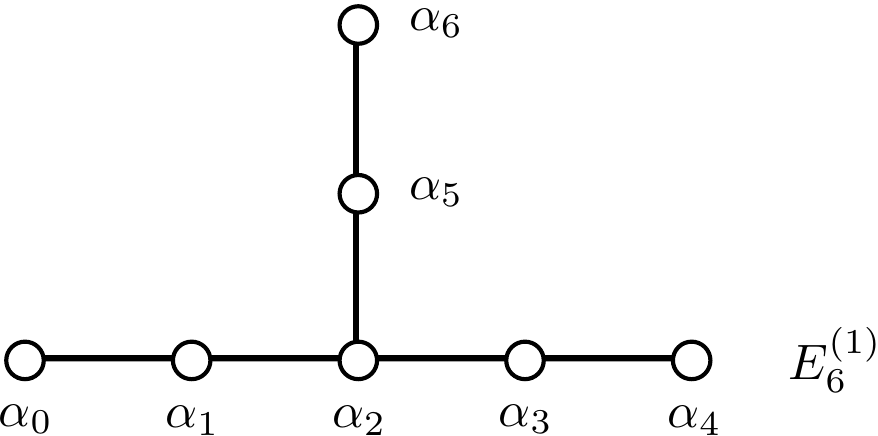}},
\end{equation*}
which is of the type $E_{6}^{(1)}$. 
Then  $\varphi_{*}$ and $\psi_{*}$ act on $R^{\perp}$ as translations in two \emph{different} directions:
\begin{align*}
	\psi_{*}: (\alpha_{0}, \alpha_{1}, \alpha_{2}, \alpha_{3}, \alpha_{4}, \alpha_{5}, \alpha_{6})&\mapsto	
	(\alpha_{0}, \alpha_{1}, \alpha_{2}, \alpha_{3}, \alpha_{4}, \alpha_{5}, \alpha_{6}) + 
	(0,0,0,-1,1,1,-1)\delta,\\
	\varphi_{*}: (\alpha_{0}, \alpha_{1}, \alpha_{2}, \alpha_{3}, \alpha_{4}, \alpha_{5}, \alpha_{6})&\mapsto	
	(\alpha_{0}, \alpha_{1}, \alpha_{2}, \alpha_{3}, \alpha_{4}, \alpha_{5}, \alpha_{6}) + 
	(0,0,0,1,0,-1,0)\delta,	
\end{align*}
where in both cases $\delta = -K_{X}$. Hence, these two dynamics are \emph{not equivalent} and we can not 
transform the elementary Schlesinger transformation dynamic into the standard form by a change of variables.

In fact, we can represent the standard Painlev\'e 
dynamic as a composition of two different Schlesinger transformations, combined with some automorphisms
of our Fuchsian system, as follows.

Let $\sigma_{i}(j,k)$ be the map that exchanges the $j$-th and the $k$-th eigenvectors of $\mathbf{A}_{i}$. 
If the eigenvalues $\theta_{i}^{j}$ and $\theta_{i}^{k}$ are non-zero, this map is just a permutation 
on the decomposition space $\mathcal{B} \times \mathcal{C}$. For this example we need $\sigma_{0}(1,3)$, 
but it also defines a mapping on $\mathcal{B} \times \mathcal{C}$, since 
$\mathbf{b}_{0}^{3} \in \operatorname{Ker}(\mathbf{C}_{0}^{\dag})$
and $\mathbf{c}_{0}^{3\dag}\in \operatorname{Ker}(\mathbf{B}_{0})$. Combining $\sigma_{0}(1,3)$ with the 
scalar gauge transformation $\rho_{0}(\theta_{0}^{1})$, where  
$\rho_{i}(s):\mathbf{A}_{i}\mapsto \mathbf{A}_{i} - s \mathbf{I},\ \mathbf{A}_{\infty}\mapsto \mathbf{A}_\infty + s \mathbf{I}$, 
see \textbf{Assumption}~\ref{assume:rank-reduce},
we get the  action $\Sigma_{0}(1,3) = \rho_{0}(\theta_{0}^{1}) \circ \sigma_{0}(1,3)$ 
on the spectral data (as encoded in the Riemann scheme) and the decomposition space:
\begin{align*}
	\Sigma_{0}(1,3) &: \left\{
	\begin{tabular}{cccc}
		$ z_0$ 		& $	z_1 $			& 	$ \infty$	\\
		$\theta_{0}^{1}$	& $\theta_{1}^{1}$		& 	$\kappa_{1}$ \\
		$\theta_{0}^{2}$	& $\theta_{1}^{2}$		& 	$\kappa_{2}$ \\
		$0$			& 	$0$					& 	$\kappa_{3}$ 
	\end{tabular}
	\right\} \overset{\sigma_{0}(1,3)}{\strut\longmapsto}
	\left\{
	\begin{tabular}{cccc}
		$ z_0$ 		& $	z_1 $			& 	$  \infty$	\\
		$0$	& $\theta_{1}^{1} $		& 	$\kappa_{1}$ \\
		$\theta_{0}^{2}$	& $\theta_{1}^{2}$		& 	$\kappa_{2}$ \\
		$\theta_{0}^{1}$			& 	$0$			 		& 	$\kappa_{3}$ 
	\end{tabular}
	\right\} 
	\overset{\rho_{0}(\theta_{0}^{1})}{\strut\longmapsto}
	\left\{
	\begin{tabular}{cccc}
		$ z_0$ 		& $	z_1 $			& 	$ \infty$	\\
		$-\theta_{0}^{1} $	& $\theta_{1}^{1} $		& 	$\kappa_{1} + \theta_{0}^{1}$ \\
		$\theta_{0}^{2}-\theta_{0}^{1}$	& $\theta_{1}^{2}$		& 	$\kappa_{2} + \theta_{0}^{1}$ \\
		$0$			& 	$0$			 		& 	$\kappa_{3} + \theta_{1}^{1}$ 
	\end{tabular}
	\right\}\\
	\Sigma_{0}(1,3) &: \left(\mathbf{b}_{0}^{1},\mathbf{b}_{0}^{2}; \mathbf{c}_{0}^{1\dag},\mathbf{c}_{0}^{2\dag};
	\mathbf{b}_{1}^{1},\mathbf{b}_{1}^{2}; \mathbf{c}_{1}^{1\dag},\mathbf{c}_{1}^{2\dag}\right) \mapsto
	\left((\mathbf{c}_{0}^{1\dag}\times \mathbf{c}_{0}^{2\dag})^{t},\mathbf{b}_{0}^{2}; 
	\frac{ - \theta_{0}^{1} (\mathbf{b}_{0}^{1}\times\mathbf{b}_{0}^{2})^{t} }{ 
	(\mathbf{c}_{0}^{1\dag}\times \mathbf{c}_{0}^{2\dag}) (\mathbf{b}_{0}^{1}\times\mathbf{b}_{0}^{2}) },\mathbf{c}_{0}^{2\dag};
	\mathbf{b}_{1}^{1},\mathbf{b}_{1}^{2}; \mathbf{c}_{1}^{1\dag},\mathbf{c}_{1}^{2\dag}\right),
\end{align*}
where we had to take into account the normalization condition $\mathbf{C}_{0}^{\dag} \mathbf{B}_{0} = \mathbf{\Theta}_{0}$.
Then, looking at the action on the Riemann scheme, we see that  
\begin{equation*}
	\text{d-$P(A_2^{(1)*})$} = 
	\Sigma_{0}(1,3)\circ\left\{\begin{smallmatrix} 1&0\\2&1\end{smallmatrix}\right\}\circ
	\Sigma_{0}(1,3)\circ\left\{\begin{smallmatrix} 1&0\\1&1\end{smallmatrix}\right\}.
\end{equation*}
Using computer algebra, we can verify directly that this composition dynamic satisfies 
equation~(\ref{eq:dpA2-st}).




\section{Conclusions} 
\label{sec:conclusions}
To summarize, we developed the theory of elementary Schlesinger transformations of 
Fuchsian systems. We found explicit \emph{discrete Schlesinger evolution equations} describing
these transformations and showed that these equations can be written in discrete 
Hamiltonian form, when considered as a dynamical system on the decomposition space 
for Fuchsian systems. We found the explicit formula for the discrete Hamiltonian 
for such systems.  We also stressed the parallel between the discrete and 
continuous cases. Finally, we used the machinery developed in the first part of this paper 
to compute two explicit 
examples of reductions from discrete Schlesinger transformations to difference
Painlev\'e equations. In doing so we emphasized the role played by the geometry of the Okamoto spaces of initial 
conditions in doing such reductions, as well as  in comparing different equations of the same
type or difference forms of the same equation. We also gave a partial answer to 
the earlier question of H.~Sakai by explicitly representing the difference Painlev\'e 
equation d-$P\big(A_{2}^{(1)*}\big)$ as such reduction. 


\small
\bibliographystyle{amsalpha}

\providecommand{\bysame}{\leavevmode\hbox to3em{\hrulefill}\thinspace}
\providecommand{\MR}{\relax\ifhmode\unskip\space\fi MR }
\providecommand{\MRhref}[2]{%
  \href{http://www.ams.org/mathscinet-getitem?mr=#1}{#2}
}
\providecommand{\href}[2]{#2}

\end{document}